\documentclass[final,authoryear,5p,times,twocolumn]{elsarticle}

\usepackage{graphicx}
\usepackage{amssymb}

\usepackage{graphicx}
\usepackage{dcolumn}
\usepackage{bm}
\usepackage{color}
\usepackage{graphicx}
\usepackage{multirow}
\usepackage{tabularx}

\begin{document}

\begin{frontmatter}

\title{Extreme value statistics and recurrence intervals of NYMEX energy futures volatility}

\author[BS,RCE,SS]{Wen-Jie Xie}
\author[BS,RCE]{Zhi-Qiang Jiang}
\author[BS,RCE,SS,ICCT]{Wei-Xing Zhou\corref{cor1}}
\cortext[cor1]{Corresponding author. Address: 130 Meilong Road, P.O. Box 114, School of Business,
              East China University of Science and Technology, Shanghai 200237, China, Phone: +86 21 64253634.}
\ead{wxzhou@ecust.edu.cn} %
\ead[url]{http://rce.ecust.edu.cn/}%

\address[BS]{School of Business, East China University of Science and Technology, Shanghai 200237, China}
\address[RCE]{Research Center for Econophysics, East China University of Science and Technology, Shanghai 200237, China}
\address[SS]{Department of Mathematics, East China University of Science and Technology, Shanghai 200237, China}
\address[ICCT]{Key Laboratory of Coal Gasification and Energy Chemical Engineering (MOE), East China University of Science and Technology, Shanghai 200237, China}

\begin{abstract}
  Energy markets and the associated energy futures markets play a crucial role in global economies. It is of great theoretical and practical significance to gain a deeper understanding of extreme value statistics of the volatility of energy futures traded on the New York Mercantile Exchange (NYMEX). We investigate the statistical properties of the recurrence intervals of daily volatility time series of four NYMEX energy futures, which are defined as the waiting times $\tau$ between consecutive volatilities exceeding a given threshold $q$. We find that the recurrence intervals are distributed as a stretched exponential $P_q(\tau)\sim e^{(a\tau)^{-\gamma}}$, where the exponent $\gamma$ decreases with increasing $q$, and there is no scaling behavior in the distributions for different thresholds $q$ after the recurrence intervals are scaled with the mean recurrence interval $\bar\tau$. These findings are significant under the Kolmogorov-Smirnov test and the Cram{\'e}r-von Mises test. We show that empirical estimations are in nice agreement with the numerical integration results for the occurrence probability $W_q(\Delta{t}|t)$ of a next event above the threshold $q$ within a (short) time interval after an elapsed time $t$ from the last event above $q$. We also investigate the memory effects of the recurrence intervals. It is found that the conditional distributions of large and small recurrence intervals differ from each other and the conditional mean of the recurrence intervals scales as a power law of the preceding interval $\bar\tau(\tau_0)/\bar\tau \sim (\tau_0/\bar\tau)^\beta$, indicating that the recurrence intervals have short-term correlations. Detrended fluctuation analysis and detrending moving average analysis further uncover that the recurrence intervals possess long-term correlations. We confirm that the ``clustering'' of the volatility recurrence intervals is caused by the long-term correlations well known to be present in the volatility. Our findings shed new lights on the behavior of large volatility and have potential implications in risk management of energy futures.
%
\end{abstract}


\end{frontmatter}


\section{Introduction}
\label{S1:Introduction}

The price behaviors of oil futures and related energy futures play a crucial role in modern economic and financial systems and our everyday life \citep{Jones-Kaul-1996-JF,Sadorsky-1999-EE}. Especially, people are more concerned with large price fluctuations than mild fluctuations for obvious reasons. Since extreme events are rare, it is a difficult task to infer their statistical properties and different mathematical tools have been developed to investigate the statistics of extreme values \citep{Kotz-Nadarajah-2000}. In this work, we adopt the recurrence interval analysis to study the behaviors of large volatility of four energy futures listed on the New York Mercantile Exchange (NYMEX).

The idea of recurrence interval analysis is as follows. Consider a time series of normalized volatility $v(t)$, as defined in Section \ref{S1:Data}. We are interested in the statistical properties of the inter-event times $\tau$ (also called recurrence intervals) between successive ``extreme'' volatilities whose values are greater than a threshold $q$. Rather than working directly on those extreme volatilities, we obtain the statistical properties of recurrence intervals above a series of small thresholds $q$ and try to uncover possible dependence of the statistical properties on the threshold $q$. If clear dependence is observed, we will be able to infer the statistical properties for extreme volatilities associated with large $q$ values.

The recurrence interval analysis has been applied to investigate the extreme value statistics of time series in many fields, such as records of climate \citep{Bunde-Eichner-Havlin-Kantelhardt-2004-PA,Bunde-Eichner-Kantelhardt-Havlin-2005-PRL}, seismic activities \citep{Saichev-Sornette-2006-PRL}, energy dissipation rates of three-dimensional turbulence \citep{Liu-Jiang-Ren-Zhou-2009-PRE}, heartbeat intervals in medicine science \citep{Bogachev-Kireenkov-Nifontov-Bunde-2009-NJP}, precipitation and river runoff \citep{Bogachev-Bunde-2012-EPL}, Internet traffic \citep{Bogachev-Bunde-2009-EPL,Cai-Fu-Zhou-Gu-Zhou-2009-EPL}, financial volatilities \citep{Yamasaki-Muchnik-Havlin-Bunde-Stanley-2005-PNAS}, equity returns \citep{Yamasaki-Muchnik-Havlin-Bunde-Stanley-2006-inPFE,Bogachev-Eichner-Bunde-2007-PRL,Bogachev-Bunde-2008-PRE,Bogachev-Bunde-2009-PRE,Ren-Zhou-2010-NJP,Ludescher-Tsallis-Bunde-2011-EPL,He-Chen-2011b-PA,Meng-Ren-Gu-Xiong-Zhang-Zhou-Zhang-2012-EPL}, and trading volumes \citep{Podobnik-Horvatic-Petersen-Stanley-2009-PNAS,Ren-Zhou-2010-PRE,Li-Wang-Havlin-Stanley-2011-PRE}.

The majority of empirical studies have been carried out on financial volatility. In early years, it is argued that the distribution of the recurrence intervals above a fixed threshold has a power-law tail:
\begin{equation}
   P_q(\tau) \sim \tau^{-(1+\gamma)}.
   \label{Eq:PDF:tau:PL}
\end{equation}
Specifically, \cite{Kaizoji-Kaizoji-2004a-PA} found that the exponent $\gamma$ decreases from 1.81 to 0.97 when the threshold increases from 0.1 to 0.9 for daily volatility for 800 stocks traded on the Tokyo Stock Exchange and decreases from 2.47 to 1.16 when the threshold increases from 0.05 to 0.3 for daily volatility of the Nikkei 225 index, \cite{Yamasaki-Muchnik-Havlin-Bunde-Stanley-2005-PNAS} reported that $\gamma\approx1.0$ for seven representative stocks and currencies which is independent of the threshold, \cite{Lee-Lee-Rikvold-2006-JKPS} observed that $\gamma\approx1.0$ for 1-min volatility of the Korean stock-market index KOSPI, and \cite{Greco-SorrisoValvo-Carbone-Cidone-2008-PA} obtained a power-law distribution for 1-min volatility of the Italian MIB30 index.

However, the overwhelming consensus is that the recurrence intervals of financial volatility are distributed as a stretched exponential:
\begin{equation}
   P_q(\tau) = c e^{-(a\tau)^\gamma},
   \label{Eq:PDF:tau:SE}
\end{equation}
which is supported by a handful of empirical evidence using daily or high-frequency data in developed or emerging stock markets \citep{Wang-Yamasaki-Havlin-Stanley-2006-PRE,Wang-Weber-Yamasaki-Havlin-Stanley-2007-EPJB,Jung-Wang-Havlin-Kaizoji-Moon-Stanley-2008-EPJB,Qiu-Guo-Chen-2008-PA,Jeon-Moon-Oh-Yang-Jung-2010-JKPS,Wang-Wang-2012-CIE}. To our knowledge, the only exception is that \cite{Zhang-Wang-Shao-2010-ACS} used an alternative function $P_q(\tau)\sim e^{-a(\ln{\tau})^\gamma}$ to fit the distribution. The adoption of stretched exponential for modelling recurrence interval distributions can be at least traced back to \cite{Bunde-Eichner-Havlin-Kantelhardt-2003-PA}. It is also interesting to note that stretched exponential distributions are ubiquitous in natural and social sciences \citep{Laherrere-Sornette-1998-EPJB}.

It is important to note that the presence of scaling in the recurrence interval distributions is also a subtle issue. Although early works favor the presence of scaling behaviors, recent studies unveil mixed results showing that some stocks possess scaling behaviors while others exhibit multiscaling behaviors in a same market \citep{Wang-Yamasaki-Havlin-Stanley-2008-PRE,Ren-Zhou-2008-EPL,Wang-Yamasaki-Havlin-Stanley-2009-PRE,Ren-Gu-Zhou-2009-PA,Ren-Guo-Zhou-2009-PA}.

If the volatility process is Poissonian, that is, there are no linear or nonlinear temporal correlations in the volatility time series, then the recurrence intervals are exponentially distributed and have no long-term memory for any threshold \citep{Kotz-Nadarajah-2000}. This condition can be fulfilled if we shuffle the volatility time series to destroy any autocorrelations. There is numerical and analytical evidence verifying that the long-term correlation of the original time series has a remarkable influence on the recurrence interval distribution \citep{Bunde-Eichner-Havlin-Kantelhardt-2003-PA,Bunde-Eichner-Kantelhardt-Havlin-2005-PRL,Altmann-Kantz-2005-PRE,Olla-2007-PRE,Santhanam-Kantz-2008-PRE,Bogachev-Eichner-Bunde-2007-PRL,Bogachev-Eichner-Bunde-2008-EPJST}. Moreover, the exponent $\gamma$ of the stretched exponential distribution of the recurrence intervals is explicitly related to the autocorrelation exponent of the original time series \citep{Bunde-Eichner-Havlin-Kantelhardt-2004-PA,Altmann-Kantz-2005-PRE,Bunde-Eichner-Kantelhardt-Havlin-2005-PRL,Livina-Havlin-Bunde-2005-PRL}.

The remainder of this paper is organized as follows. Section~\ref{S1:Data} describes the data sets of NYMEX energy futures prices we shall investigate and the basic properties of the recurrence interval time series. Section \ref{S1:PDF} investigates the probability distributions of the recurrence intervals of the volatility. We find that there is no scaling in the distribution and the empirical distributions can be well fitted by stretched exponentials. Section \ref{S1:Memory} investigates the memory effects of the recurrence intervals. We show that there are both short-term and long-term correlations in the recurrence interval time series. We summarize our findings in Section \ref{S1:Conclusion}.

\section{Data description}
\label{S1:Data}

We retrieve the daily prices of four NYMEX futures of crude oil (Light-Sweet, Cushing, Oklahoma), reformulated regular gasoline (New York Harbor), No. 2 heating oil (New York Harbor), and propane (Mont Belvieu, Texas). The prices of crude oil are in dollars per barrel, while all others in dollars per gallon. The raw data sets are downloaded from the web site of the U.S. Energy Information Administration. For each energy's futures, we choose ``contract 1'' for analysis. The time periods of the records are 4 April 1983 - 2 October 2012 for crude oil,
3 October 2005 - 2 October 2012 for gasoline, 2 January 1980 - 2 October 2012 for heating oil, and 17 December 1993 - 18 September 2009 for propane, each containing 7401, 5512, 8214 and 3941 data points.

\begin{figure}[htb]
  \centering
  \includegraphics[width=8.6cm]{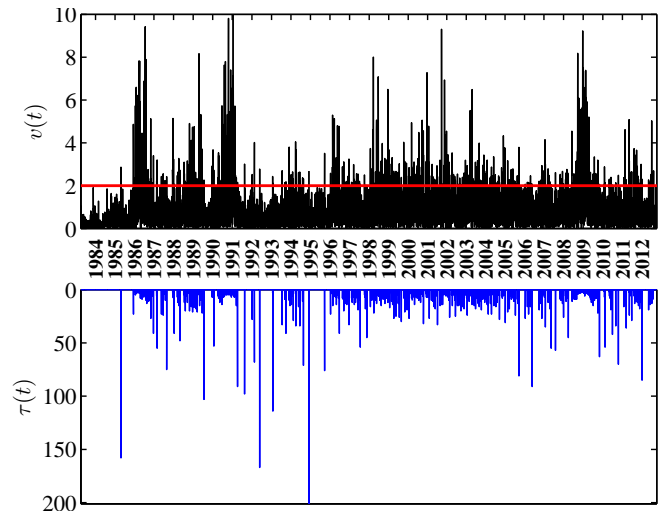}
  \caption{\label{Fig:NYMEX:Energy:RI:Data} (Color online.) The upper panel illustrates the normalized volatility $v(t)$ of the daily crude oil futures prices and the lower panel shows the recurrence intervals $\tau$ between successive normalized volatilities that are larger than a threshold $q=2$.}
\end{figure}

Denote $Y(t)$ the futures price at time $t$. The volatility is defined as the absolute value of the logarithmic return
\begin{equation}
  R(t)=|\ln Y(t)-\ln Y(t-1)|,
  \label{Eq:Volatility}
\end{equation}
and the normalized volatility is determined as follows:
\begin{equation}
  v(t)=\frac{R(t)}{[\langle R(t)^2 \rangle-\langle R(t)\rangle^2]^{1/2}}.
  \label{Eq:NormalizedVolatility}
\end{equation}
The upper panel of Fig.~\ref{Fig:NYMEX:Energy:RI:Data} illustrates the time series of the normalized volatility of the crude oil. There is clear evidence of the volatility clustering phenomenon indicating long-term correlations well documented in \cite{Tabak-Cajueiro-2007-EE}, \cite{Elder-Serletis-2008-RFE} and \cite{Cunado-GilAlana-PerezDeGracia-2010-JFutM}.

\begin{figure*}[htb]
  \centering
  \includegraphics[width=4.5cm]{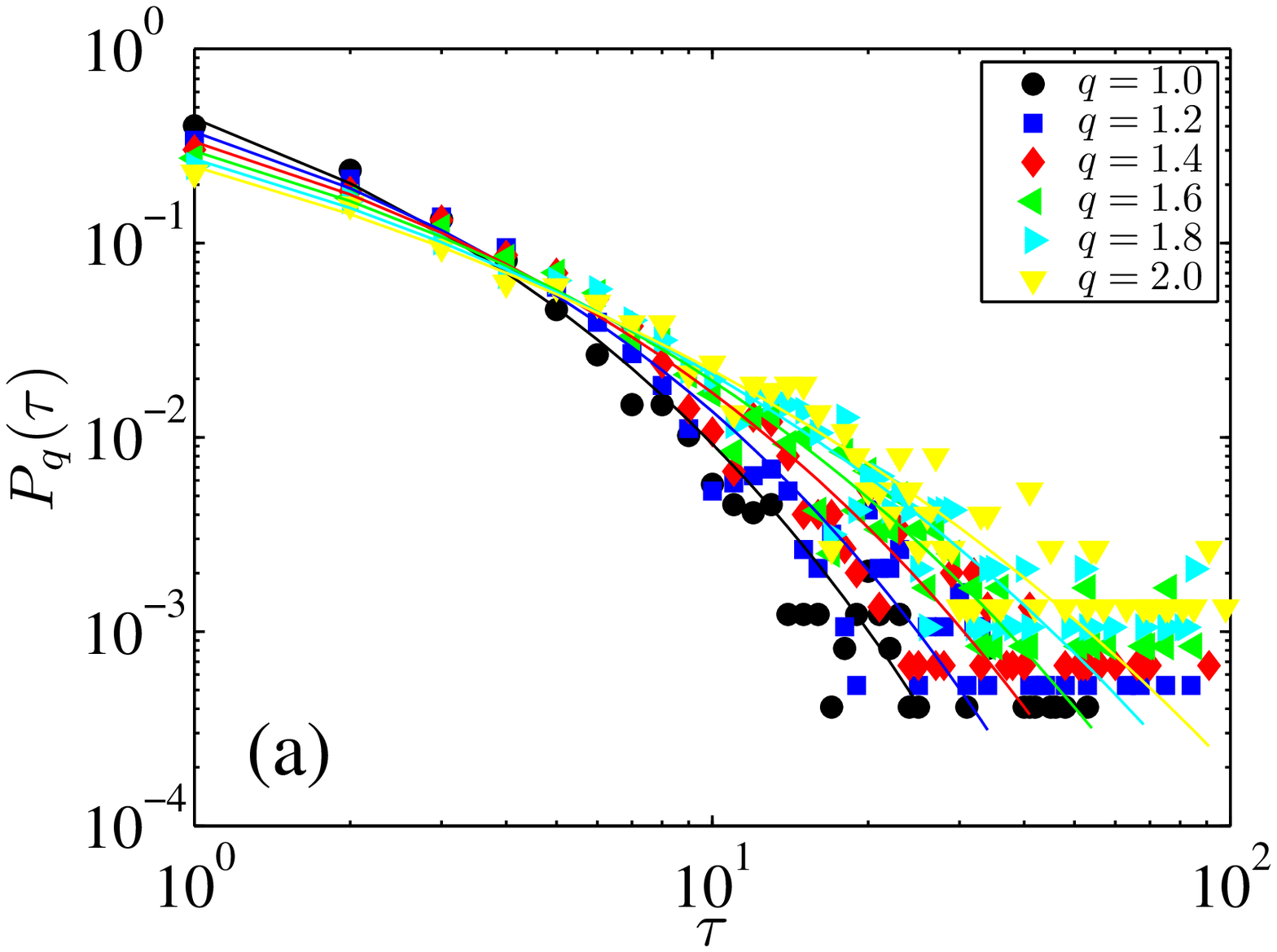}
  \includegraphics[width=4.5cm]{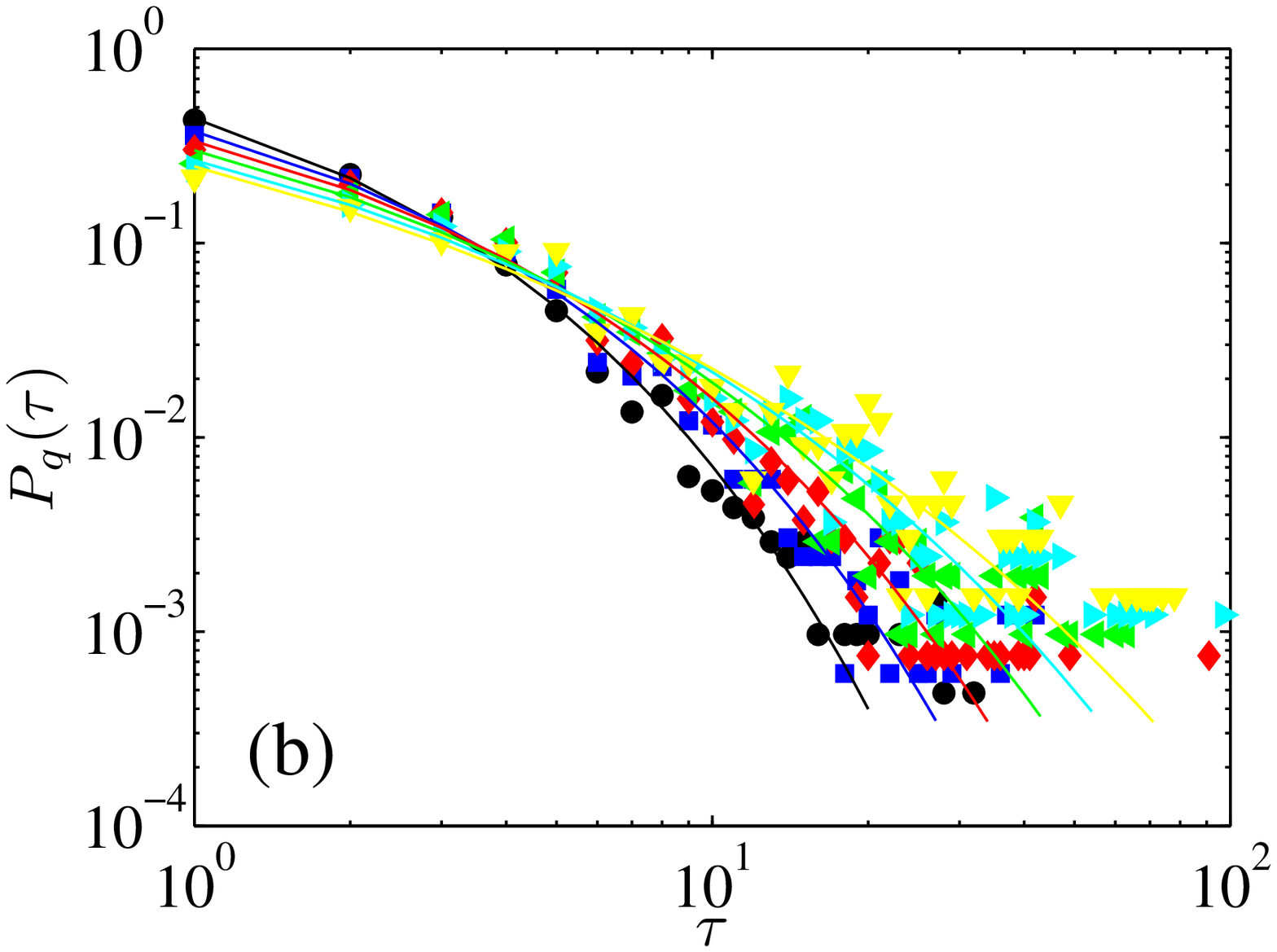}
  \includegraphics[width=4.5cm]{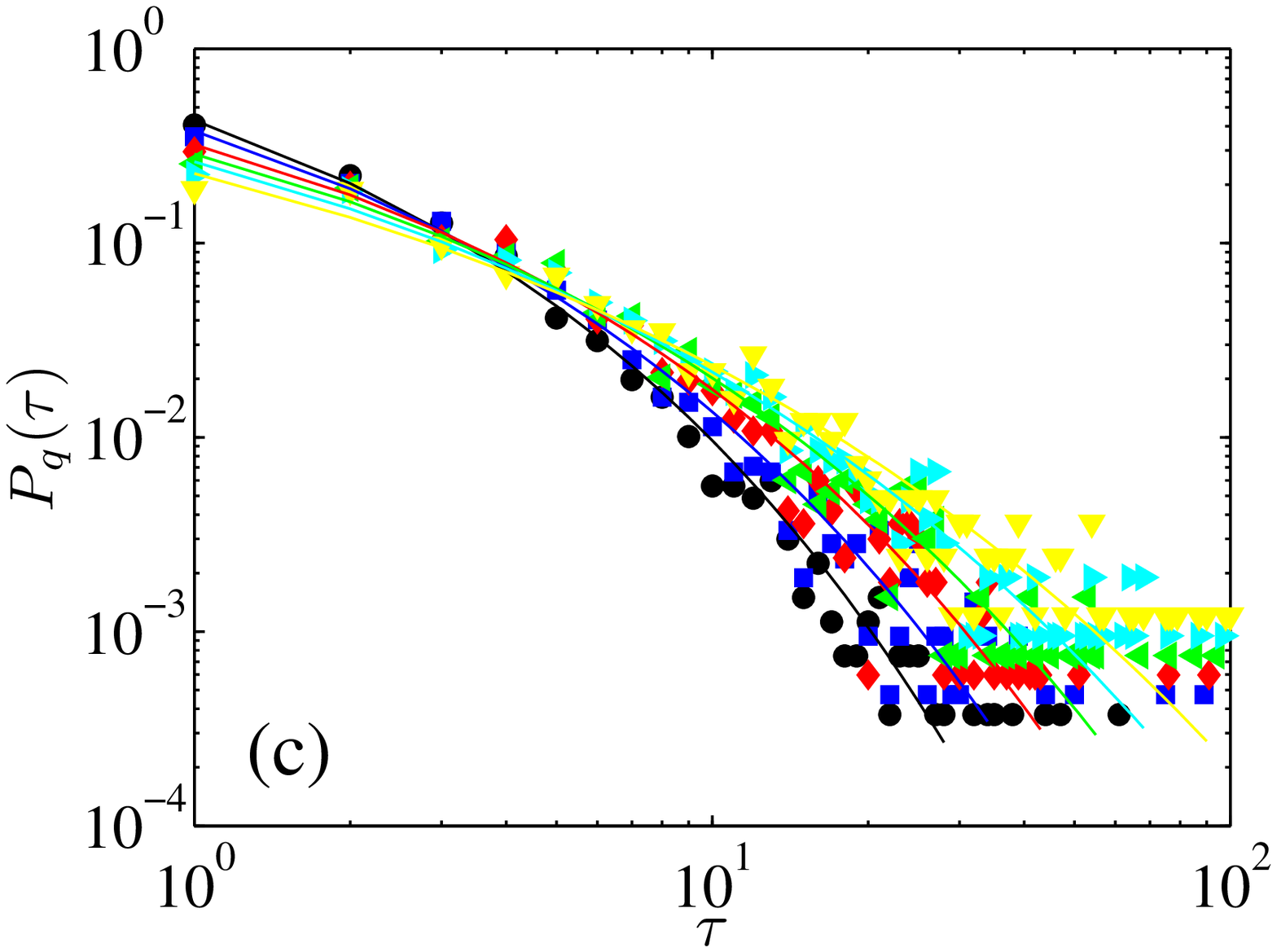}
  \includegraphics[width=4.5cm]{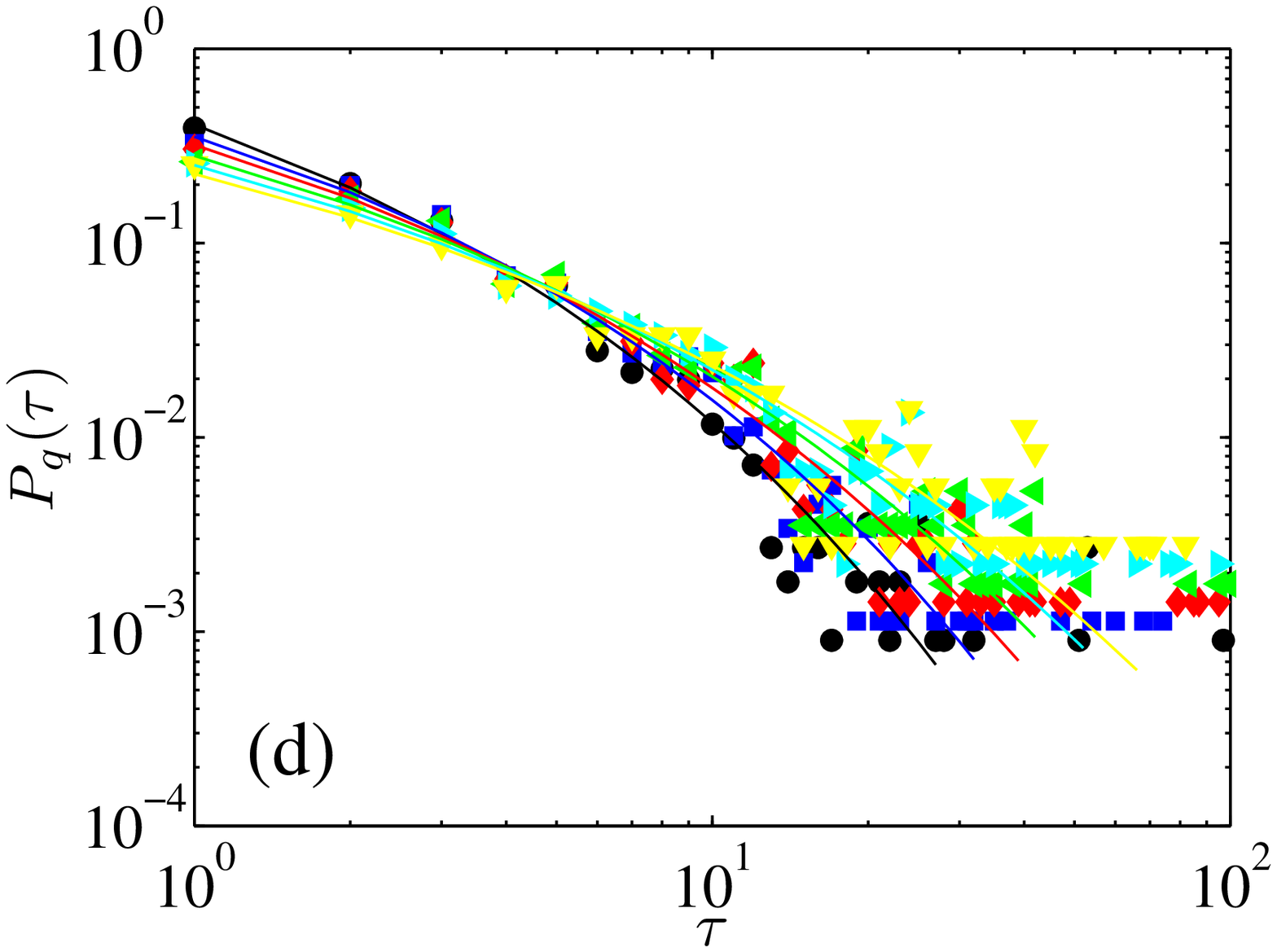}
  \includegraphics[width=4.5cm]{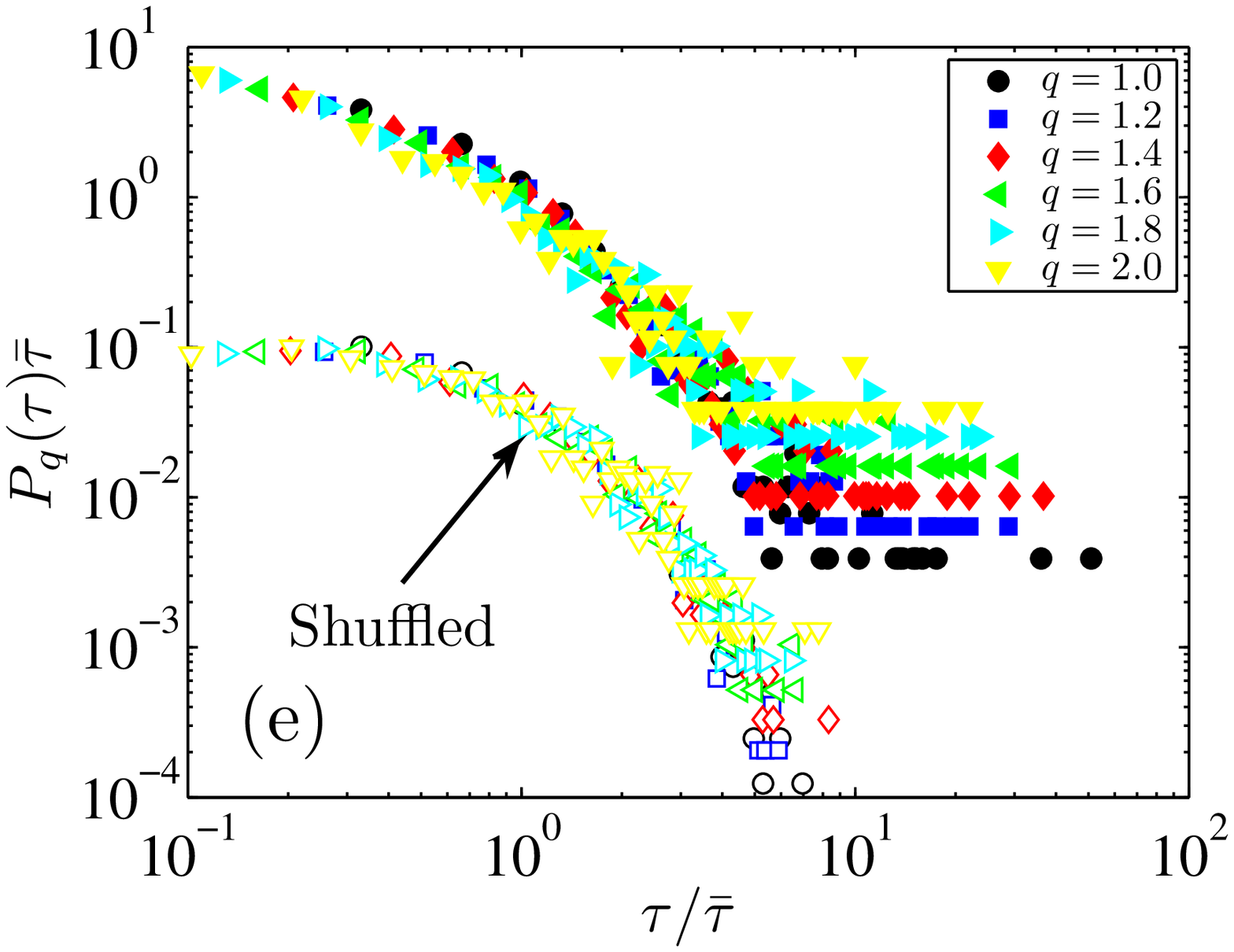}
  \includegraphics[width=4.5cm]{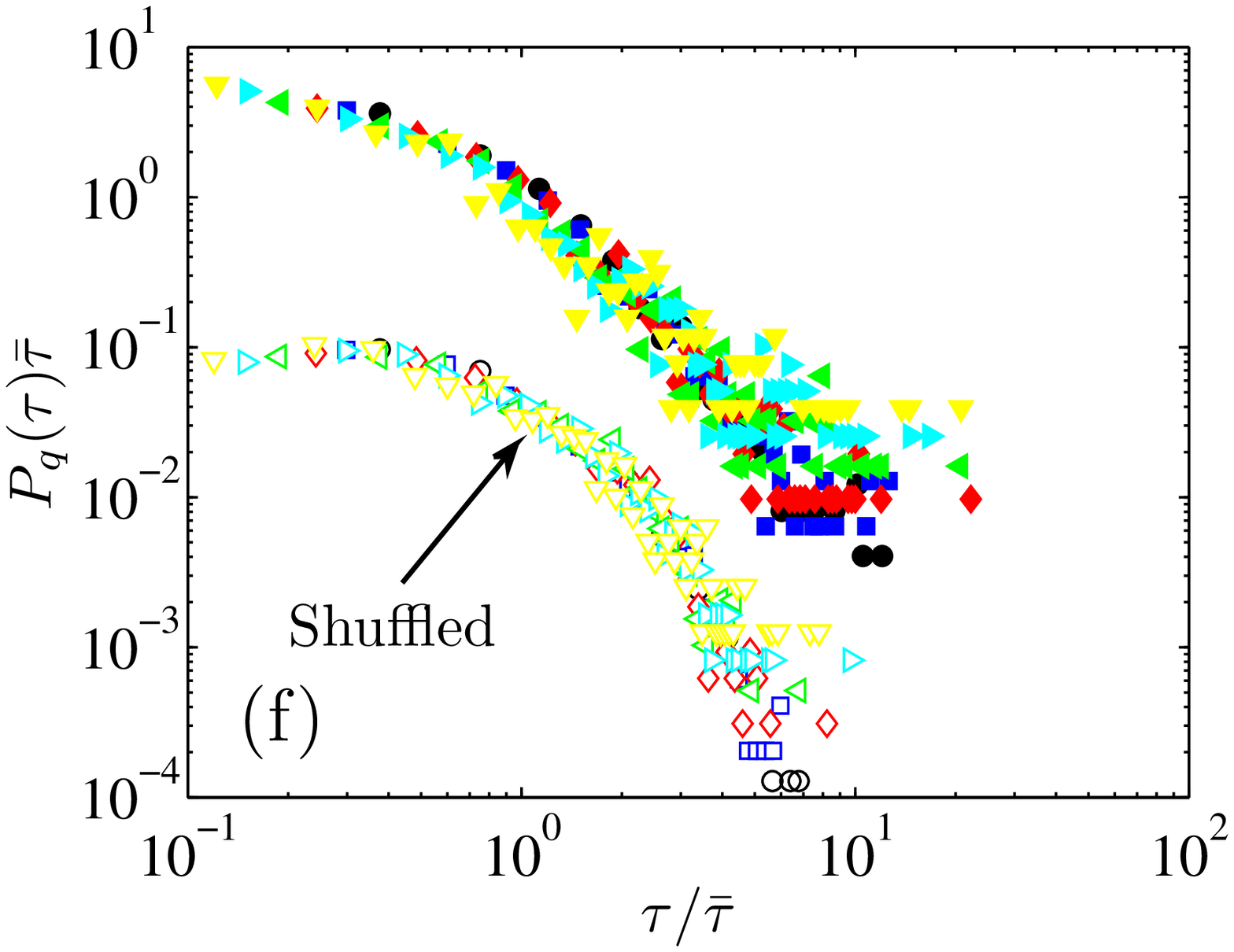}
  \includegraphics[width=4.5cm]{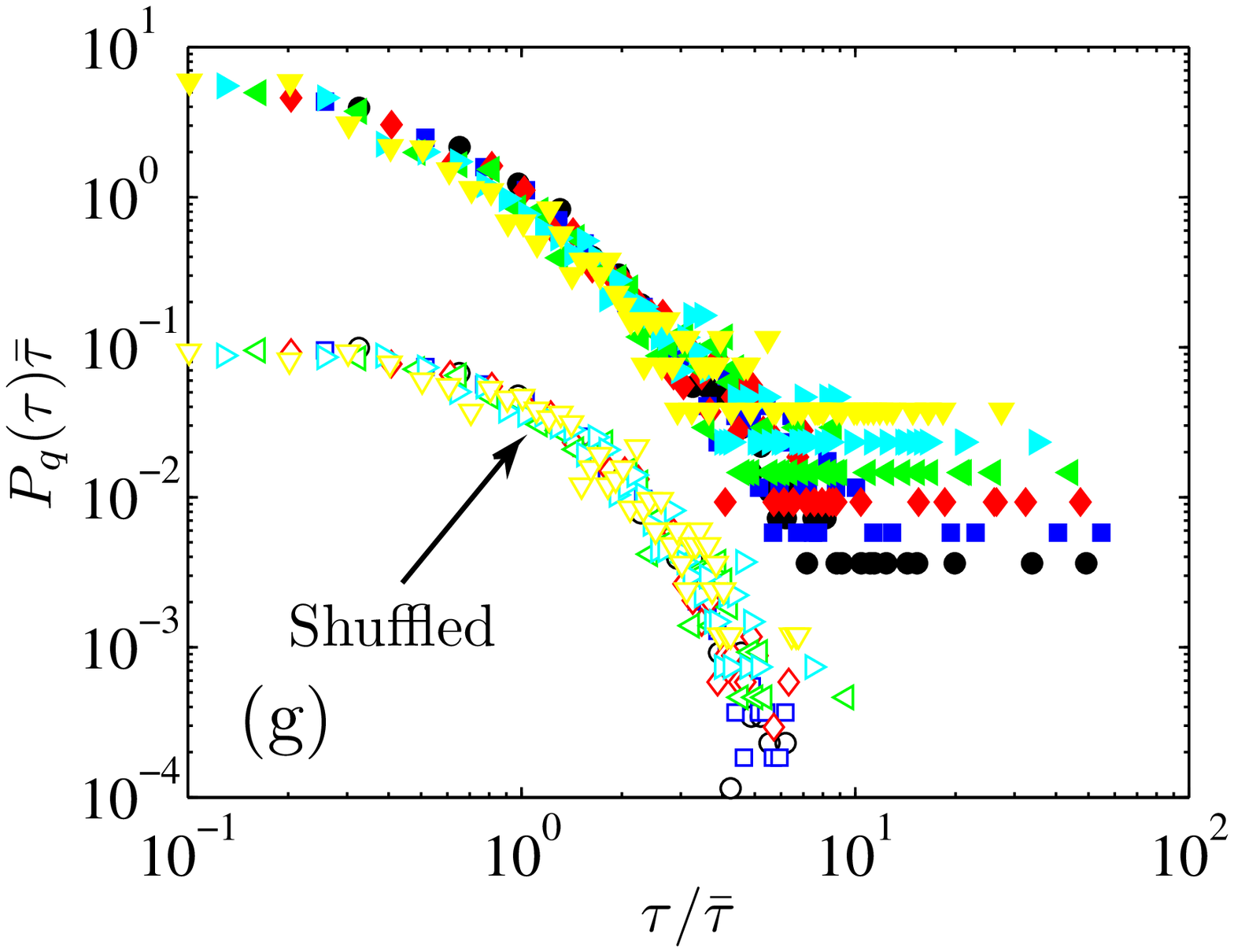}
  \includegraphics[width=4.5cm]{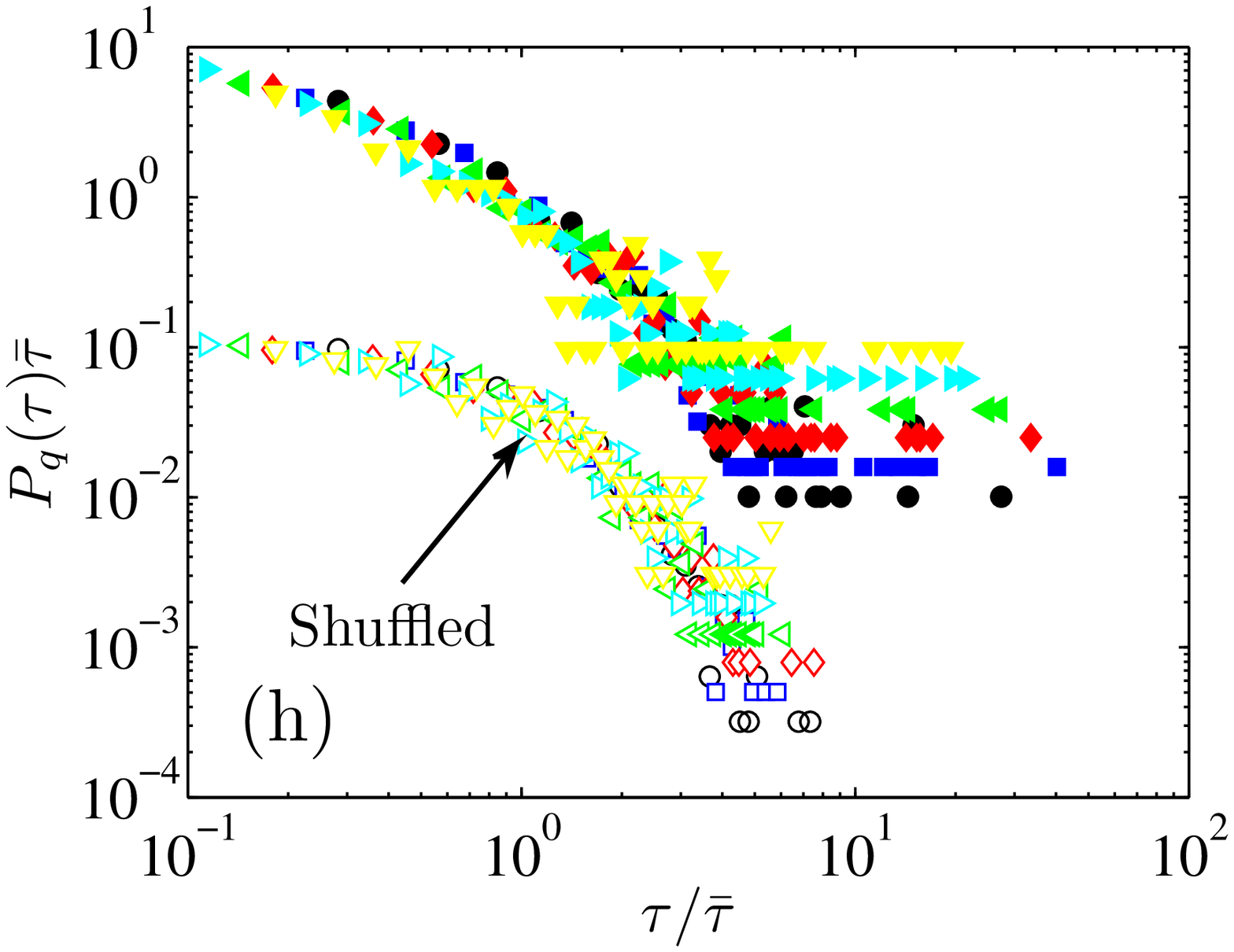}
  \caption{\label{Fig:NYMEX:Energy:RI:PDF} (Color online.) (Top panel) Empirical probability distributions $P_q(\tau)$ of the volatility recurrence intervals  $\tau$ with different thresholds $q$ for the four futures: (a) Crude oil, (b) Gasoline, (c) Heating oil, and (d) Propane. The solid lines are the best fits to the stretched exponential function. (Bottom panel) Scaled probability distributions $P_q(\tau)\bar{\tau}$ of the volatility recurrence intervals $\tau/\bar{\tau}$ with different $q$ values for the four futures: (e) Crude oil, (f) Gasoline, (g) Heating oil, and (h) Propane. Also shown in plots (e-h) are the scaled probability distributions for the shuffled volatility time series, which have been shifted downwards for clarity.}
\end{figure*}

The recurrence intervals $\tau$ are then determined as the waiting times between successive volatilities exceeding a given threshold $q$. Assume that the volatility at day $t$ is greater than $q$: $v(t)>q$. Then the recurrence interval is
\begin{equation}
  \tau(t) = \min\{t'-t: v(t')>q, t'>t\}.
\end{equation}
The lower panel of Fig.~\ref{Fig:NYMEX:Energy:RI:Data} shows the recurrence interval time series obtained for $q=2$, where each recurrence interval $\tau(t)$ is drawn at the time $t$ of the first volatility $v(t)$. During volatile time periods when large volatilities cluster, the recurrence intervals are small and dense. In contrast, during calm periods with small volatility, the recurrence intervals are large and sparse. These observations are consistent with the fact that the temporal structure of the original volatility time series has significant impacts on the temporal structure and distribution of the recurrence interval time series.

\section{Empirical probability distributions}
\label{S1:PDF}

\subsection{Tests of scaling in scaled probability distributions}

For each threshold $q$, we can obtain a sequence of recurrence intervals for each volatility time series. The empirical probability distribution of the sequence of recurrence intervals can be determined for each $q$. Figure~\ref{Fig:NYMEX:Energy:RI:PDF}(a) illustrates the empirical probability distributions $P_q(\tau)$ of the volatility recurrence intervals  $\tau$ associated with different thresholds $q$ for the WTI crude oil futures. Similarly, the empirical PDFs for gasoline, heating oil and propane futures are shown in Fig.~\ref{Fig:NYMEX:Energy:RI:PDF}(b-d). With the increase of the threshold there are more large recurrence intervals and less small recurrence intervals. It is also found that all the curves in Fig.~\ref{Fig:NYMEX:Energy:RI:PDF}(a-d) for different thresholds and different equities have a similar shape.

We are interested in the presence of possible scaling behaviors in the PDFs. Following \cite{Yamasaki-Muchnik-Havlin-Bunde-Stanley-2005-PNAS}, we investigate the scaled recurrence intervals $\tau/\bar\tau$, whose probability distribution $f_q(\tau/\bar\tau)$ is as follows
\begin{equation}
  f_q(\tau/\bar\tau)=P_q(\tau)\bar\tau,
  \label{Eq:Pq:f}
\end{equation}
where $\bar\tau$ is the mean return interval that depends on the threshold $q$. The scaled distributions are presented in Fig.~\ref{Fig:NYMEX:Energy:RI:PDF}(e-h). We do observe that the curves of $f_q(\tau/\bar\tau)$ have a better collapsing than the original curves of $P_q(\tau)$. However, it is not clear if there is a scaling behavior. If we shuffle the volatility time series, the resulting recurrence intervals are exponentially distributed for any threshold, as depicted in the lower panel of Fig.~\ref{Fig:NYMEX:Energy:RI:PDF}.

Theoretically, if $f_q(\tau/\bar\tau)$ is independent of $q$, the scaled PDFs are regarded to exhibit scaling behaviors. Expressing alternatively, $f_q$ has a scaling behavior if and only if for any pair of $q_i$ and $q_j$ the two corresponding sequences of recurrence intervals have the same distribution, that is, $f_{q_i}=f_{q_j}$. In this work, we adopt the well established Kolmogorov-Smirnov (KS) test to check if $f_{q_i}=f_{q_j}$.

The standard KS test is designed to test the hypothesis that the distribution of the empirical data is equal to a particular distribution by comparing their cumulative distribution functions, which can also be applied to test if two samples have the same probability distribution. Denote $F_{q_i}$ the cumulative probability function of $f_{q_i}$. We calculate the KS statistic by comparing the two CDFs in the overlapping region:
\begin{equation}
   KS = \max\left(|F_{q_i}-F_{q_j}|\right),~~ q_i\neq q_j~.
   \label{Eq:KS}
\end{equation}
When the KS statistic is less than a critical value $CV$, the hypothesis is accepted. The critical value is
\begin{equation}
  CV=c_\alpha \sqrt{(m+n)/mn},
\end{equation}
where $m$ and $n$ are the numbers of recurrence interval samples for $q_i$ and $q_j$ \citep{Darling-1957-AMS,Stephens-1974-JASA}, and the threshold is $c_\alpha=1.36$ at the significance level of $\alpha=5\%$ \citep{Smirnov-1948-AMS,Young-1977-JHcCc}. The null hypothesis is that the two samples are drawn from the same distribution and it is rejected if $KS>CV$ at the significance level of 5\%.

We have performed KS tests for each pair of samples $(\tau_{q_i},\tau_{q_j})$ for the four futures. The results are presented in Table \ref{TB:Scaling:KS}. One can find that $KS$ is much greater than $CV$ in all cases. It means that there is no scaling in the distribution of recurrence intervals for different thresholds $q$.

\setlength\tabcolsep{2.5pt}
\begin{table}[htb]
  \centering 
  \caption{\label{TB:Scaling:KS} Two-sample Kolmogorov-Smirnov test of possible scaling behaviors in the return interval distributions by comparing the statistic $KS$ with the critical value $CV$ at significance level of $\alpha=5\%$. The null hypothesis that the two samples have the same distribution is rejected if $KS>CV$.}
  \medskip
\begin{tabular}{ccccccccccccccccccc}
  \hline\hline
  &&&\multicolumn{2}{c}{Crude oil} && \multicolumn{2}{c}{Gasoline} && \multicolumn{2}{c}{Heating oil} && \multicolumn{2}{c}{Propane}\\  %
     \cline{4-5} \cline{7-8} \cline{10-11} \cline{13-14}
    $q_i$ & $q_j$ && $KS$ & $CV$ && $KS$ & $CV$ && $KS$ & $CV$ && $KS$ & $CV$\\
  \hline
  1.0   &  1.2   &&  0.34   &  0.04   &&  0.36   &  0.04   &&  0.35   &  0.04   &&  0.33   &  0.06   \\
  1.0   &  1.4   &&  0.30   &  0.04   &&  0.30   &  0.05   &&  0.30   &  0.04   &&  0.30   &  0.07   \\
  1.0   &  1.6   &&  0.27   &  0.05   &&  0.26   &  0.05   &&  0.26   &  0.05   &&  0.26   &  0.07   \\
  1.0   &  1.8   &&  0.42   &  0.05   &&  0.40   &  0.06   &&  0.41   &  0.05   &&  0.41   &  0.08   \\
  1.0   &  2.0   &&  0.39   &  0.06   &&  0.47   &  0.06   &&  0.47   &  0.05   &&  0.49   &  0.08   \\
  1.2   &  1.4   &&  0.30   &  0.05   &&  0.30   &  0.05   &&  0.30   &  0.04   &&  0.30   &  0.07   \\
  1.2   &  1.6   &&  0.27   &  0.05   &&  0.26   &  0.05   &&  0.26   &  0.05   &&  0.26   &  0.07   \\
  1.2   &  1.8   &&  0.25   &  0.05   &&  0.24   &  0.06   &&  0.23   &  0.05   &&  0.26   &  0.08   \\
  1.2   &  2.0   &&  0.39   &  0.06   &&  0.37   &  0.06   &&  0.38   &  0.06   &&  0.40   &  0.09   \\
  1.4   &  1.6   &&  0.27   &  0.05   &&  0.26   &  0.06   &&  0.26   &  0.05   &&  0.26   &  0.08   \\
  1.4   &  1.8   &&  0.25   &  0.06   &&  0.24   &  0.06   &&  0.23   &  0.05   &&  0.26   &  0.08   \\
  1.4   &  2.0   &&  0.23   &  0.06   &&  0.22   &  0.06   &&  0.38   &  0.06   &&  0.25   &  0.09   \\
  1.6   &  1.8   &&  0.25   &  0.06   &&  0.24   &  0.06   &&  0.23   &  0.06   &&  0.26   &  0.09   \\
  1.6   &  2.0   &&  0.23   &  0.06   &&  0.22   &  0.07   &&  0.22   &  0.06   &&  0.25   &  0.09   \\
  1.8   &  2.0   &&  0.23   &  0.07   &&  0.22   &  0.07   &&  0.19   &  0.06   &&  0.25   &  0.10   \\
  \hline\hline
\end{tabular}
\end{table}

\subsection{Fitting the PDFs}

We use the stretched exponential, expressed in Eq.~(\ref{Eq:PDF:tau:SE}) to fit the empirical recurrence interval distributions of financial volatility. Considering Eq.~(\ref{Eq:Pq:f}), we have
\begin{equation}
  f_q(x)=f(x)=c\bar\tau e^{- (a\bar{\tau}x)^{\gamma}},
  \label{Eq:StrExp}
\end{equation}
where $x=\tau/\bar\tau$, $c$ and $a$ are two parameters and $\gamma$ is the correlation exponent characterizing the long-term memory of volatilities. Due to the definition of $x$, the parameters $a$ and $c$ are dependent of $\gamma$ \citep{Altmann-Kantz-2005-PRE,Wang-Yamasaki-Havlin-Stanley-2008-PRE}. The normalization condition states that
\begin{equation}
 1 = \int_0^\infty f_q(x) dx = c\bar{\tau} \int_0^\infty e^{- (a\bar{\tau}x)^{\gamma}}dx.
 \label{eq:SE:normalization}
\end{equation}
Let $y=(a\bar{\tau}x)^{\gamma}$ and notice that $\int_0^\infty t^{z-1}e^{-t}dt=\Gamma(z)$. We have
\begin{equation}
 c\Gamma(1/\gamma) = a\gamma.
 \label{eq:SE:normalization:Gamma}
\end{equation}
In addition, since the mean of $x$ is 1 by definition we have
\begin{equation}
 1 = \int_0^\infty x f_q(x) dx = c\bar{\tau} \int_0^\infty x e^{- (a\bar{\tau}x)^{\gamma}}dx.
 \label{eq:SE:mean}
\end{equation}
Similarly, we obtain
\begin{equation}
 c\Gamma(1/\gamma)^2 = a\Gamma(2/\gamma).
 \label{eq:SE:mean:Gamma}
\end{equation}
Solving Eqs.~(\ref{eq:SE:normalization:Gamma}) and (\ref{eq:SE:mean:Gamma}), we have
\begin{equation}
 a=\Gamma(2/\gamma)/\Gamma(1/\gamma)
 \label{eq:SE:a}
\end{equation}
and
\begin{equation}
 c=\gamma\Gamma(2/\gamma)/\Gamma(1/\gamma)^2.
 \label{eq:SE:c}
\end{equation}
Therefore there is only one free parameter $\gamma$ in the fitting of the normalized recurrence intervals $x=\tau/\bar\tau$.

However, the situation is more complicated. First of all, the recurrence intervals are integers and $\tau$ is thus discrete. It means that the derivation above is not accurate and might be biased in capturing the ``true'' distribution. In addition, in most cases, one can fit only part of the distribution using chosen functions \citep{Clauset-Shalizi-Newman-2009-SIAMR}. Hence, we use the three-parameter stretched exponential in Eq.~(\ref{Eq:PDF:tau:SE}) to fit the recurrence intervals that are longer than some minimal value $\tau_{\min}$. The idea of the method is the same as in \cite{Clauset-Shalizi-Newman-2009-SIAMR} and \cite{Jiang-Xie-Li-Podobnik-Zhou-Stanley-2013-PNAS}, which we describe below.

\setlength\tabcolsep{4.4pt}
\begin{table}[htb]
  \centering
  \caption{\label{TB:goodness-of-fit} Estimates of $\tau_{\min}$, $a$, $c$ and $\gamma$ of the stretched exponential expressed in Eq.~(\ref{Eq:PDF:tau:SE}) using maximal likelihood estimation and the $p$-values of the goodness-of-fit tests using KS statistic and CvM statistic.}
  \medskip
\begin{tabular}{lcccccccccccccccc}
  \hline\hline
Futures
  &  $q$ & $\tau_{\min}$  & $c$ & $a$ & $\gamma$ & $p_{\mathrm{KS}}$ & $p_{\mathrm{CvM}}$  \\
  \hline
Crude oil
  &  $1.0$   &  $2$   &   $37.24$   &  $37.04$   &  $0.35$   &  $0.14$   &  $0.23$ \\
  &  $1.2$   &  $2$   &   $23.04$   &  $38.50$   &  $0.33$   &  $0.11$   &  $0.08$ \\
  &  $1.4$   &  $3$   &   $24.88$   &  $37.95$   &  $0.32$   &  $0.11$   &  $0.26$ \\
  &  $1.6$   &  $2$   &   $10.66$   &  $40.00$   &  $0.30$   &  $0.21$   &  $0.18$  \\
  &  $1.8$   &  $1$   &  $~~2.82$   &  $14.35$   &  $0.32$   &  $0.26$   &  $0.32$   \\
  &  $2.0$   &  $1$   &  $~~3.16$   &  $25.10$   &  $0.28$   &  $0.23$   &  $0.59$  \\
  \hline
Gasoline
  &  $1.0$   &  $3$   &   $98.64$   &  $33.75$   &  $0.37$   &  $0.59$   &  $0.64$   \\
  &  $1.2$   &  $3$   &   $50.15$   &  $33.35$   &  $0.35$   &  $0.46$   &  $0.66$   \\
  &  $1.4$   &  $3$   &   $35.15$   &  $38.40$   &  $0.33$   &  $0.15$   &  $0.37$   \\
  &  $1.6$   &  $3$   &   $19.24$   &  $32.95$   &  $0.32$   &  $0.35$   &  $0.33$   \\
  &  $1.8$   &  $2$   &  $~~4.88$   &  $14.60$   &  $0.33$   &  $0.63$   &  $0.50$   \\
  &  $2.0$   &  $2$   &  $~~5.82$   &  $31.85$   &  $0.28$   &  $0.47$   &  $0.43$   \\
  \hline
Heating oil
  &  $1.0$   &  $2$   &   $30.26$   &  $33.25$   &  $0.35$   &  $0.31$   &  $0.18$    \\
  &  $1.2$   &  $3$   &   $38.10$   &  $40.00$   &  $0.33$   &  $0.21$   &  $0.29$   \\
  &  $1.4$   &  $1$   &  $~~1.77$   & $~~3.50$   &  $0.43$   &  $0.09$   &  $0.06$   \\
  &  $1.6$   &  $2$   &   $10.66$   &  $40.00$   &  $0.30$   &  $0.48$   &  $0.37$   \\
  &  $1.8$   &  $2$   &  $~~8.04$   &  $40.00$   &  $0.28$   &  $0.09$   &  $0.46$   \\
  &  $2.0$   &  $2$   &  $~~6.05$   &  $40.00$   &  $0.28$   &  $0.23$   &  $0.18$    \\
  \hline
Propane
  &  $1.0$   &  $2$   &   $23.41$   &  $38.85$   &  $0.33$   &  $0.83$   &  $0.65$   \\
  &  $1.2$   &  $2$   &   $16.55$   &  $37.25$   &  $0.32$   &  $0.68$   &  $0.61$   \\
  &  $1.4$   &  $1$   &  $~~3.78$   &  $14.00$   &  $0.34$   &  $0.62$   &  $0.45$   \\
  &  $1.6$   &  $1$   &  $~~2.96$   &  $13.05$   &  $0.33$   &  $0.59$   &  $0.42$   \\
  &  $1.8$   &  $1$   &  $~~4.33$   &  $39.25$   &  $0.28$   &  $0.31$   &  $0.84$   \\
  &  $2.0$   &  $1$   &  $~~3.49$   &  $40.00$   &  $0.27$   &  $0.14$   &  $0.23$   \\
   \hline\hline
\end{tabular}
\end{table}

The approach is based on the maximum likelihood estimation (MLE) and KS tests. We assume that the intervals larger than a truncated value $\tau_{\min}$ are described by the three-parameter stretched exponential in Eq.~(\ref{Eq:PDF:tau:SE}). We also need to determine the lowest boundary $\tau_{\min}$ as an additional parameter. A truncated sample is obtained by discarding the
recurrence intervals less than $\tau_{\min}$ in the original interval sample. Following \cite{Clauset-Shalizi-Newman-2009-SIAMR}, we search for the distribution parameters $a$, $c$ and $\gamma$ of the truncated sample using MLE and the associated KS statistic for different lowest boundaries. The optimal $\tau_{\min}$ is determined as the one corresponding to the truncated sample with the smallest KS value.

The optimal value of $\tau_{\min}$ and the corresponding estimates of $a$, $c$ and $\gamma$ for the samples of recurrence intervals above six thresholds of the four NYMEX energy futures contracts are presented in Table \ref{TB:goodness-of-fit}. We notice that only a few smallest recurrence intervals are excluded in the fitting and the stretched exponential fits most of the data points. For parameter $a$, no clear dependence on $q$ can be observed. Interestingly, we find that both $c$ and $\gamma$ exhibit a decreasing trend with increasing threshold $q$, except that the case of heating oil with $q=1.4$ is an outlier. The decreasing trend of $\gamma$ is also observed for stock prices and stock indices when there is no scaling behavior in $P_q(\tau)$ and the values of $\gamma$ of the energy futures are close to those of stock prices or stock indices \citep{Ren-Zhou-2008-EPL,Ren-Gu-Zhou-2009-PA}.

\subsection{Goodness-of-fit}

For each sample, we use the one-sample Kolmogorov-Smirnov test and Cram{\'e}r-von Mises (CvM) test to assess the goodness-of-Fit after the optimal $\tau_{\min}$ and the corresponding distribution parameters are obtained. The null hypothesis $H_0$ for our KS test and CvM test is that the left-truncated data ($\tau > \tau_{\min}$) are drawn from the fitted stretched exponential distribution. The methods are described below, which are similar but not the same as in \cite{Ren-Zhou-2008-EPL}.

In the one-sample situation, the KS statistic defined in Eq.~(\ref{Eq:KS}) for the real sample of recurrence intervals is calculated as follows
\begin{equation}
   KS = \max\left(|F_q-F_{\rm{SE}}|\right),
   \label{Eq:KS2}
\end{equation}
where $F_{\rm{SE}}=\int_0^tce^{-(a\tau)^{-\gamma}}d\tau$ is the cumulative distribution of $P_q(\tau)$ and the parameters $a$, $c$ and $\gamma$ are associated with the given $q$. Then the bootstrapping approach is adopted to determine the distribution of the KS statistic, $p(KS_{\rm{sim}})$. We generate 10000 synthetic samples from the best fitted distribution and calculate the KS statistic for each synthetic sample in reference to the fitted distribution as follows
\begin{equation}
   KS_{\rm{sim}} = \max\left(|F_{\rm{sim}}-F_{\rm{SE}}|\right).
   \label{Eq:KS2:sim}
\end{equation}
The $p$-value is determined by the proportion that $KS_{\rm{sim}}>KS$, which is the area enclosed by three curves $p(KS_{\rm{sim}})$, $KS_{\rm{sim}}=KS$, and $p(KS_{\rm{sim}})=0$. Hence, a small $KS$ value corresponds to a large $p$-value and high goodness-of-fit.

\begin{figure}[htb]
\centering
\includegraphics[width=4.3cm,height=3.2cm]{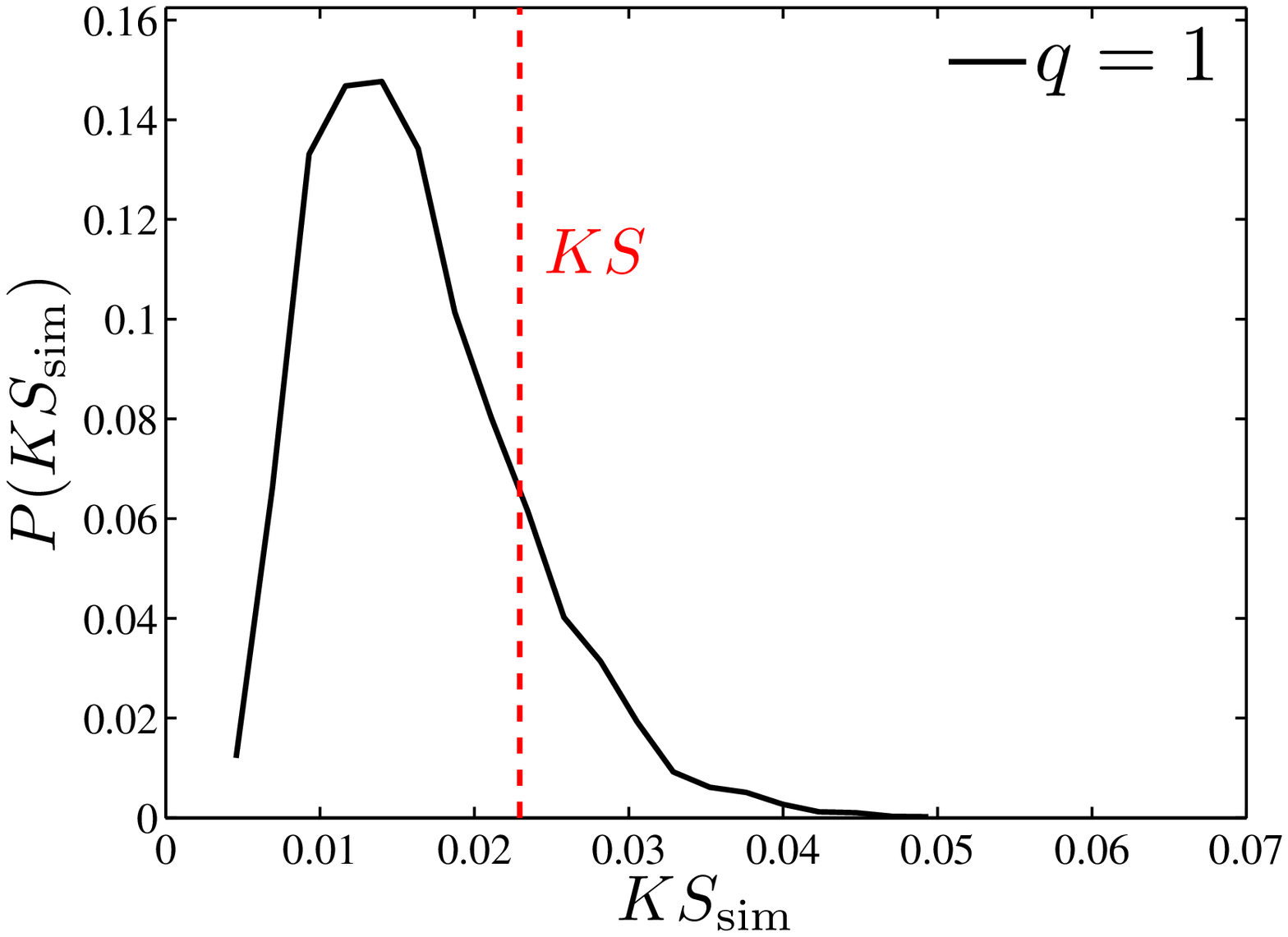}
\includegraphics[width=4.3cm,height=3.2cm]{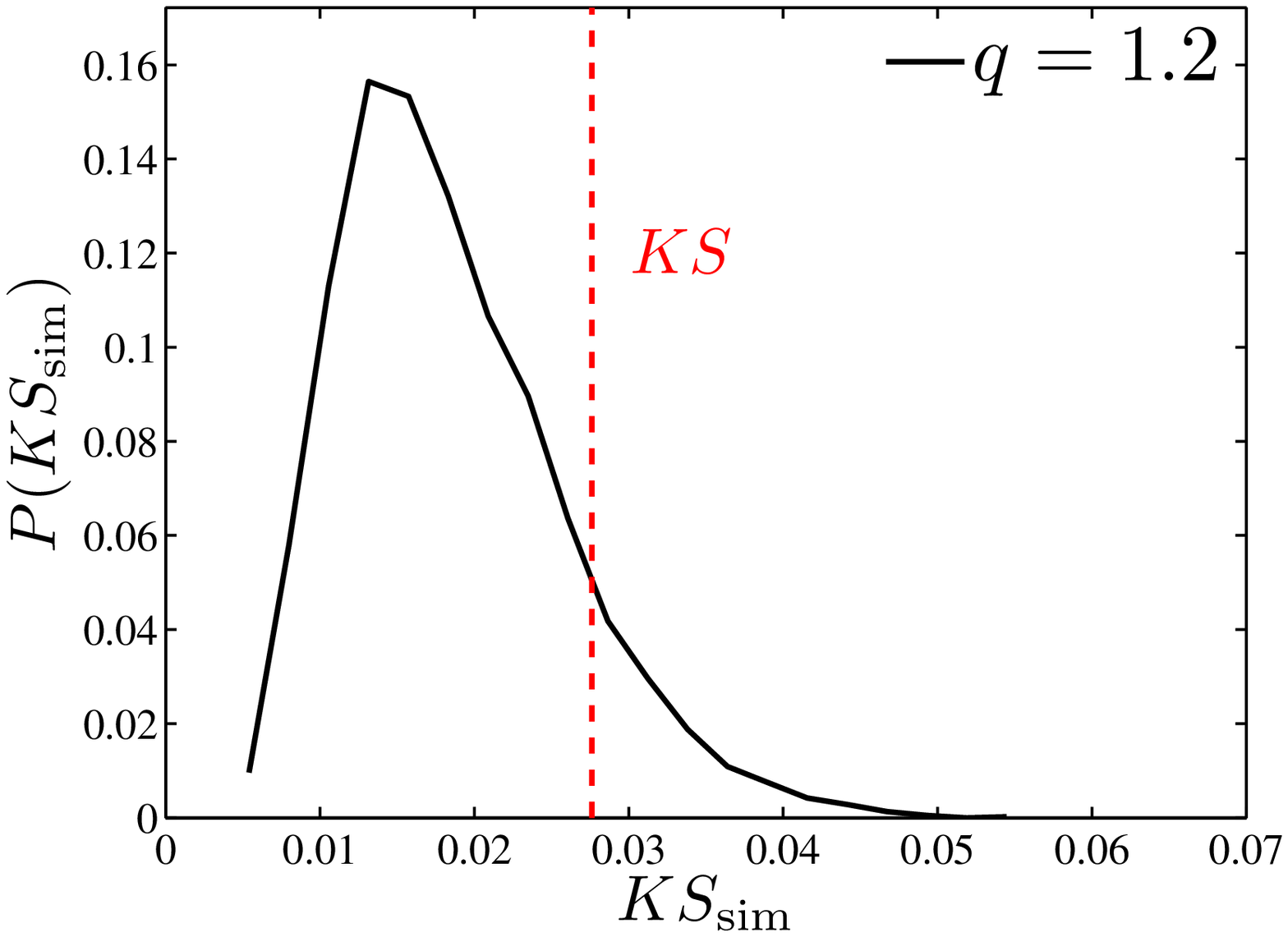}
\includegraphics[width=4.3cm,height=3.2cm]{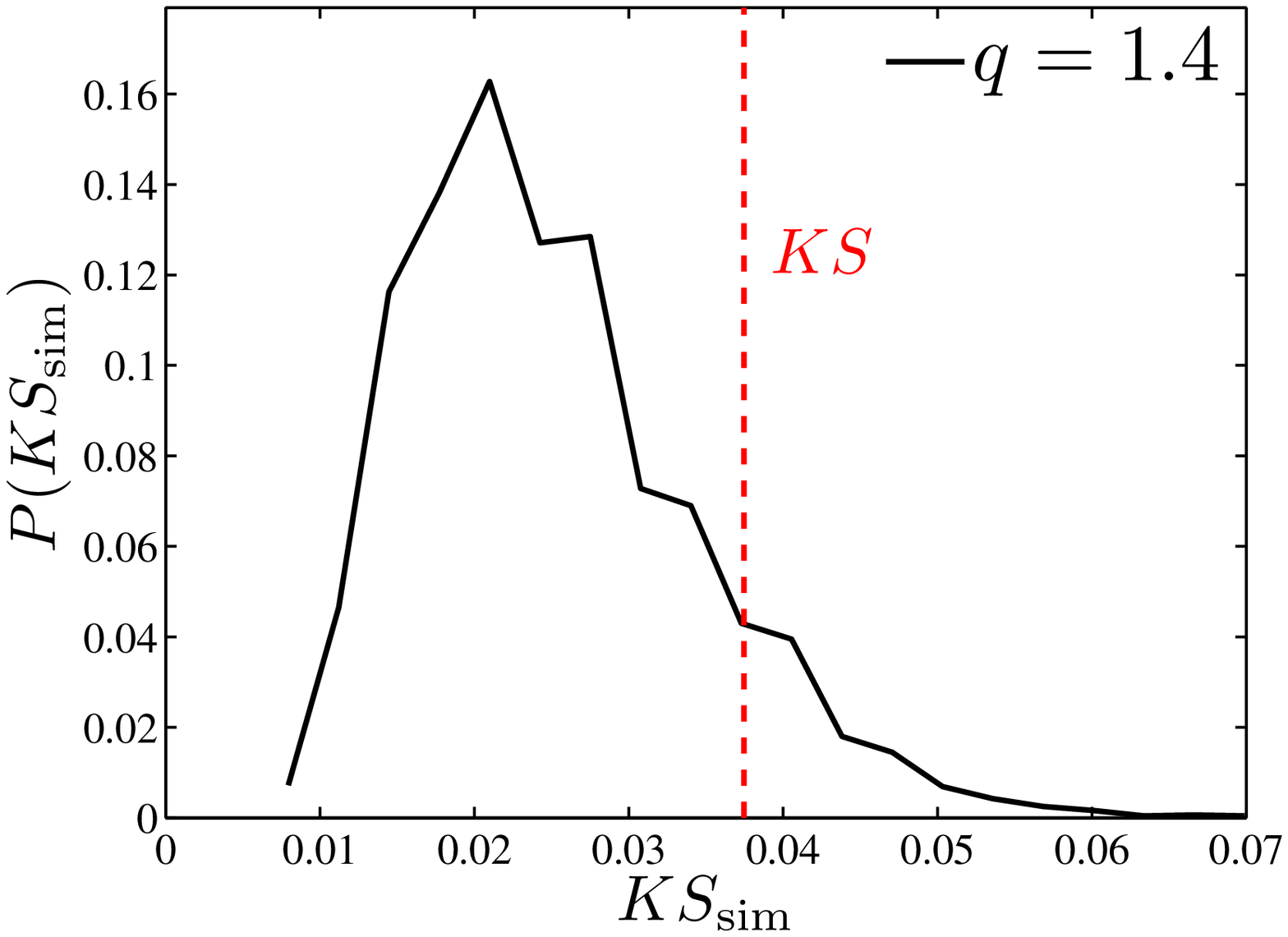}
\includegraphics[width=4.3cm,height=3.2cm]{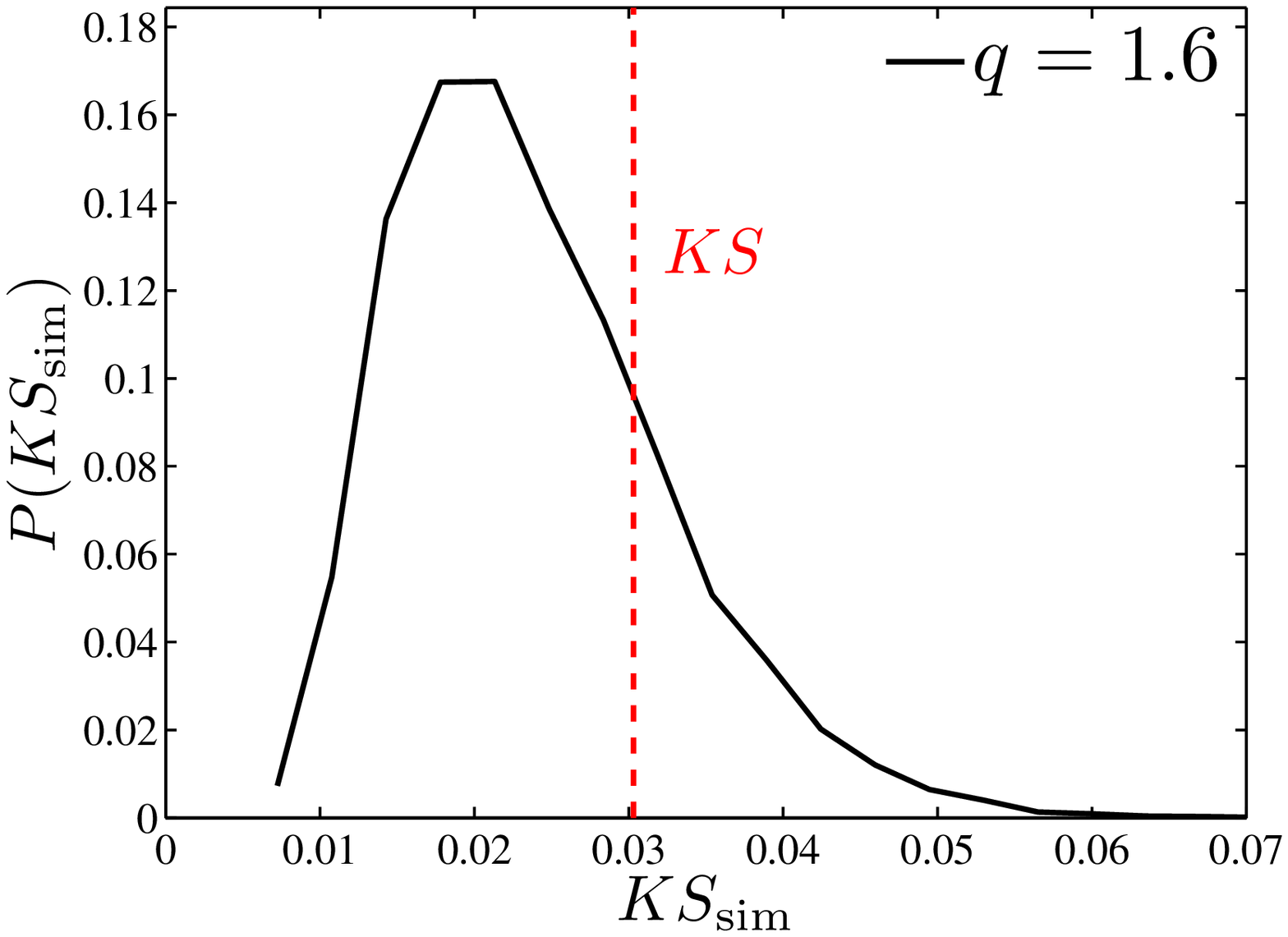}
\includegraphics[width=4.3cm,height=3.2cm]{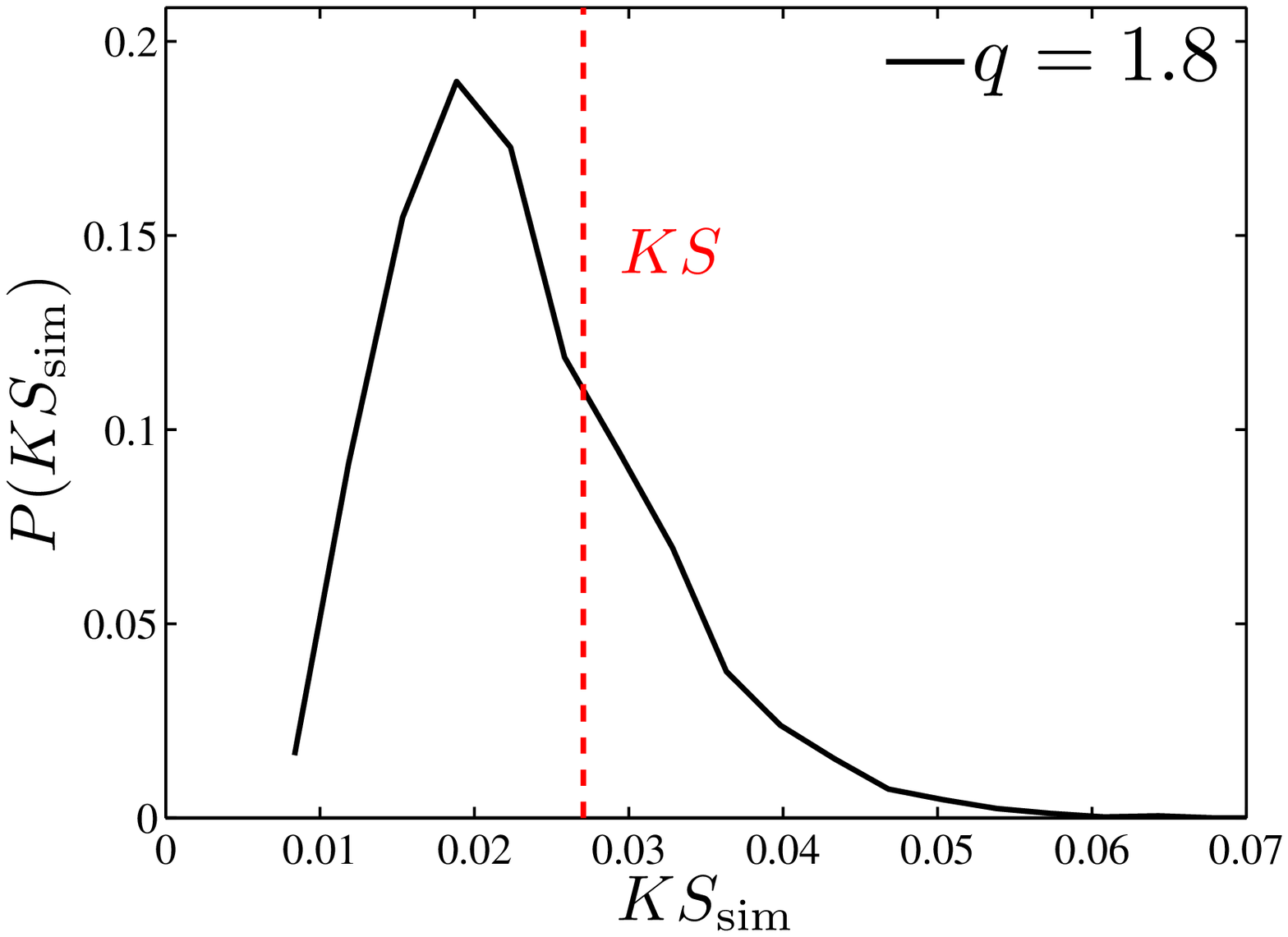}
\includegraphics[width=4.3cm,height=3.2cm]{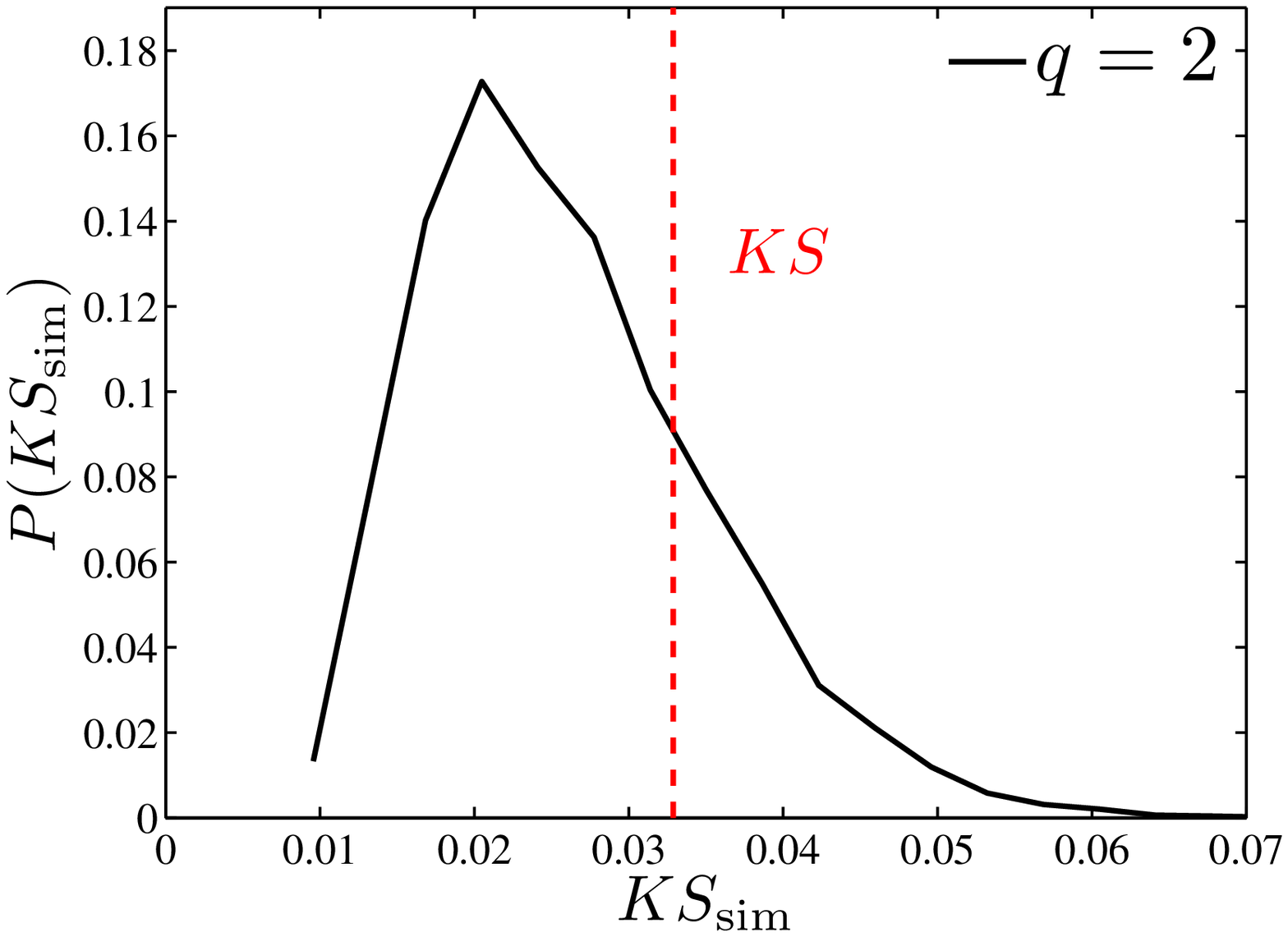}
\caption{\label{Fig:RI:KS} (Color online.) Goodness-of-fit tests of the stretched exponential distributions using Kolmogorov-Smirnov statistic for different thresholds $q$ for crude oil futures. In each plot, the black solid curve stands for the probability distribution $p(KS_{\rm{sim}})$ and the red dashed line is the value of $KS$ for the real sample. The $p$-value is the area enclosed by $p(KS_{\rm{sim}})$, $KS_{\rm{sim}}=KS$, and $p(KS_{\rm{sim}})=0$.}
\end{figure}

Figure \ref{Fig:RI:KS} illustrates the results of the goodness-of-fit tests using KS statistic for crude oil futures. The six distributions $p(KS_{\rm{sim}})$ have similar shapes. With the increase of $q$, the distribution becomes broader and most probable value that corresponds to the maximum of the distribution increases. The estimated $p$-values are presented in Table \ref{TB:goodness-of-fit}, all greater than 5\%. The results for gasoline, heating oil and propane are also listed in Table \ref{TB:goodness-of-fit}. The minimal $p$-value is given by the recurrence intervals of heating oil with $q=1.4$, which is identified as an outlier due to its {\textit{abnormally}} large value of $\gamma$. We conclude that the stretched exponential can be used to model the recurrence interval distributions under the KS tests.

As shown in \cite{Pearson-Stephens-1962-Bm}, \cite{Stephens-1964-Bm} and \cite{Stephens-1970-JRSSB}, the CvM statistic is determined as follows:
\begin{equation}
   W^2 = N \int_{-\infty}^{\infty} ( F_q-F_{\rm{SE}})^2 d F_{\rm{SE}},
   \label{Eq:CvM}
\end{equation}
where $F_q$ is the CDF of empirical data for threshold $q$ and $N$ is the total number of the recurrence interval samples. Then the bootstrapping approach is adopted to determine the distribution of the CvM statistic, $p(W^2_{\rm{sim}})$. We generate 10000 synthetic samples from the best fitted distribution and calculate the CvM statistic for each synthetic sample in reference to the fitted distribution as follows:
\begin{equation}
   W^2_{\rm{sim}} = N \int_{-\infty}^{\infty} ( F_{\rm{sim}}-F_{\rm{SE}})^2 d F_{\rm{SE}},
   \label{Eq:CvM:sim}
\end{equation}
The $p$-value is determined by the proportion that $W^2_{\rm{sim}}>W^2$, which is the area enclosed by three curves $p(W^2_{\rm{sim}})$, $W^2_{\rm{sim}}=W^2$, and $p(W^2_{\rm{sim}})=0$. Hence, a small $W^2$ value corresponds to a large $p$-value and high goodness-of-fit.

\begin{figure}[htb]
\centering
\includegraphics[width=4.3cm,height=3.1cm]{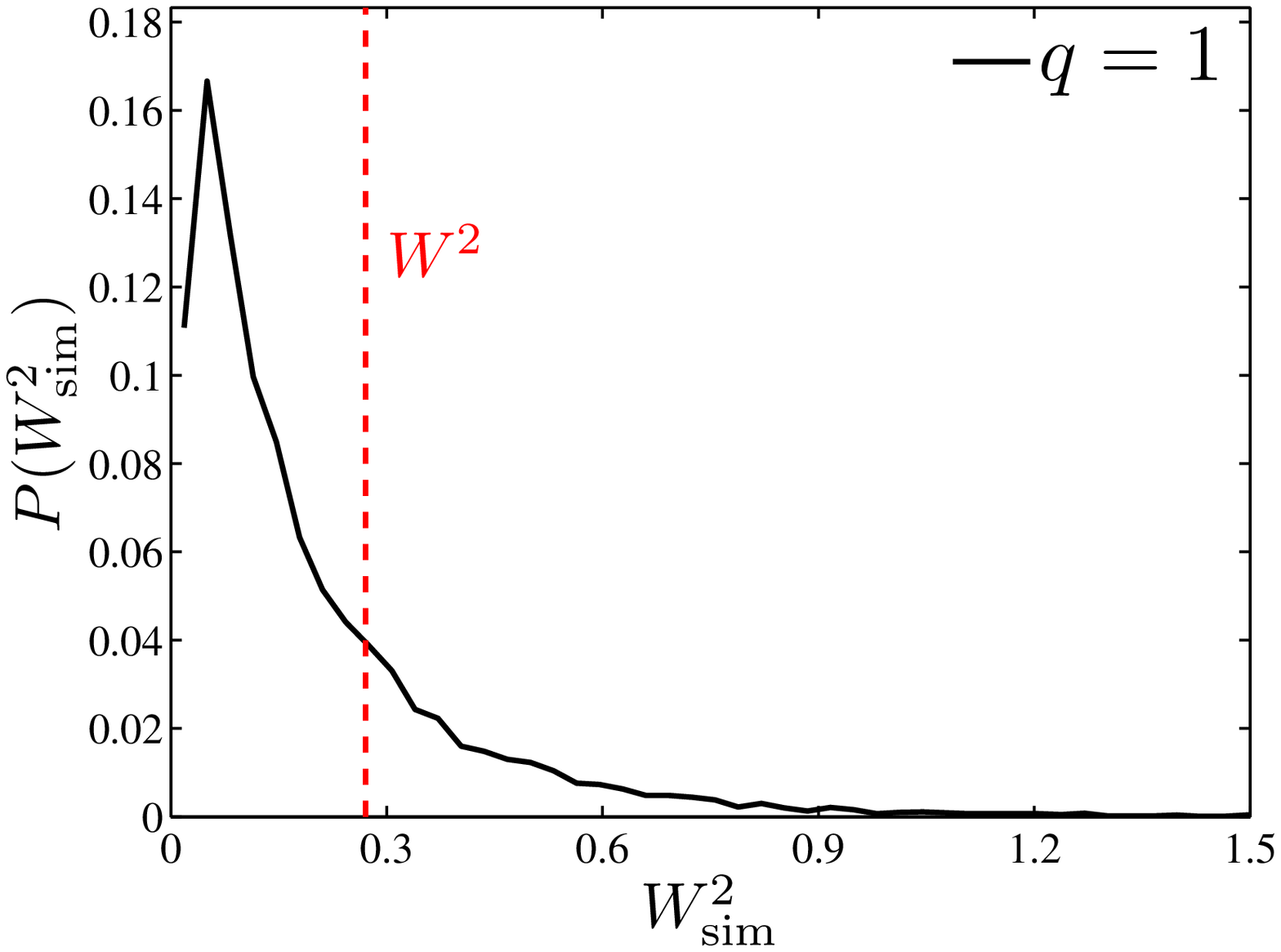}
\includegraphics[width=4.3cm,height=3.1cm]{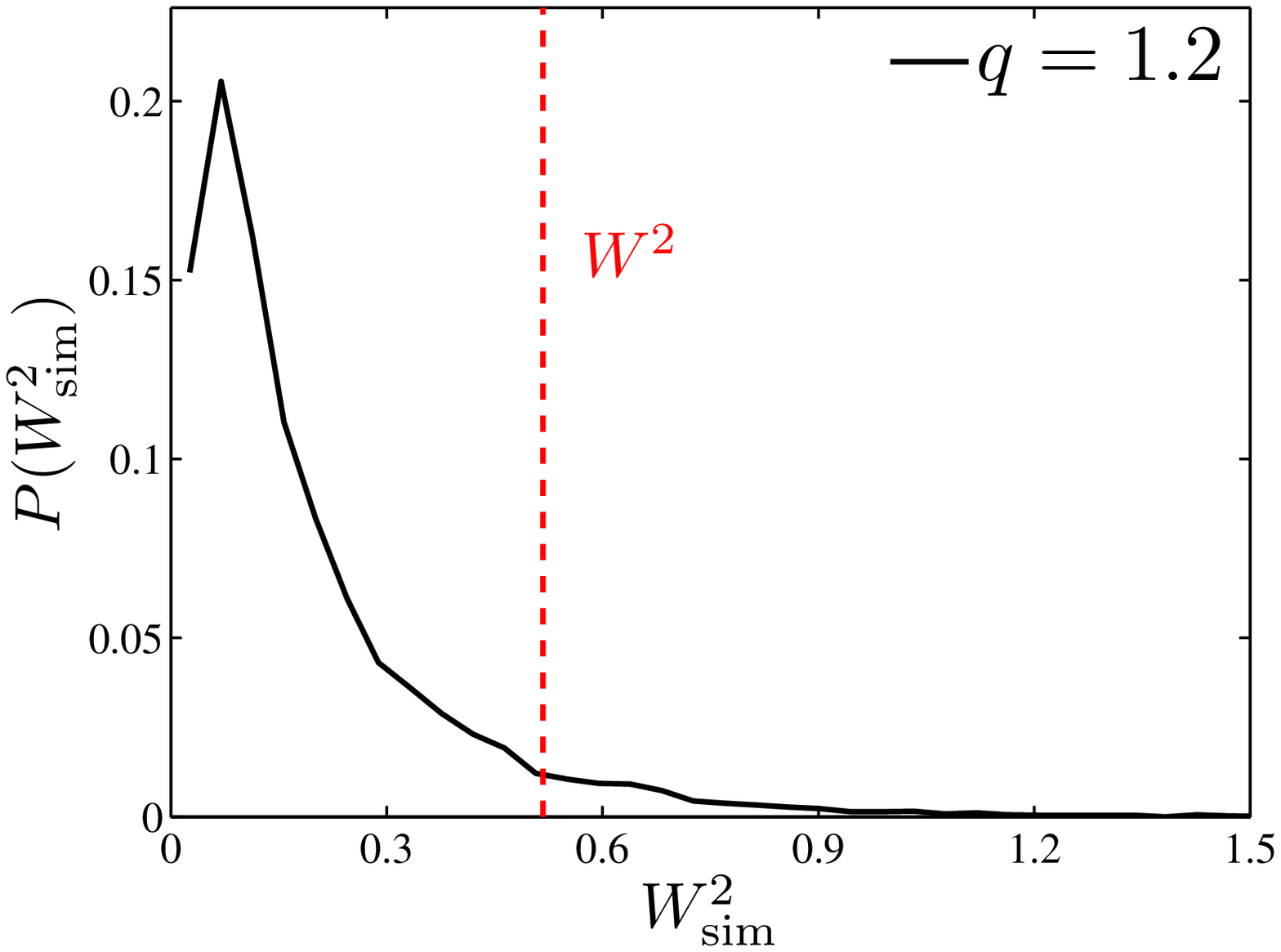}
\includegraphics[width=4.3cm,height=3.1cm]{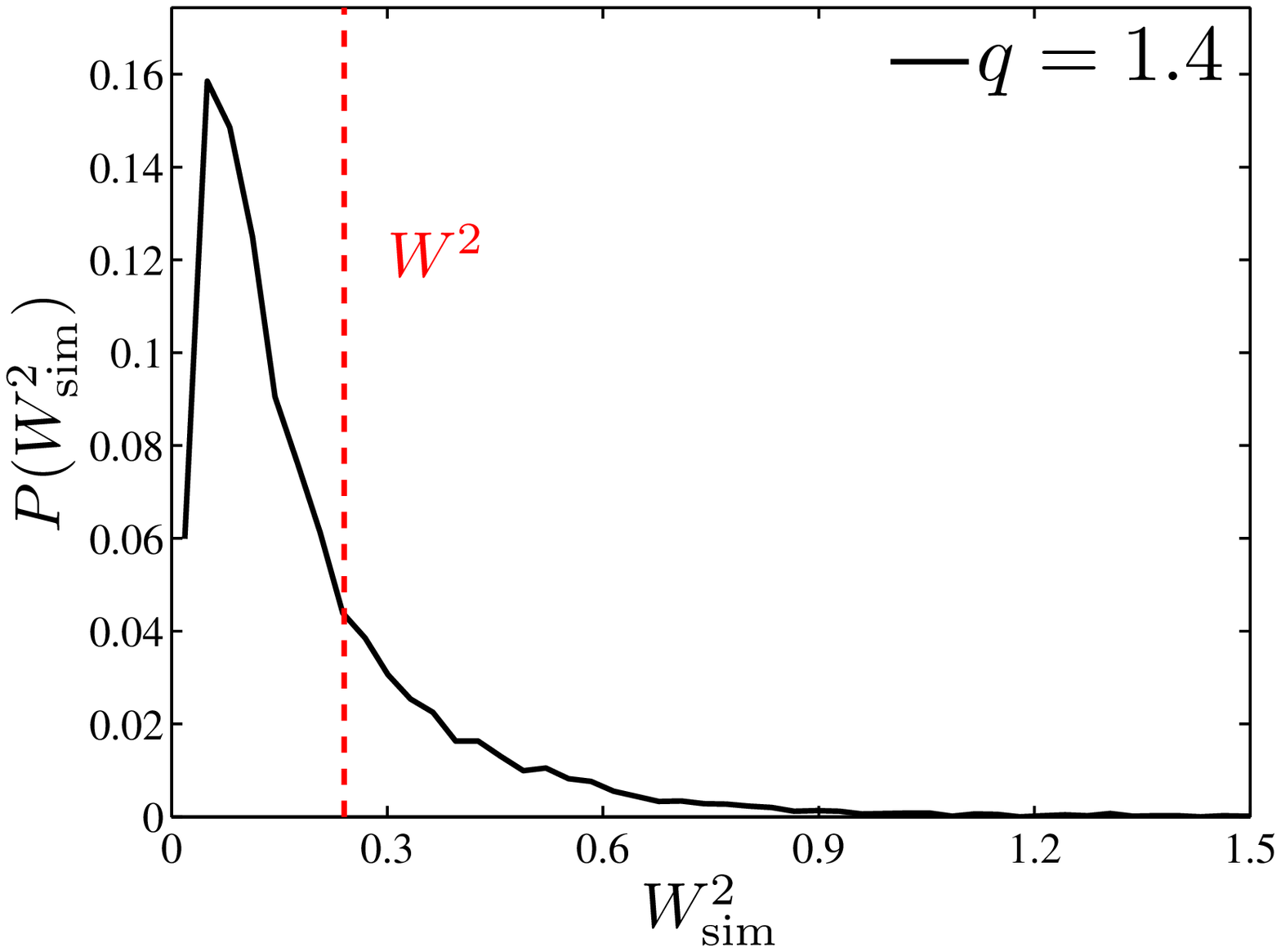}
\includegraphics[width=4.3cm,height=3.1cm]{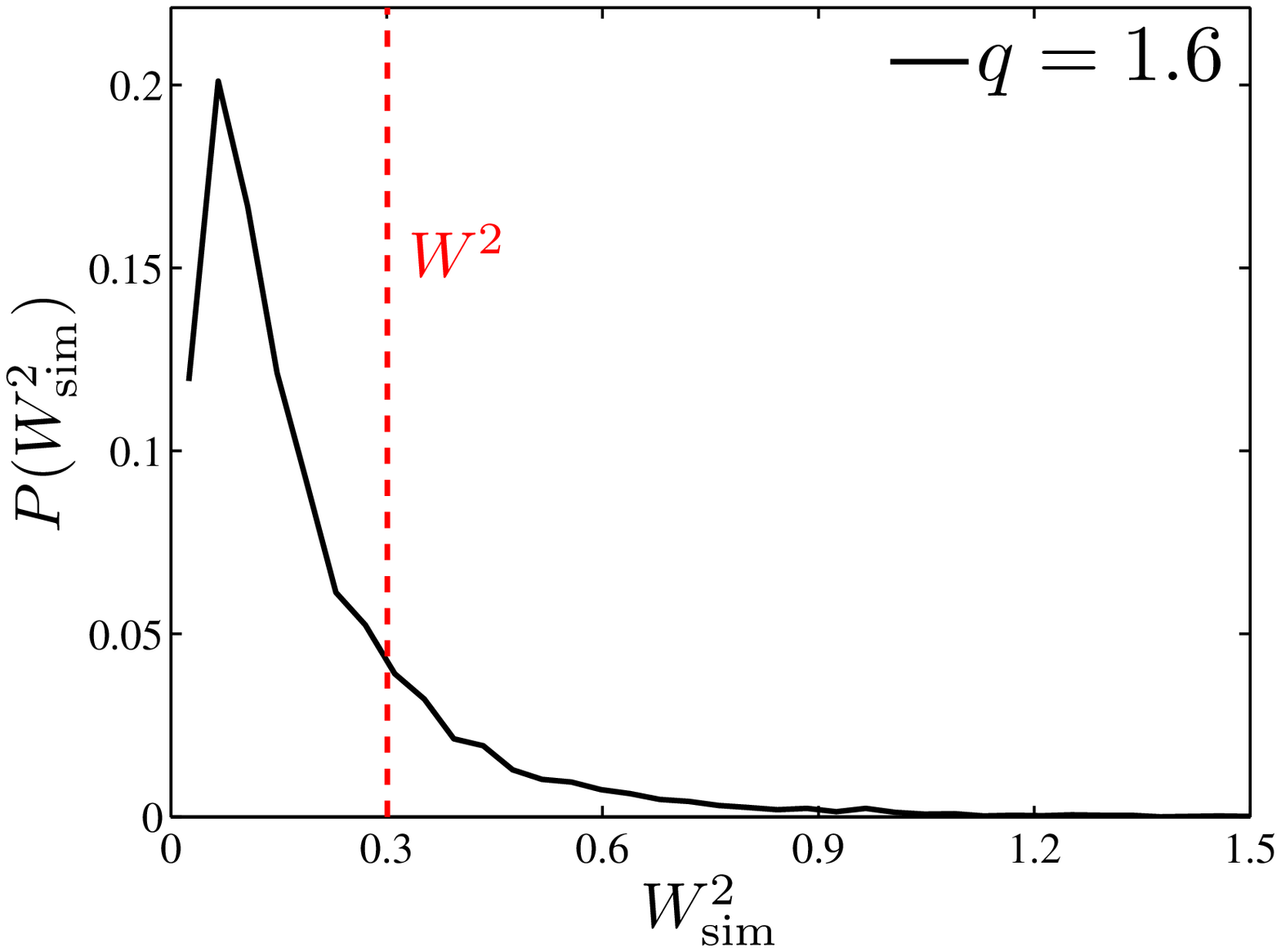}
\includegraphics[width=4.3cm,height=3.1cm]{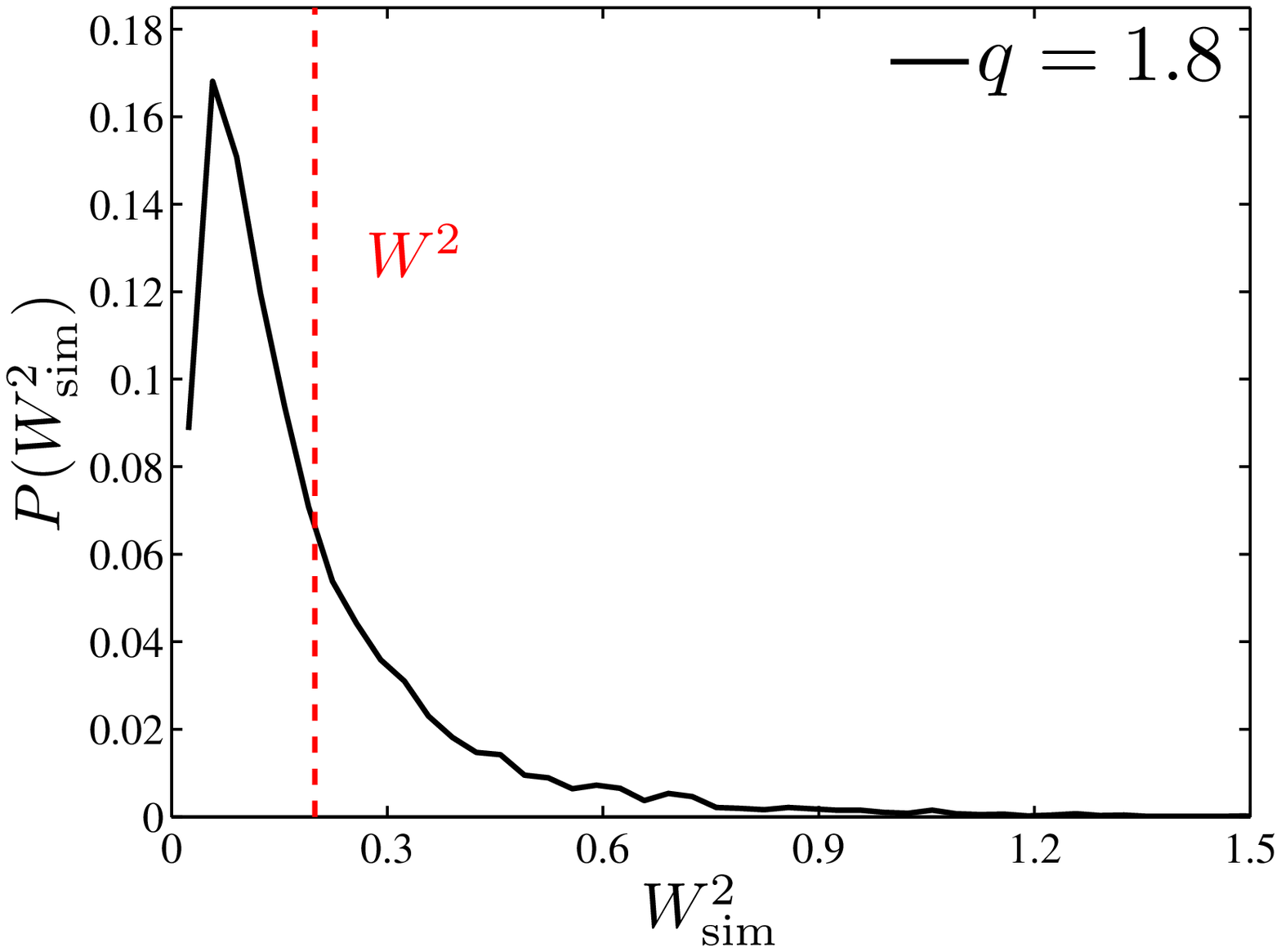}
\includegraphics[width=4.3cm,height=3.1cm]{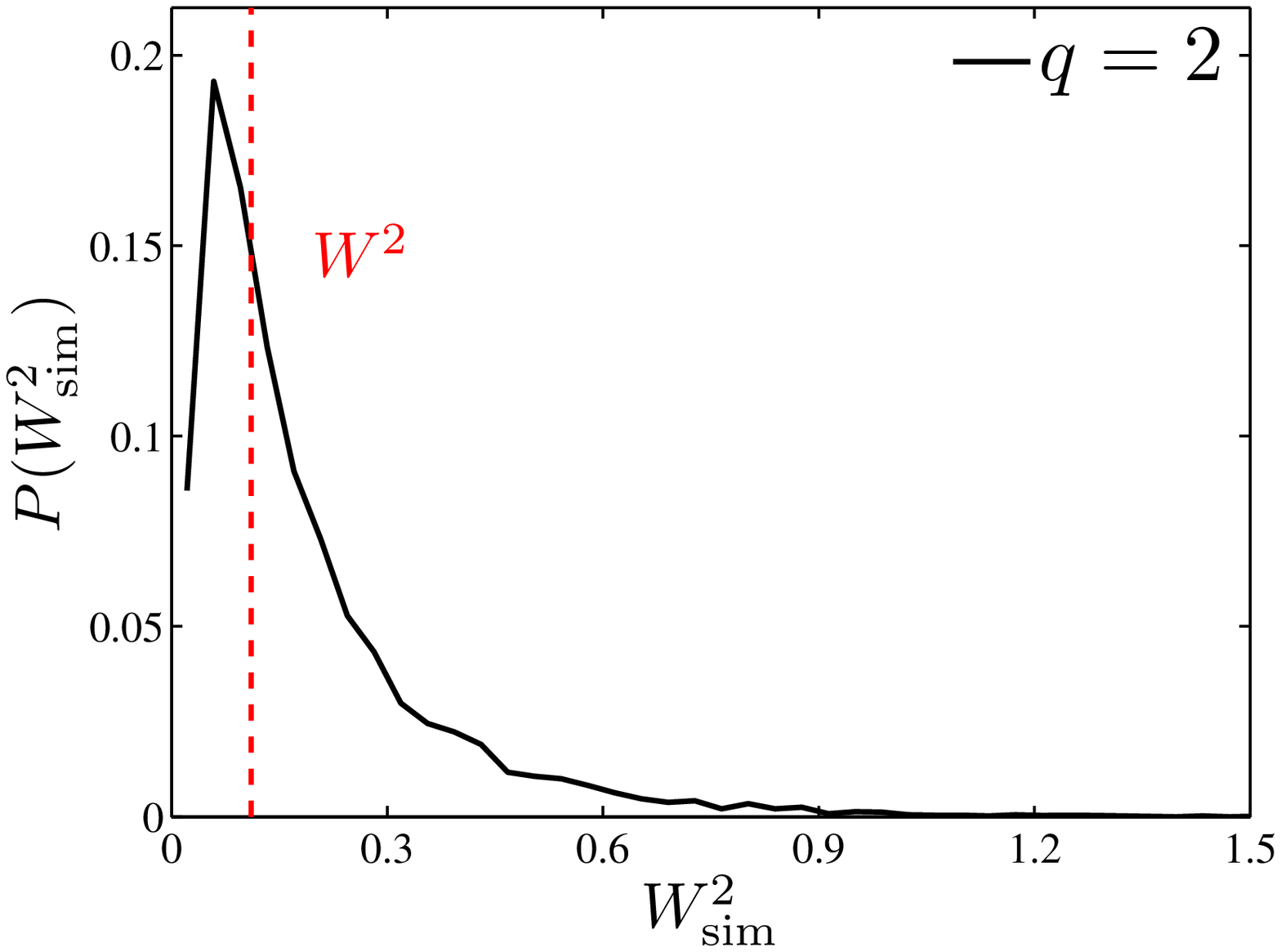}
\caption{\label{Fig:RI:CvM} (Color online.) Goodness-of-fit tests of the stretched exponential distributions using Cram{\'e}r-von Mises statistic for different thresholds $q$ for crude oil futures. In each plot, the black solid curve stands for the probability distribution $p(W^2_{\rm{sim}})$ and the red dashed line is the value of $W^2$ for the real sample. The $p$-value is the area enclosed by $p(W^2_{\rm{sim}})$, $W^2_{\rm{sim}}=KS$, and $p(W^2_{\rm{sim}})=0$.}
\end{figure}

Figure \ref{Fig:RI:CvM} illustrates the results of the goodness-of-fit tests using CvM statistic for crude oil futures. The six distributions $p(W^2_{\rm{sim}})$ have very similar shapes and the quantitative differences between them are much smaller than in Fig.~\ref{Fig:RI:CvM}. The estimated $p$-values are also presented in Table \ref{TB:goodness-of-fit}, as well as the results for gasoline, heating oil and propane. We find that all the $p$-values are greater than 5\% and the minimal $p$-value is 6\% for the outlier. Hence, the distributions of recurrence intervals above a fixed threshold $q$ can be well fitted by stretched exponentials.

\subsection{Risk assessment}

Consider the situation that $t$ trading days have elapsed since the last large volatility greater than $q$. We are interested in the hazard probability that a new volatility greater than $q$ will occur within $\Delta{t}$ trading days. The hazard probability gives a quantitative estimation of risk. Mathematically, the hazard probability can be expressed as follows \citep{Bogachev-Eichner-Bunde-2007-PRL}:
\begin{equation}
   W_Q(\Delta{t}|t)=\frac{\int_t^{t+\Delta{t}}P_q(t)dt}{\int_t^{\infty}P_q(t)dt}.
   \label{Eq:Wq}
\end{equation}
Since each distribution $P_q(t)$ has been fitted with a stretched exponential with the parameters given in Table \ref{TB:goodness-of-fit}, we can obtain the theoretical function of hazard probability $W_q(\Delta{t}|t)$ for a given value of $\Delta{t}$ through numerical integration.

In order to determine the $W_q(\Delta{t}|t)$ values empirically, we can rewrite Eq.~(\ref{Eq:Wq}) in the following way:
\begin{equation}
   W_q(\Delta{t}|t)=\frac{\#(t<\tau_q\leq{t+\Delta{t}})}{\#(\tau_q>t)},
   \label{Eq:Wq:Emp}
\end{equation}
where the denominator $\#(\tau_q>t)$ is the number of recurrence intervals greater than $t$ and the numerator $\#(t<\tau_q\leq{t+\Delta{t}})$ is the number of intervals greater than $t$ and not greater than $t+\Delta{t}$ for a given $q$.

\begin{figure}[htb]
  \centering
  \includegraphics[width=8.6cm]{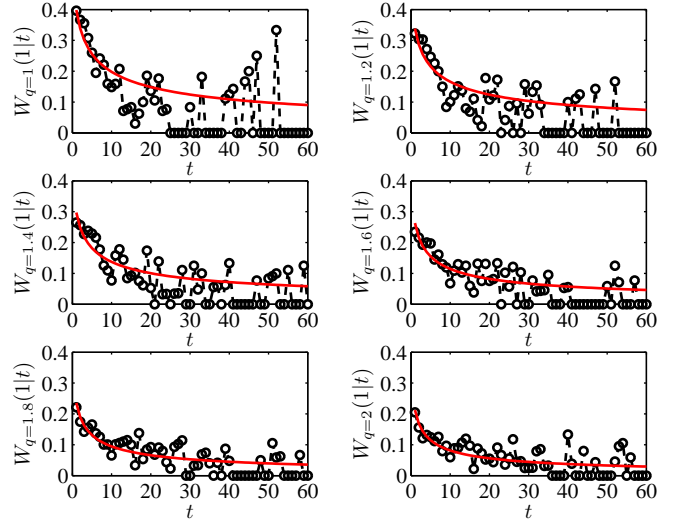}
  \caption{\label{Fig:RI:WQ} (Color online.) Comparison of $W_q(\Delta{t}=1|t)$ estimated empirically using Eq.~(\ref{Eq:Wq:Emp}) and numerically using Eq.~(\ref{Eq:Wq}) for different $q$ values for crude oil futures. The black circles and red lines correspond respectively to empirical and numerical results.}
\end{figure}

Figure \ref{Fig:RI:WQ} illustrates the dependence of the hazard probability $W_q(\Delta{t}|t)$ on $t$ when fixing $\Delta{t}=1$ for crude oil futures. It is observed that the theoretical curves approximate the empirical curves very nicely. It is very important to notice that the discrepancy between theoretical and empirical curves decreases when $q$ increases. Another intriguing observation is that $W_q(\Delta{t}|t)$ decreases with increasing $t$, which is simply a consequence of the fact that recurrence intervals exhibit clustering behaviors as shown in Fig.~\ref{Fig:NYMEX:Energy:RI:Data}(b) or long-term correlations that will be confirmed in Section \ref{S1:Memory}. We also observe that, for small $t$ values,
\begin{equation}
   W_{q_1}(\Delta{t}|t)>W_{q_2}(\Delta{t}|t),
   \label{Eq:Wq:q1q2}
\end{equation}
if $q_1>q_2$. This is consistent with the intuition that the hazard probability decreases when the condition approaches to extreme and there are less recurrence intervals.

We have also studied two other cases in which $\Delta{t}=5$ and $\Delta{t}=10$. The results are qualitatively the same as for $\Delta{t}=1$. A new finding is that $W_q(\Delta{t}|t)$ increases with $\Delta{t}$ for given $t$ and $q$, which is actually trivial and can be derived from Eq.~(\ref{Eq:Wq}) directly. In general, $W_q(\Delta{t}|t)$ is a monotonously increasing function of $\Delta{t}$. All these findings apply to other three energy futures investigated in this work.

\section{Memory effects}
\label{S1:Memory}

Previous works have reported that the recurrence intervals of stock and stock index volatility possess both short-term memory and long-term memory. It is interesting to investigate the memory effects of volatility recurrence intervals of energy futures prices. Note that the methods used in this work were proposed by \cite{Yamasaki-Muchnik-Havlin-Bunde-Stanley-2005-PNAS}.

\subsection{Short-term memory}

To investigate possible short-term correlations in the recurrence intervals, we compute and compare the conditional probability distributions $P_q(\tau|\tau_0)$, which is the distribution of recurrence interval $\tau$ conditioned on the value of its preceding recurrence interval $\tau_0$. If there is no short-term memory, $P_q(\tau|\tau_0)$ is independent of $\tau_0$. However, it is hard to determine $P_q(\tau|\tau_0)$ for a single value of $\tau_0$ since the size of the interval sample is not sufficiently large. We thus adopt an alternative approach, which employs the idea of coarse graining.

For a given $q$, the set $\mathbf{T}$ of all recurrence intervals is partitioned into four non-overlapping subsets:
\begin{equation}
  \mathbf{T} = {\mathbf{T}}_1 \cup {\mathbf{T}}_2  \cup {\mathbf{T}}_3  \cup {\mathbf{T}}_4,
  \label{Eq:RI:T1:T4}
\end{equation}
where ${\mathbf{T}}_i \cap {\mathbf{T}}_j =\Phi$ for $i\neq j$. In the partitioning procedure, all recurrence intervals in $\mathbf{T}$ are sorted in an increasing order and then assigned in turn into ${\mathbf{T}}_1$, ${\mathbf{T}}_2$, ${\mathbf{T}}_3$ and ${\mathbf{T}}_4$ such that their sizes are approximately identical. We estimate the empirical conditional probability density functions $P_q(\tau|{\mathbf{T}}_i)=P_q(\tau|\tau_0\in{\mathbf{T}}_i)$ of recurrence intervals that immediately follow a recurrence interval $\tau_0$ belonging to ${\mathbf{T}}_i$. If recurrence intervals have no short-term memory, we should find that $P_q(\tau|{\mathbf{T}}_i)=P_q(\tau|{\mathbf{T}}_j)$ for any $i \neq j$.

\begin{figure}[htb]
  \centering
  \includegraphics[width=4.3cm]{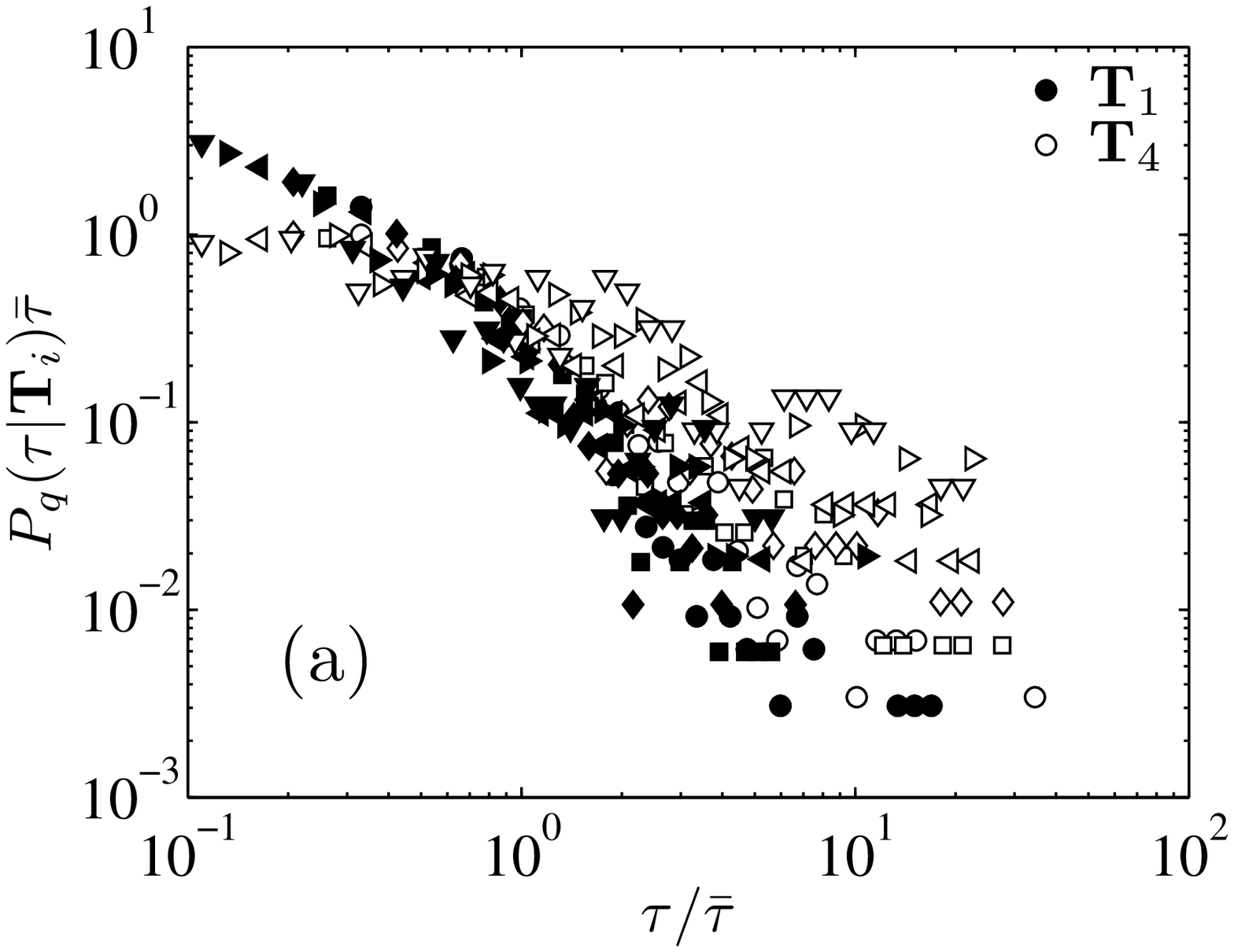}
  \includegraphics[width=4.3cm]{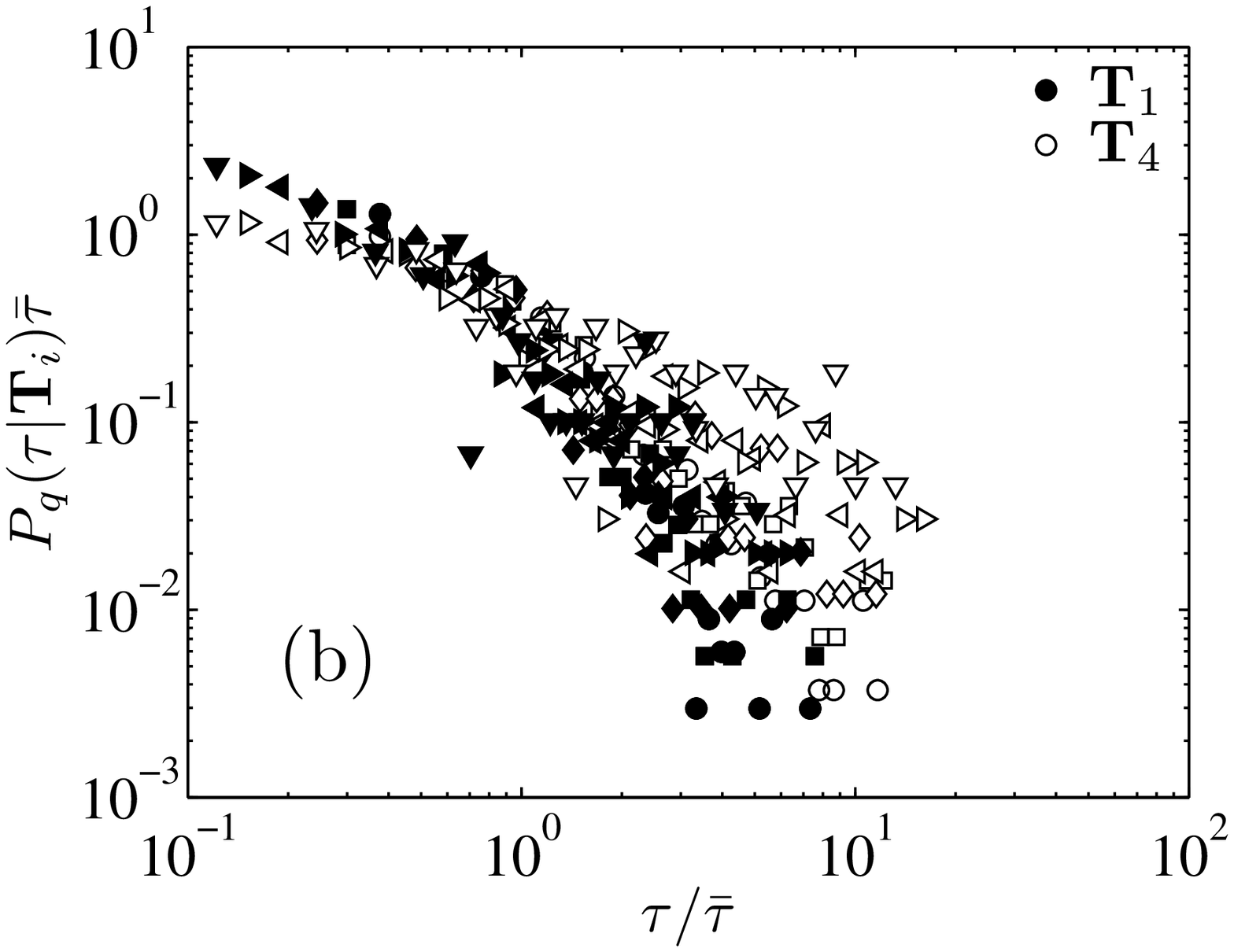}
  \includegraphics[width=4.3cm]{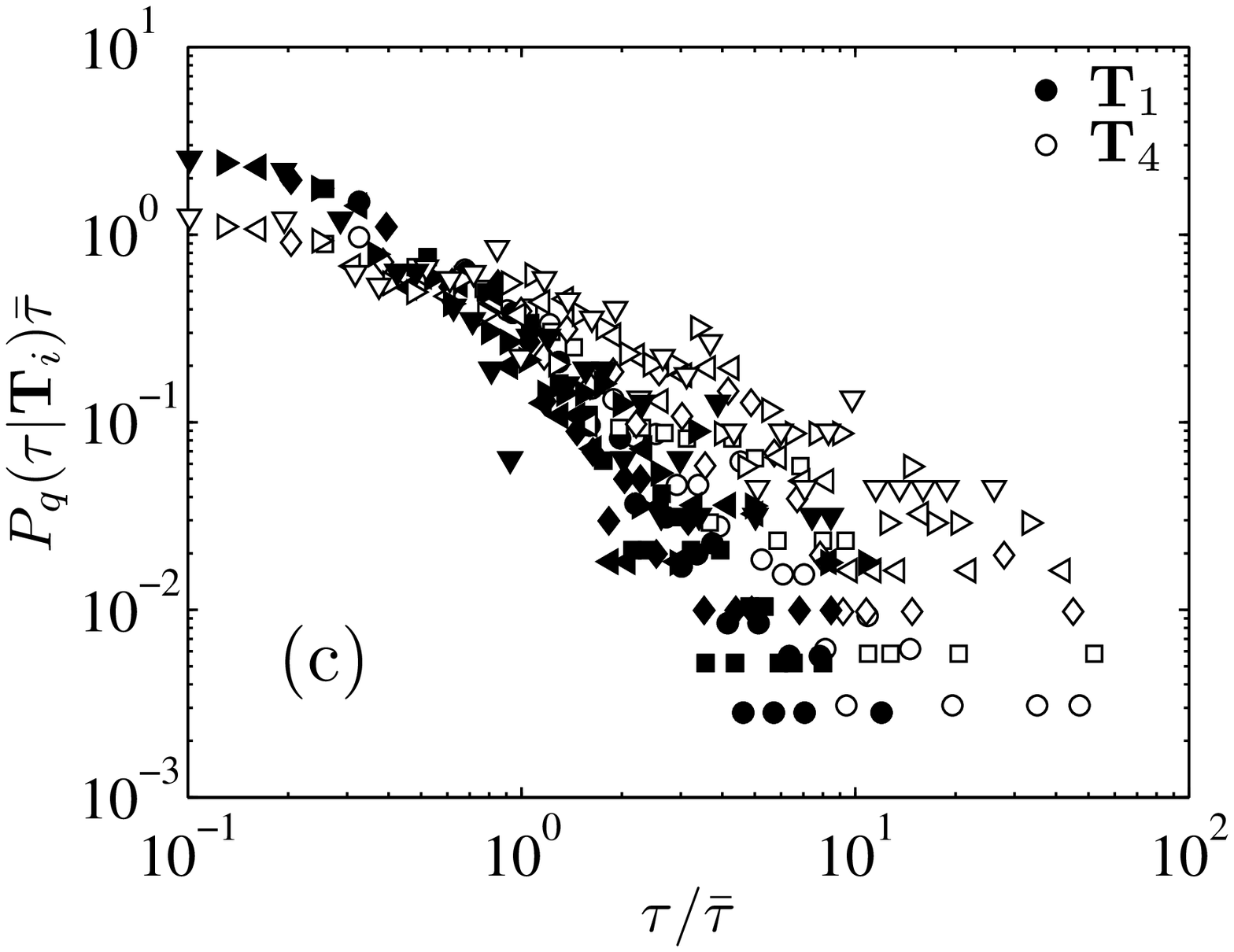}
  \includegraphics[width=4.3cm]{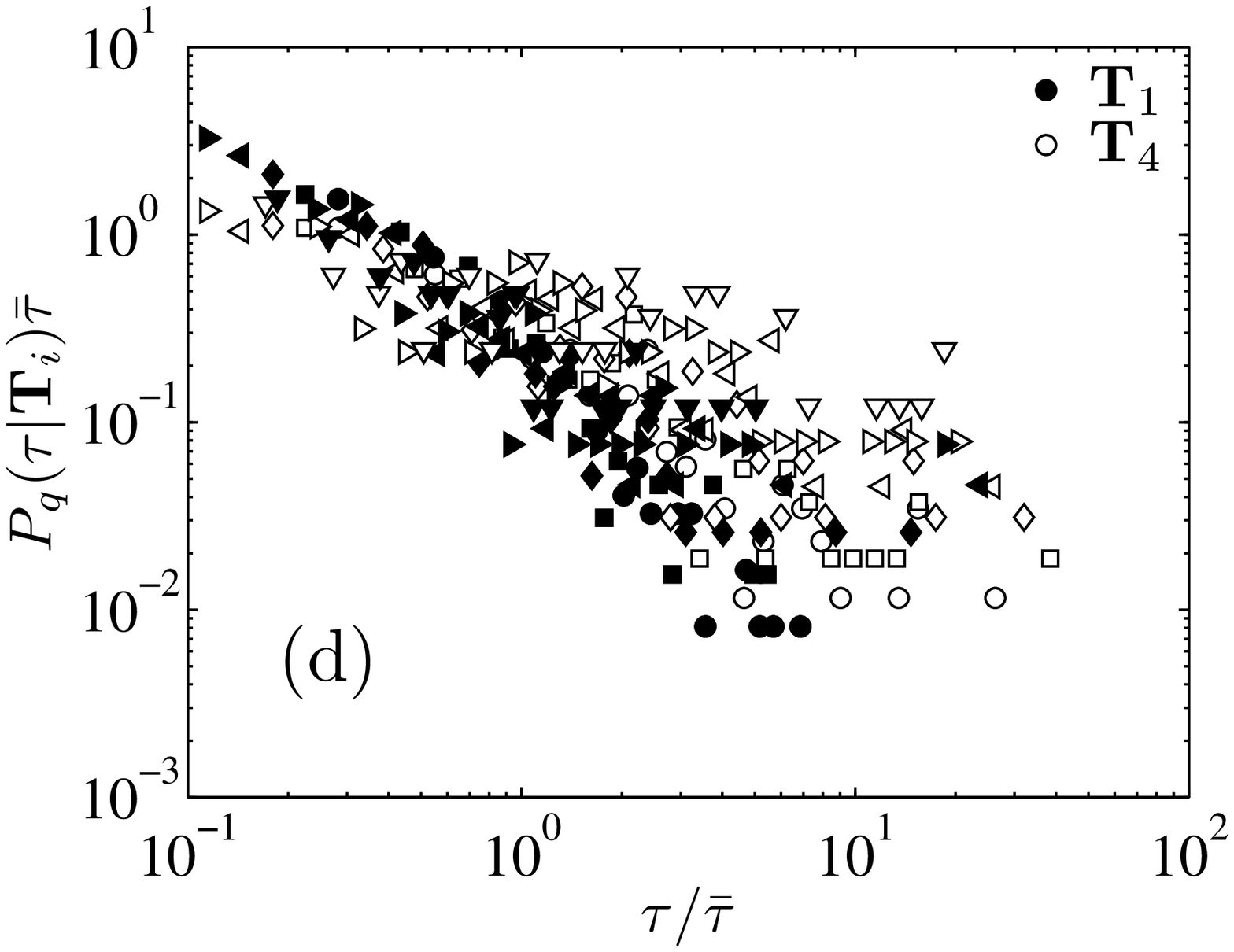}
  \caption{\label{Fig:NYMEX:Energy:RI:CondProb} Conditional distributions $P_q(\tau|{\mathbf{T}}_i)$ of recurrence intervals in the smallest set $\mathbf{T}_1$ (filled symbols) and the largest set $\mathbf{T}_4$ (open symbols) for different thresholds $q$ for the four energy futures: (a) Crude oil, (b) gasoline, (c) heating oil, and (d) propane.}
\end{figure}

The results for $\mathbf{T}_1$ and $\mathbf{T}_4$ are illustrated in Fig.~\ref{Fig:NYMEX:Energy:RI:CondProb}. We find that there is no scaling in the distributions $P_q(\tau|{\mathbf{T}}_i)$ in regard to different $q$ value for any ${\mathbf{T}}_i$ set. It is also evident that $P_q(\tau|{\mathbf{T}}_1){\neq}P_q(\tau|{\mathbf{T}}_4)$ and small (large) recurrence intervals are more probably followed by small (large) recurrence intervals, indicating that there is short-term memory in the recurrence interval time series.

\begin{figure}[htb]
\centering
\includegraphics[width=4.3cm]{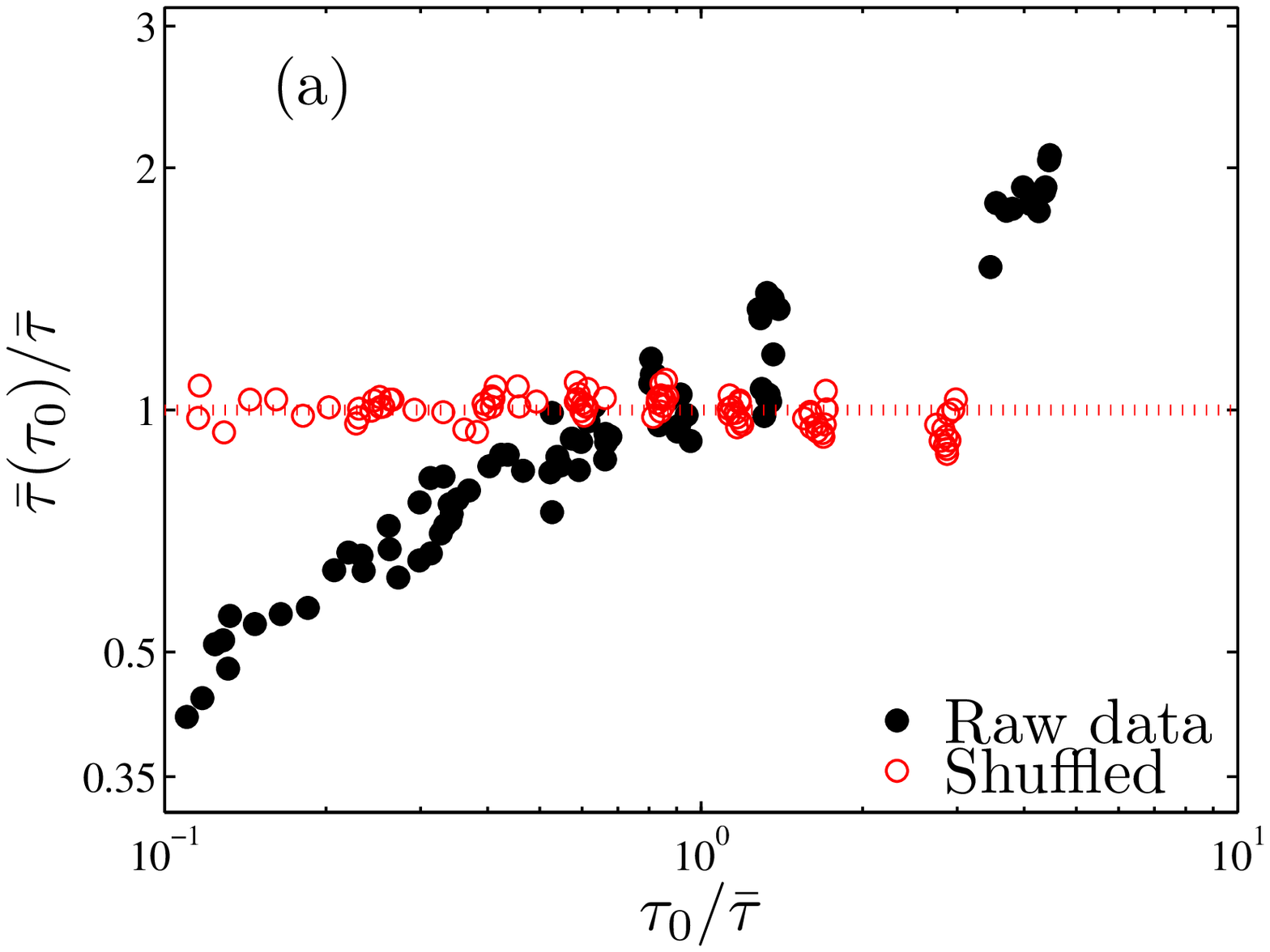}
\includegraphics[width=4.3cm]{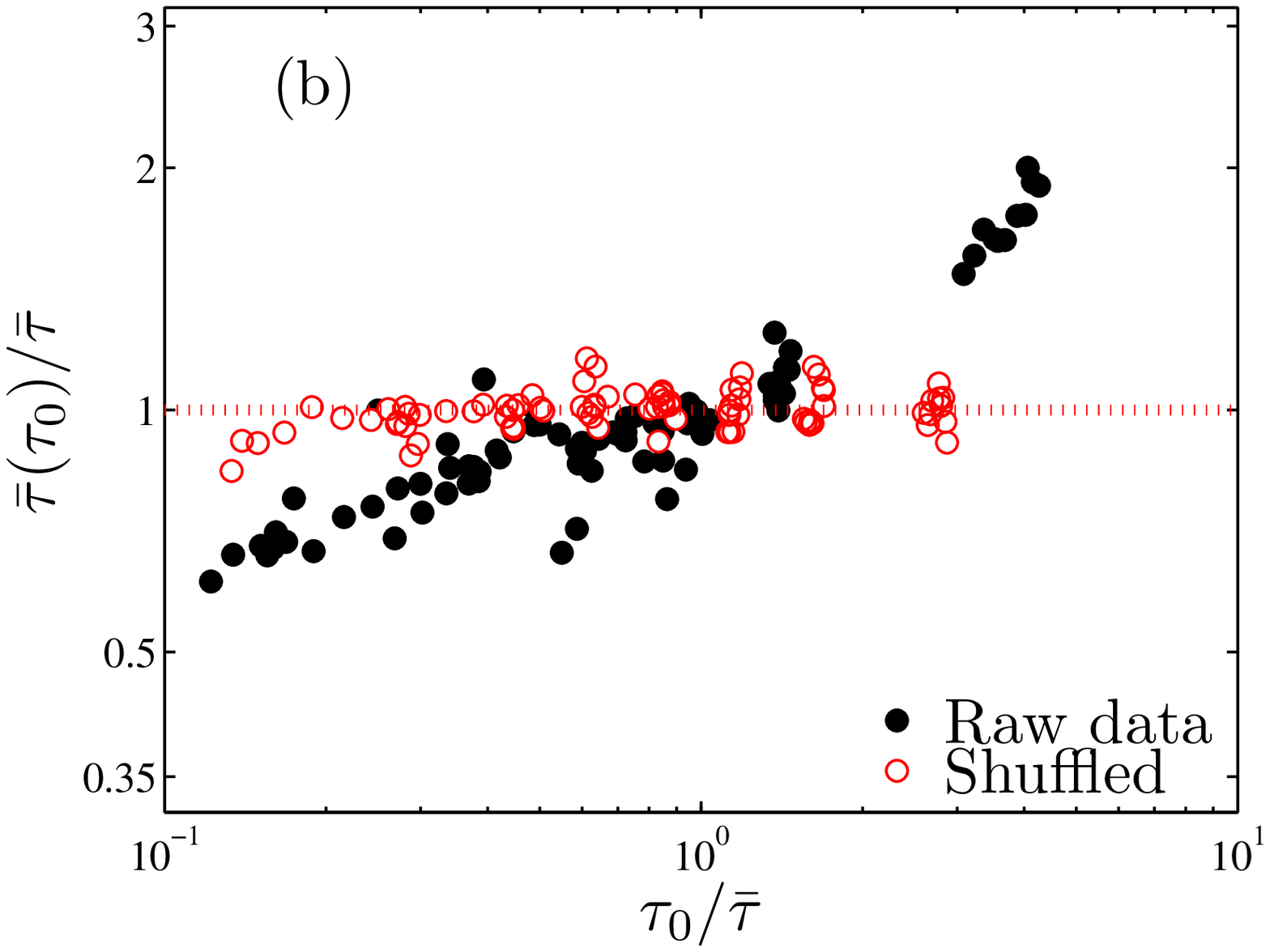}
\includegraphics[width=4.3cm]{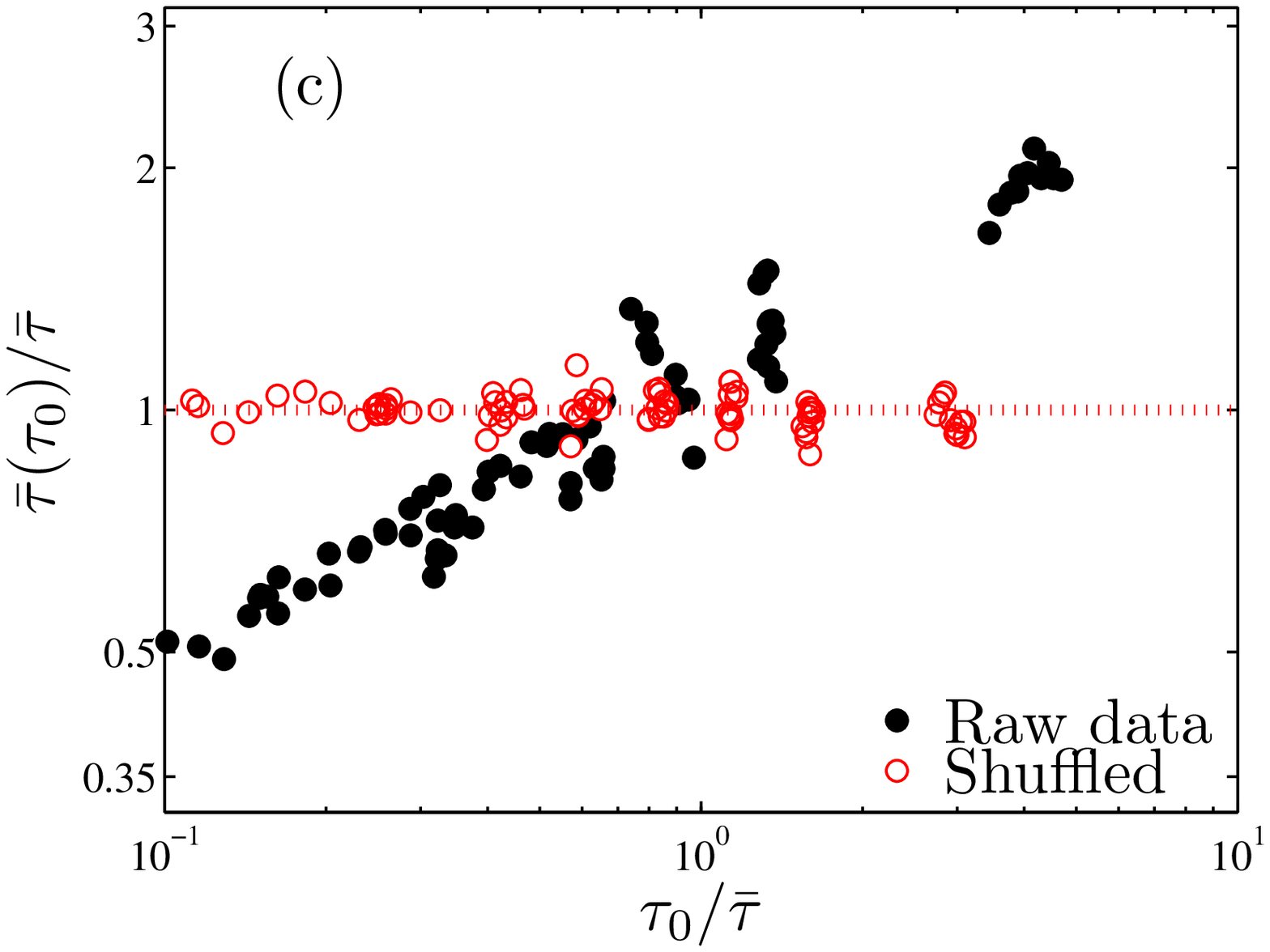}
\includegraphics[width=4.3cm]{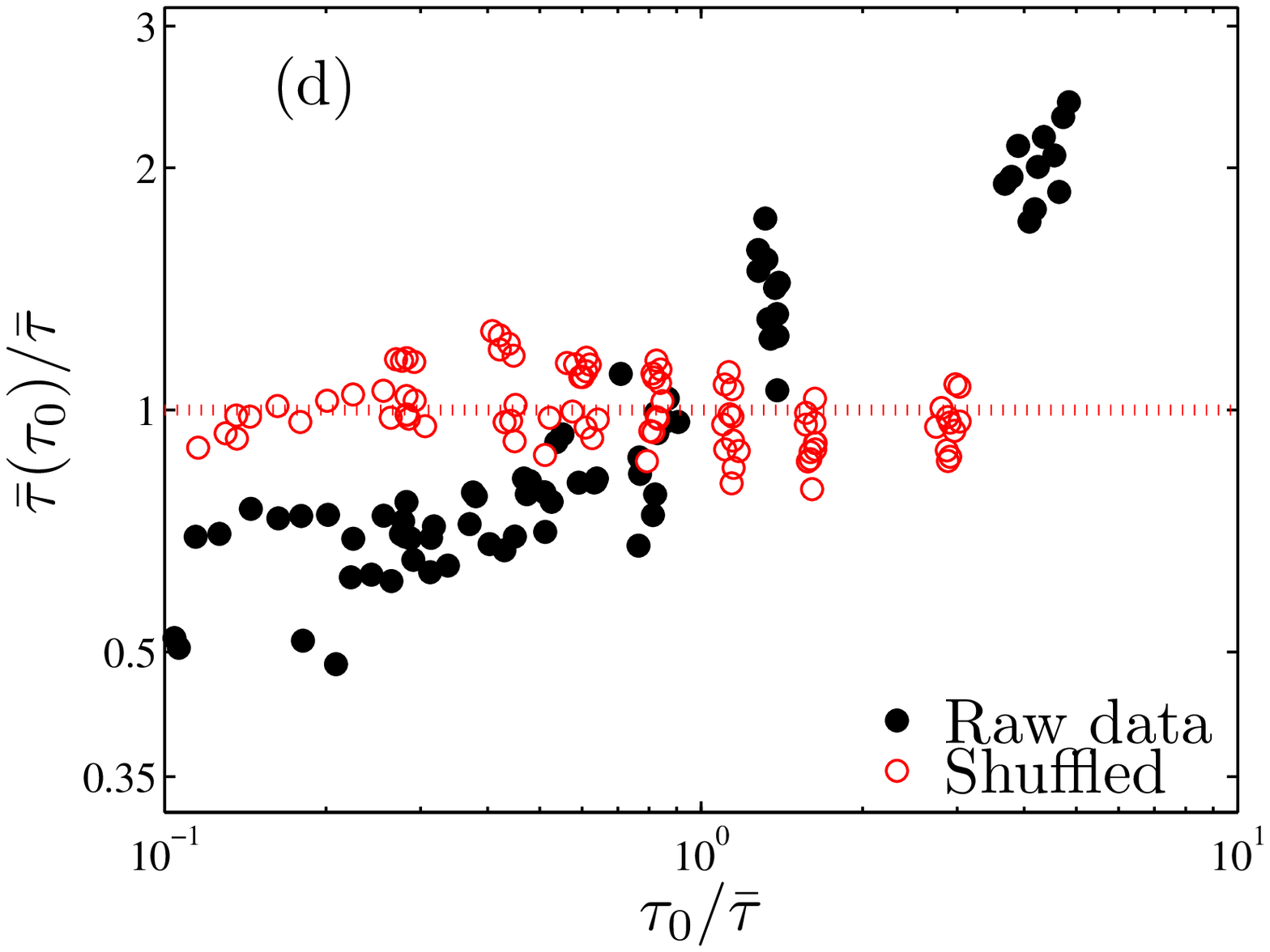}
\caption{\label{Fig:NYMEX:Energy:RI:CondMean} (Color online.) Scaled conditional means $\bar\tau(\tau_0)/\bar\tau$ of recurrence intervals in eight subsets for different thresholds $q$ for the four energy futures: (a) Crude oil, (b) gasoline, (c) heating oil, and (d) propane. Black filled circles and red open circles correspond respectively to the results of the original volatility sereis and the shuffled volatility series.}
\end{figure}

We further investigate the mean of the conditional recurrence intervals $\bar\tau(\tau_0)$. Each whole sample for a given $q$ is divided into eight subsets and each subset is characterized by its mean recurrence interval $\tau_0$. The conditional mean $\bar\tau(\tau_0)$ is then determined for each subset. For every $q$, the scatter plots of $\bar\tau(\tau_0)/\bar\tau$ against $\tau_0/\bar\tau$ for the four energy futures are shown in Fig.~\ref{Fig:NYMEX:Energy:RI:CondMean}, where $\bar\tau$ is the mean of the whole sample of recurrence intervals for $q$. Hence, each plot in Fig.~\ref{Fig:NYMEX:Energy:RI:CondMean} contains $8\times6$ points, where ``8'' is the number of subsets for each $q$ and ``6'' is the number of $q$ values. We observe a nice power-law dependence
\begin{equation}
  \bar\tau(\tau_0)/\bar\tau \sim (\tau_0/\bar\tau)^\beta
  \label{Eq:RI:CondMean:PL}
\end{equation}
with a positive power-law exponent $\beta$, which also holds for stock volatility. In contrast, when we shuffle the original volatility time series and repeat the above procedure, the conditional means are constant as shown in Fig.~\ref{Fig:NYMEX:Energy:RI:CondMean}, {\textit{i.e.}}, $\bar\tau(\tau_0) = \bar\tau$ for each $q$. Therefore, Fig.~\ref{Fig:NYMEX:Energy:RI:CondMean} further confirms the presence of short-term correlations in the recurrence intervals.

\subsection{Long-term memory}

To investigate the possible long-term correlations in the recurrence intervals, we adopt the detrended fluctuation analysis (DFA) and the detrending moving average (DMA) analysis, which are regarded as ``The Methods of Choice'' in determining the Hurst index of time series \citep{Shao-Gu-Jiang-Zhou-Sornette-2012-SR}. Indeed, a lot of numerical experiments have unveiled that the performance of the DMA method are comparable to the DFA method with slightly differences under different situations \citep{Xu-Ivanov-Hu-Chen-Carbone-Stanley-2005-PRE,Bashan-Bartsch-Kantelhardt-Havlin-2008-PA,Shao-Gu-Jiang-Zhou-Sornette-2012-SR}.

\begin{figure*}[htb]
\centering
\includegraphics[width=4.5cm]{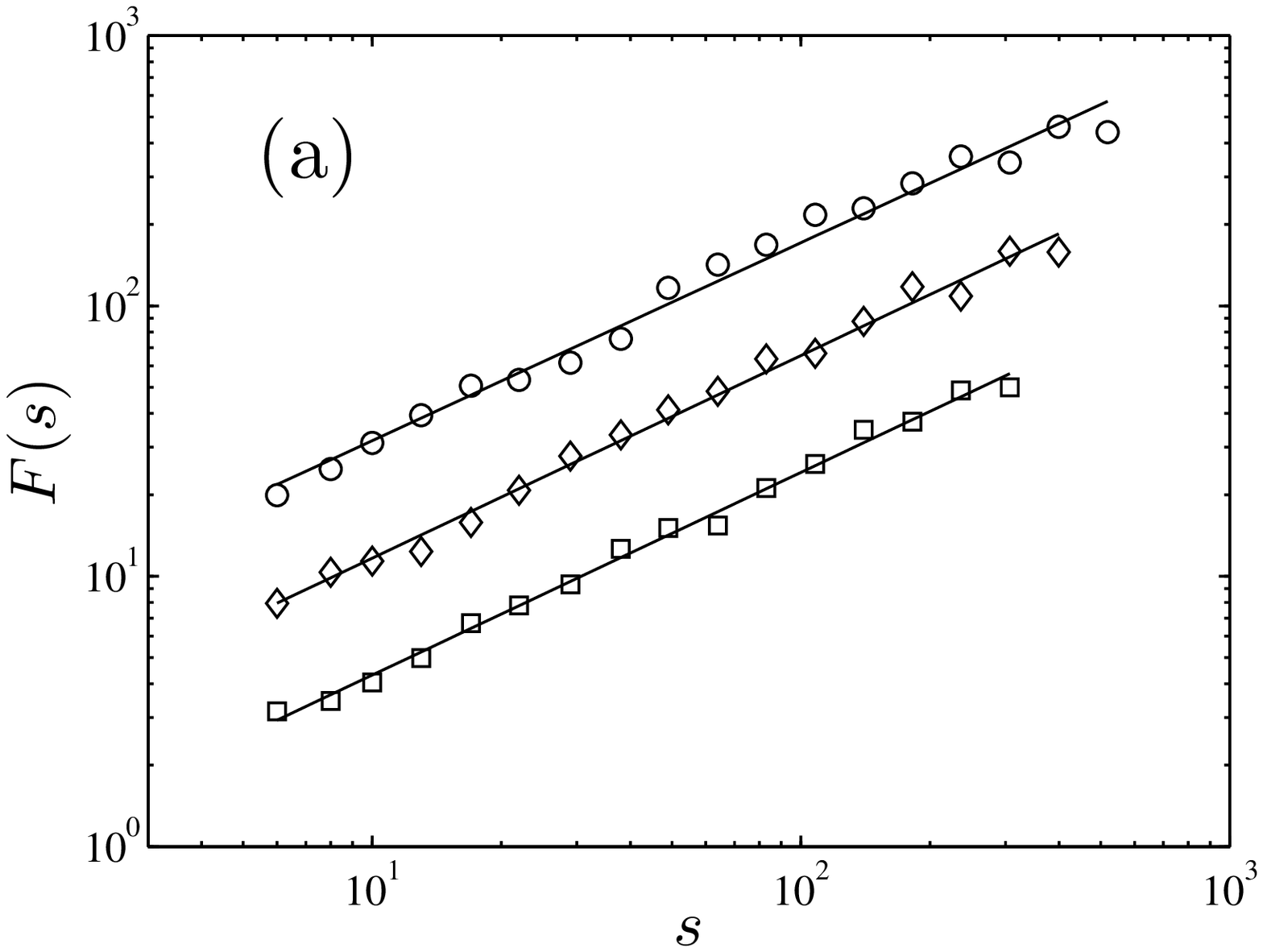}
\includegraphics[width=4.5cm]{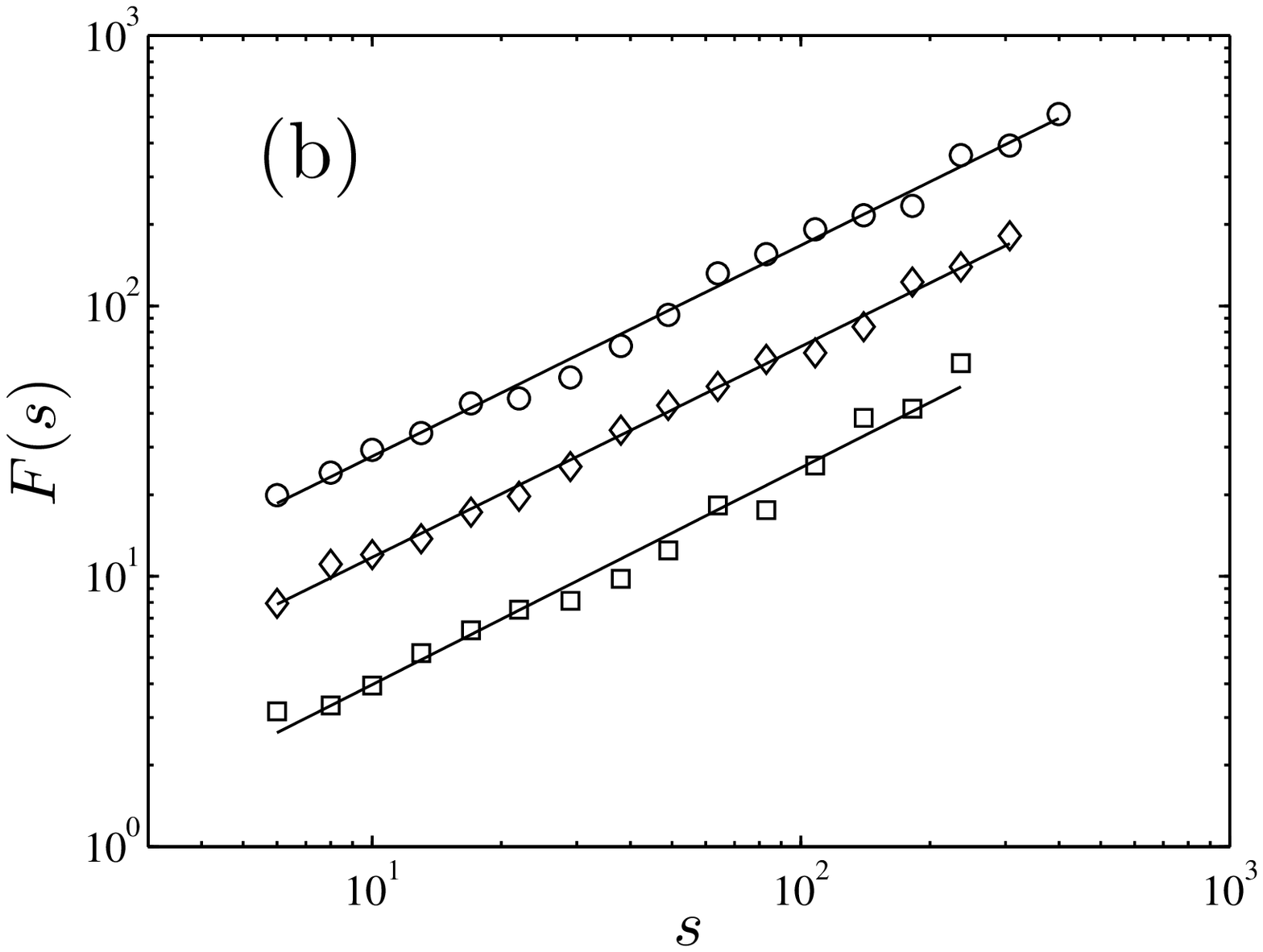}
\includegraphics[width=4.5cm]{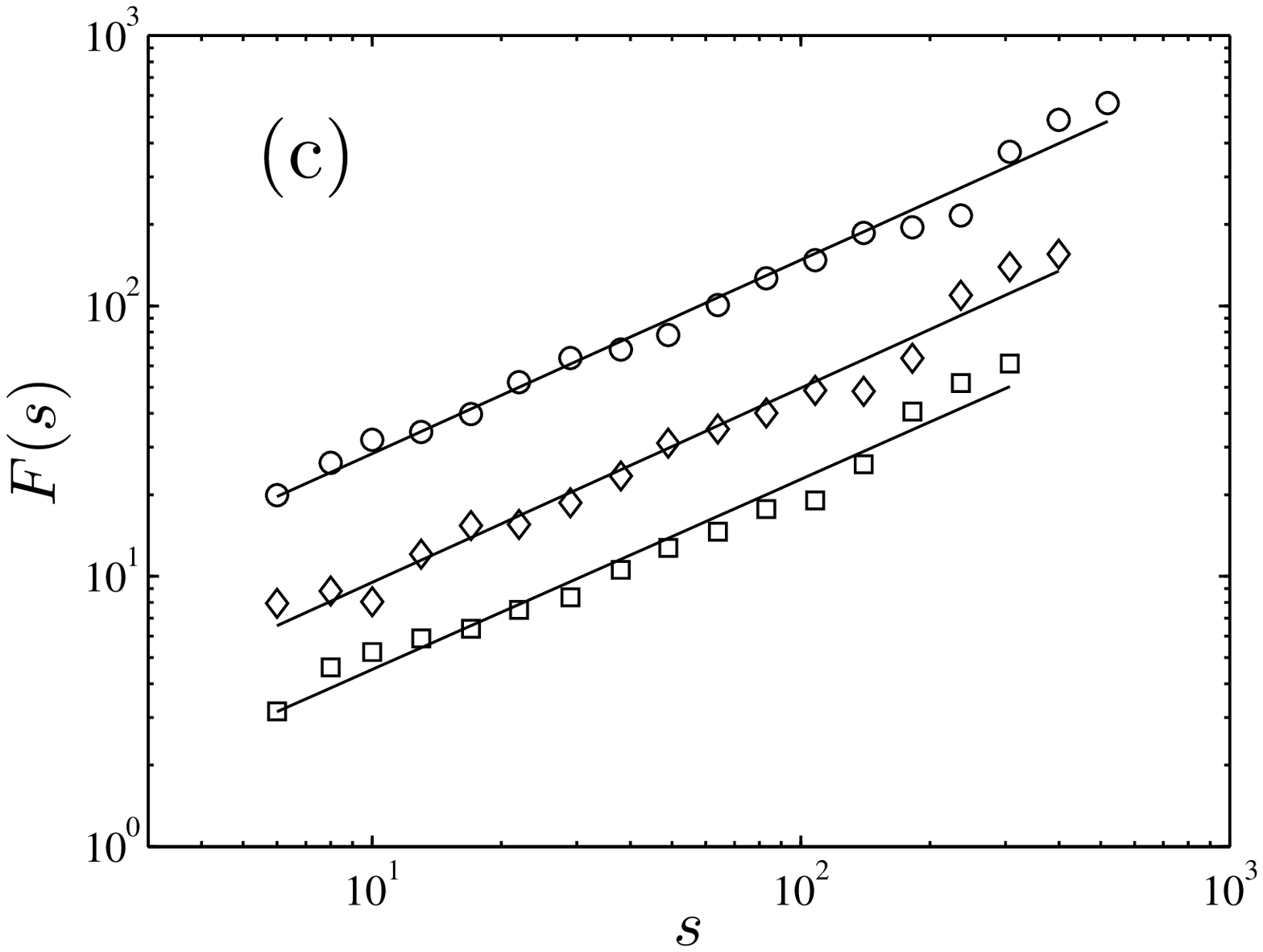}
\includegraphics[width=4.5cm]{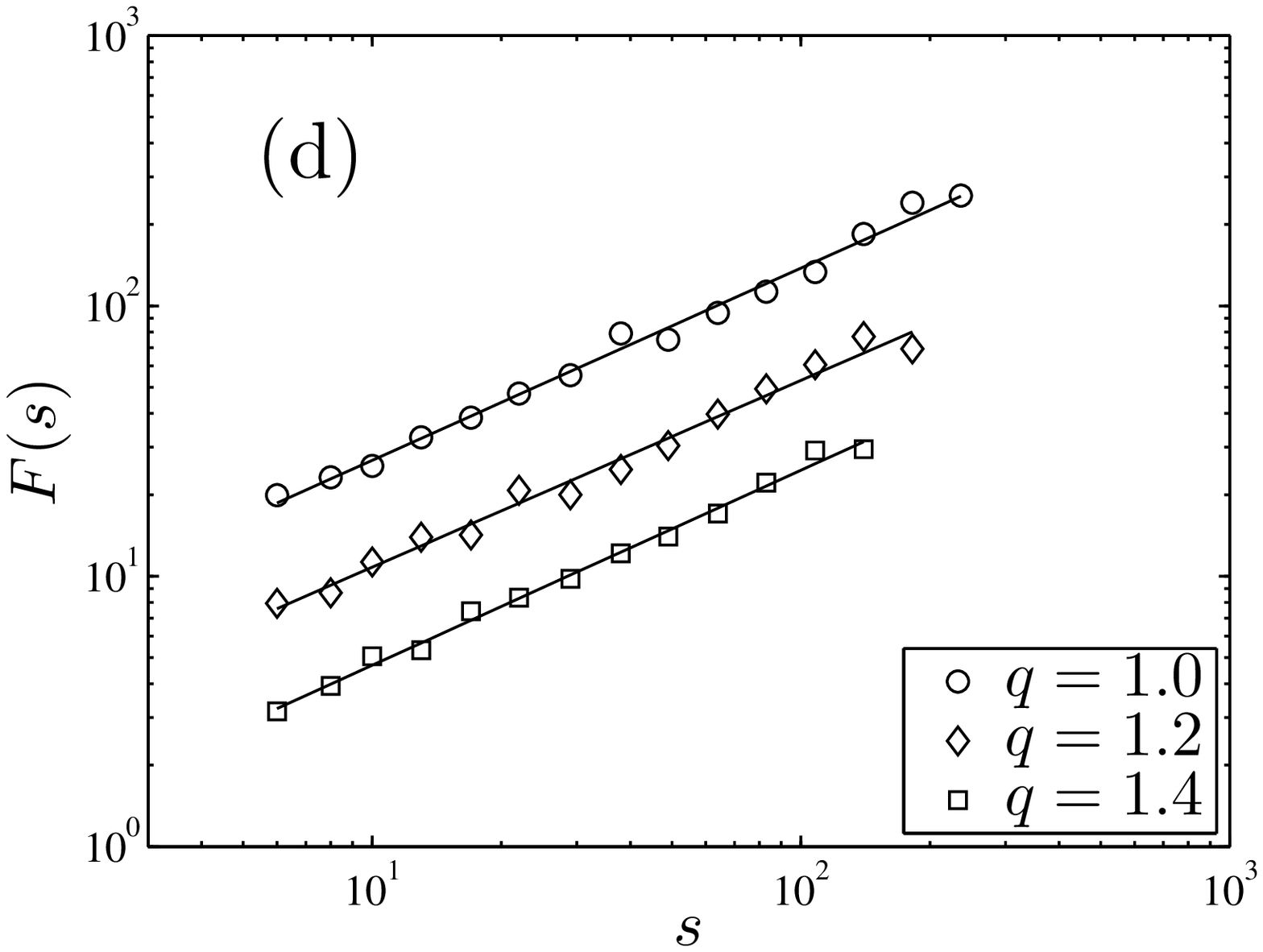}
\includegraphics[width=4.5cm]{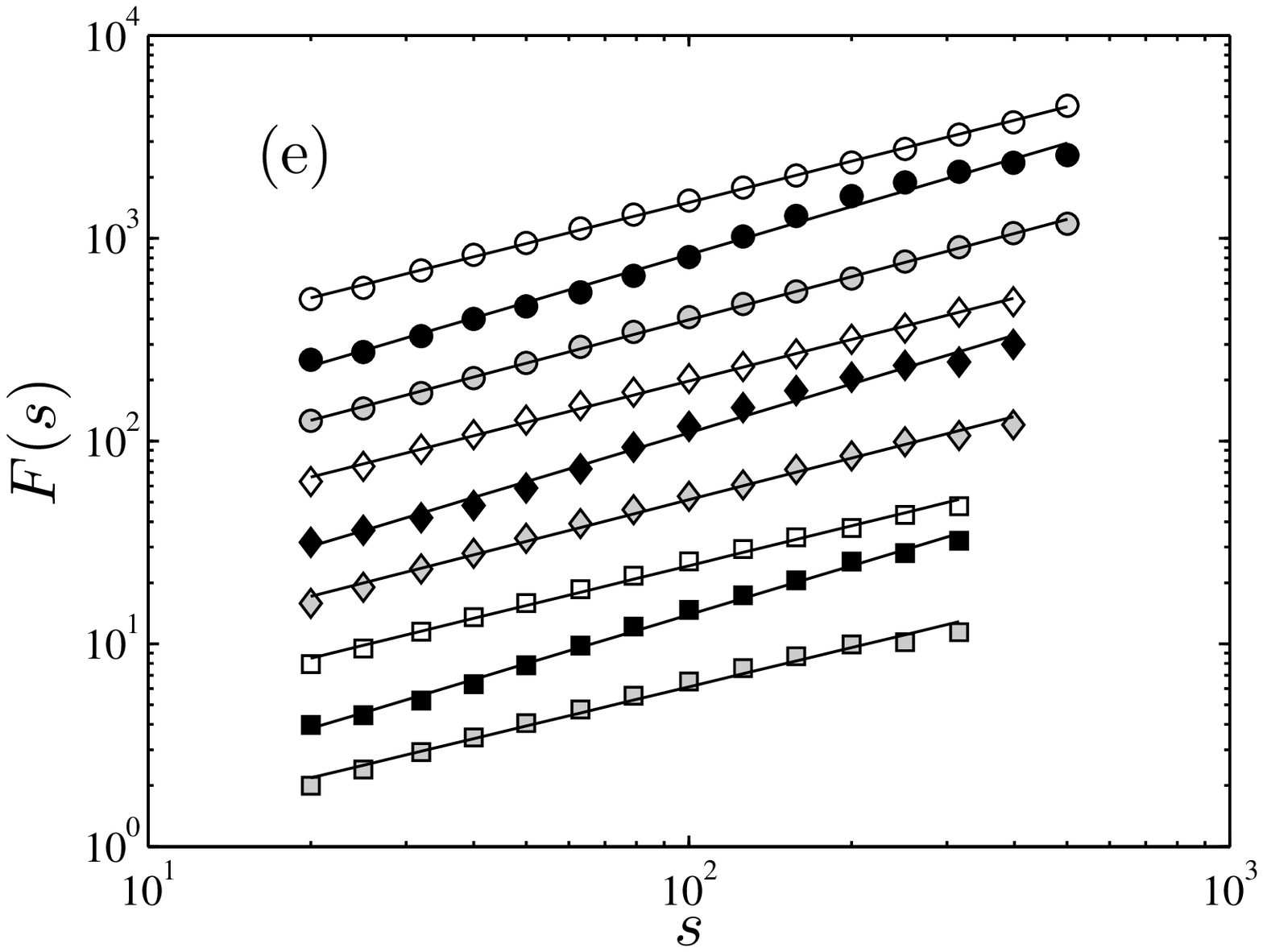}
\includegraphics[width=4.5cm]{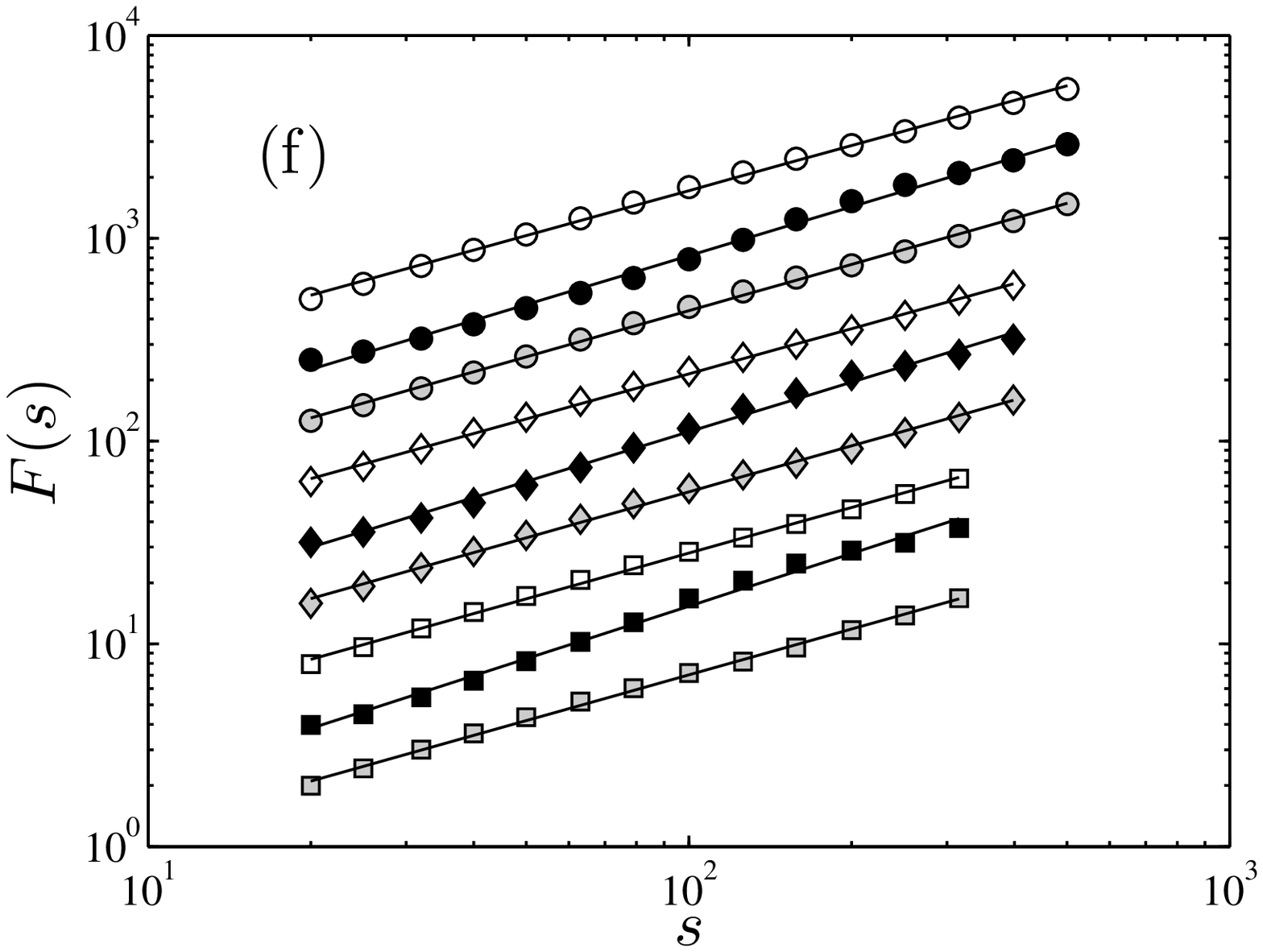}
\includegraphics[width=4.5cm]{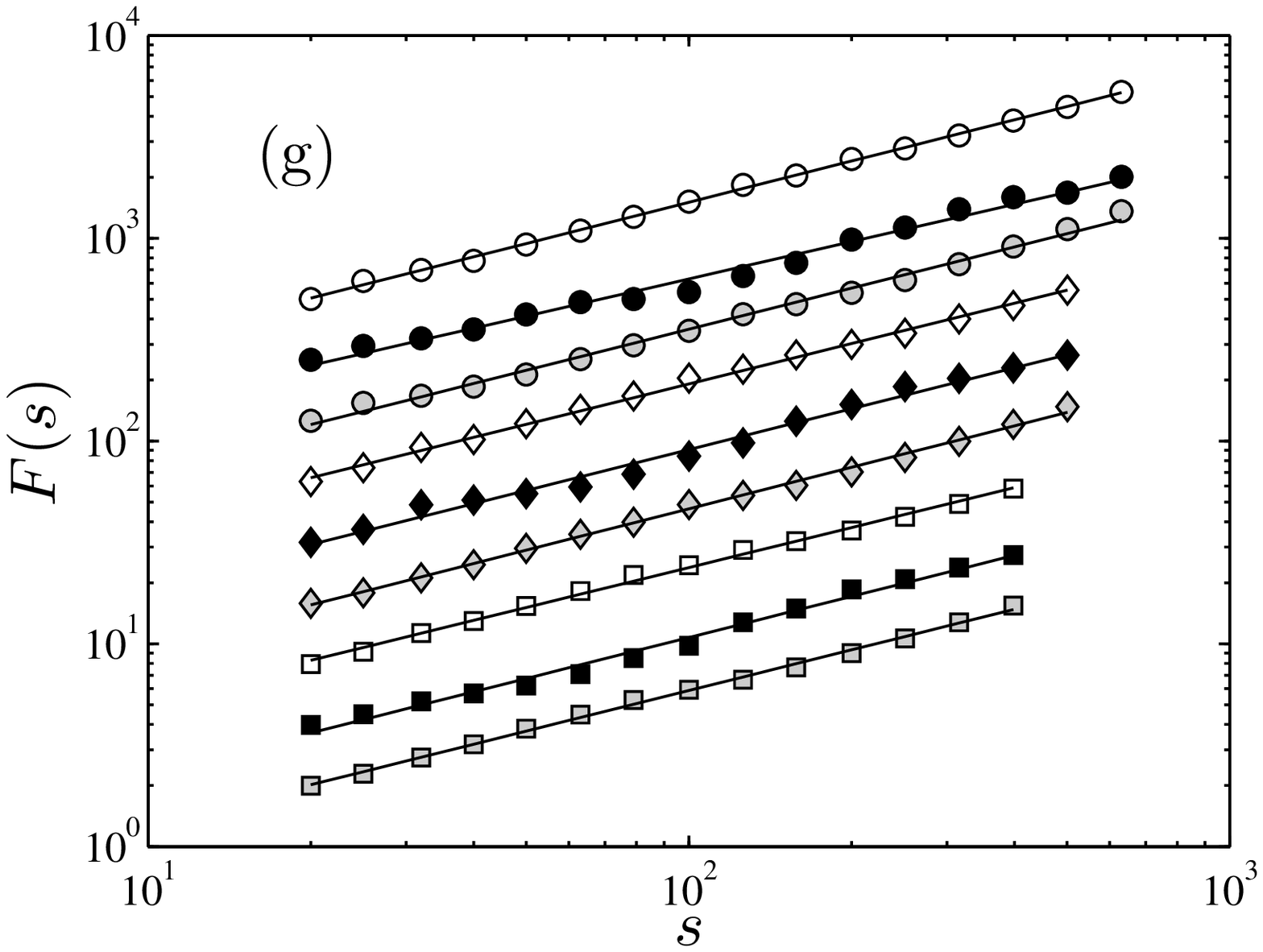}
\includegraphics[width=4.5cm]{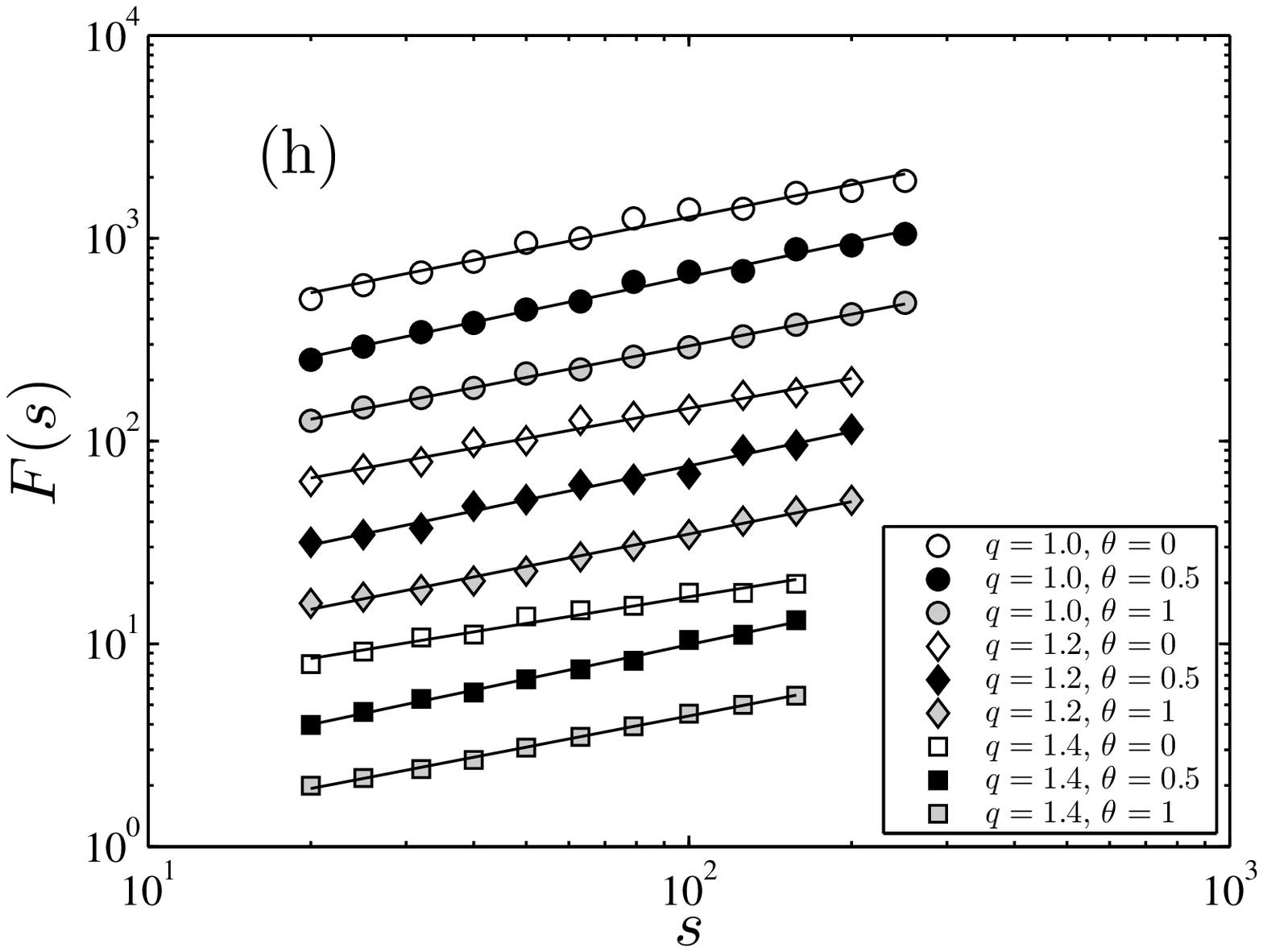}
\caption{\label{Fig:NYMEX:Energy:RI:tau:DFA:DMA} Detrended fluctuation analysis (top panel) and detrending moving average analysis (bottom panel) of the recurrence intervals of energy futures price volatility: (a,e) Crude oil, (b,f) gasoline, (c,g) heating oil, and (d,h) propane. Only three thresholds $q=1.0$, 1.2 and 1.4 are considered, since larger thresholds result in shorter recurrence interval time series. In the plots of the bottom panel, the detrending moving average analyses with $\theta=0$ (backward DMA), $\theta=0.5$ (centred DMA) and $\theta=1$ (forward DMA) have been adopted.}
\end{figure*}

The DFA approach was invented by \cite{Peng-Buldyrev-Havlin-Simons-Stanley-Goldberger-1994-PRE} to investigate the long-range dependence in DNA nucleotide sequences. The properties of DFA have been extensively studied \citep{Hu-Ivanov-Chen-Carpena-Stanley-2001-PRE,Chen-Ivanov-Hu-Stanley-2002-PRE,Chen-Hu-Carpena-BernaolaGalvan-Stanley-Ivanov-2005-PRE,Shang-Lin-Liu-2009-PA,Xu-Shang-Kamae-2009-CSF,Ma-Bartsch-BernaolaGalvan-Yoneyama-Ivanov-2010-PRE}. The DMA approach is based on the moving average technique \citep{Carbone-2009-IEEE} and developed by \cite{Vandewalle-Ausloos-1998-PRE} and \cite{Alessio-Carbone-Castelli-Frappietro-2002-EPJB}. The DMA method has been widely applied to the analysis of real-world time series \citep{Carbone-Castelli-2003-SPIE,Carbone-Castelli-Stanley-2004-PA,Carbone-Castelli-Stanley-2004-PRE,Varotsos-Sarlis-Tanaka-Skordas-2005-PRE,Serletis-Rosenberg-2007-PA,Arianos-Carbone-2007-PA,Matsushita-Gleria-Figueiredo-Silva-2007-PLA,Serletis-Rosenberg-2009-CSF} and synthetic signals \citep{Carbone-Stanley-2004-PA,Xu-Ivanov-Hu-Chen-Carbone-Stanley-2005-PRE,Serletis-2008-CSF}.

The DFA and DMA methods are briefly described below. In the first step, the cumulative summation series $y_i$ is determined as follows
\begin{equation}
  y_i = \sum_{j=1}^{i} \left[\tau_i-\langle{\tau}\rangle\right],~~i = 1, 2, \cdots, N,
  \label{Eq:cumsum}
\end{equation}
where $\langle{r}\rangle$ is the sample mean of the return series. Next, we determine the local trend $\widetilde{y}_i(s)$. The residual sequence $\epsilon_i$ is obtained by removing local trend function $\widetilde{y}_i$ from $y_i$:
\begin{equation}
  \epsilon_i(s)=y_i-\widetilde{y}_i(s).
  \label{Eq:1ddma:epsilon}
\end{equation}
We can then calculate the overall fluctuation function $F(s)$ as follows
\begin{equation}
  \left[F(s)\right]^2 = \frac{1}{N}\sum_{i=1}^{N} \left[\epsilon_i(s)\right]^2.
  \label{Eq:F2:s}
\end{equation}
As the box size $s$ varies in the range of $[20,N/4]$, one can determine the power law relationship between the overall fluctuation function $F(s)$ and the box size $s$,
\begin{equation}
  F(s) \sim s^H,
  \label{Eq:Hurst}
\end{equation}
where $H$ signifies the DFA or DMA scaling exponent. It is necessary to emphasize that the scaling exponent $H$ obtained from Eq.~(\ref{Eq:Hurst}) should be called ``DFA exponent'' or ``DMA exponent'' rather than ``Hurst index'' in the rigorous sense \citep{Jiang-Xie-Zhou-2013-EE}.

The main difference between the DFA and DMA algorithms is the determination of the ``local trend function'' $\widetilde{y}_i$, which is dependent of box size $s$. In the DFA algorithm, the local trend is determined by polynomial fits in boxes \citep{Peng-Buldyrev-Havlin-Simons-Stanley-Goldberger-1994-PRE}. In the DMA approach, one calculates the moving average function $\widetilde{y}_i$ in a moving window \citep{Arianos-Carbone-2007-PA},
\begin{equation}
  \widetilde{y}_i(s)=\frac{1}{s}\sum_{k=-k_{\rm{B}}}^{k_{\rm{F}}}y_{i-k},
  \label{Eq:1ddma:y1}
\end{equation}
where $k_{\rm{B}}=\lfloor(s-1)\theta\rfloor$, $k_{\rm{F}}=\lceil(s-1)(1-\theta)\rceil$, $s$ is the window size, $\lfloor{x}\rfloor$ is the largest integer not greater than $x$, $\lceil{x}\rceil$ is the smallest integer not smaller than $x$, and $\theta$ is the position parameter with the value varying in the range $[0,1]$. Hence, the moving average function considers $\lceil(s-1)(1-\theta)\rceil$ data points in the past and $\lfloor(s-1)\theta\rfloor$ points in the future. Three special cases are adopted in this paper. The first case $\theta=0$ refers to the backward moving average resulting in the backward DMA analysis (BDMA), in which the moving average function $\widetilde{y}_i$ is calculated over all the past $s-1$ data points of the signal. The second case $\theta=0.5$ corresponds to the centered moving average resulting in the centred DMA analysis (CDMA), where $\widetilde{y}_i$ contains half past and half future information in each window. The third case $\theta=1$ is called the forward moving average resulting in the forward DMA analysis (FDMA), where $\widetilde{y}_i$ considers the trend of $s-1$ data points in the future.

\setlength\tabcolsep{5pt}
\begin{table*}[htb]
  \centering 
  \caption{\label{TB:HurstIndex} Estimated DFA and DMA exponents of the recurrence intervals with $q=1.0$, 1.2 and 1.4 for the four futures, which are the slopes of the power laws in Fig.~\ref{Fig:NYMEX:Energy:RI:tau:DFA:DMA} obtained with linear least-squares regressions. $N_q$ is the size of the recurrence interval sample for $q$.}
  \medskip
\begin{tabular}{ccccccccccccccccccc}
  \hline\hline
   &&\multicolumn{3}{c}{Crude oil} && \multicolumn{3}{c}{Gasoline} && \multicolumn{3}{c}{Heating oil} && \multicolumn{3}{c}{Propane}\\  %
     \cline{3-5} \cline{7-9} \cline{11-13} \cline{15-17}
  $q$    && 1.0 & 1.2 & 1.4 &&  1.0 & $1.2$ & $1.4$ &&  $1.0$ & $1.2$ & $1.4$ &&  $1.0$ & $1.2$ & $1.4$ \\
  $N_q$  && 2446 & 1893 & 1498 && 2073 & $1644$ & $1333$ && $2674$ & $2115$ & $1672$ && $1111$ & $884$ & $705$ \\
  \hline
  DFA    &&  0.73   &  0.75   &  0.75   && 0.78   &  0.78   &  0.80   && 0.72   &  0.72   &  0.70   && 0.71   &  0.69   &  0.72  \\
  BDMA   &&  0.67   &  0.68   &  0.65   && 0.74   &  0.74   &  0.75   && 0.68   &  0.66   &  0.66   && 0.53   &  0.49   &  0.43  \\
  CDMA   &&  0.79   &  0.80   &  0.80   && 0.80   &  0.81   &  0.86   && 0.61   &  0.67   &  0.67   && 0.56   &  0.56   &  0.56  \\
  FDMA   &&  0.71   &  0.68   &  0.64   && 0.76   &  0.75   &  0.75   && 0.67   &  0.68   &  0.67   && 0.52   &  0.53   &  0.51  \\
\hline\hline
\end{tabular}
\end{table*}

Figure \ref{Fig:NYMEX:Energy:RI:tau:DFA:DMA} shows the power-law dependence of the fluctuation function $F(s)$ with respect to the box size $s$. Note that only three thresholds $q=1.0$, 1.2 and 1.4 are considered, since larger thresholds result in shorter recurrence interval time series. With the increase of $q$, the length of the recurrence interval time series decreases and the scaling range becomes narrower. A comparison of the four corresponding power-law curves of DFA, BDMA, CDMA and FDMA for each recurrence interval time series unveils that some CDMA curves exhibit low-frequency trends around the power laws, and BDMA and CDMA curves have smaller fluctuations around the power laws than DFA curves. It is noteworthy to mention that crossover phenomena have been observed in the DFA fluctuation functions of recurrence intervals for stocks \citep{Wang-Yamasaki-Havlin-Stanley-2006-PRE,Ren-Gu-Zhou-2009-PA,Ren-Guo-Zhou-2009-PA}, but not for the energy futures studied in this work.

We fit the power-law curves in Fig.~\ref{Fig:NYMEX:Energy:RI:tau:DFA:DMA} using linear least-squares regressions between $\ln{F(s)}$ and $\ln{s}$. The estimated DFA and DMA exponents are presented in Table \ref{TB:HurstIndex}. For each futures contract and for each DFA/DMA estimator, no evident dependence of the scaling exponent $H$ on the threshold $q$ is observed. For crude oil and gasoline, we find roughly that
\begin{equation}
  H_{\rm{CDMA}} > H_{\rm{DFA}} > H_{\rm{BDMA}} \approx H_{\rm{FDMA}}.
  \label{Eq:Hs:CO:GL}
\end{equation}
For heating oil, we have
\begin{equation}
  H_{\rm{DFA}} > H_{\rm{BDFA}} \approx H_{\rm{CDMA}} \approx H_{\rm{FDMA}}.
  \label{Eq:Hs:HO}
\end{equation}
For propane, we find
\begin{equation}
  H_{\rm{DFA}} > H_{\rm{CDFA}} > H_{\rm{BDMA}} \approx H_{\rm{FDMA}}.
  \label{Eq:Hs:PP}
\end{equation}
Despite of these discrepancies, there is no doubt that the recurrence intervals of crude oil, gasoline and heating oil futures possess long-term correlations. However, the results for propane are ambiguous: The DFA approach suggests the presence of long-term correlations, while the DMA methods do not.

We have also performed DFA and DMA on the recurrence interval time series of the shuffled volatility series. We find that the estimated DFA and DMA exponents are close to 0.5. It means that the long-term correlations in the recurrence intervals originate from the long-term correlation in volatility. This finding is the same as for stock and index volatilities \citep{Yamasaki-Muchnik-Havlin-Bunde-Stanley-2005-PNAS}.

\section{Conclusion}
\label{S1:Conclusion}

In this work, we have studies the properties of the recurrence intervals above different thresholds of the daily volatility time series for four NYMEX energy futures (crude oil, gasoline, heating oil and propane), attempting to understand the behaviors of large volatilities. Specifically, the distributions and memory effects of recurrence intervals are investigated.

We found that there is no scaling behavior in the distributions for different thresholds $q$ after the recurrence intervals are scaled with the mean recurrence interval $\bar\tau$, which has been verified by the two-sample Kolmogorov-Smirnov test. We also found that the recurrence intervals are distributed as a stretched exponential $P_q(\tau)\sim e^{(a\tau)^{-\gamma}}$ and the exponent $\gamma$ decreases with increasing $q$, which are significant under the Kolmogorov-Smirnov test and the Cram{\'e}r-von Mises test. We showed that the empirical estimations are in nice agreement with the numerical integration results for the occurrence probability $W_q(\Delta{t}|t)$ of a next large volatility above the threshold $q$ within a short time interval after an elapsed time $t$ from the last large volatility above $q$.

We also investigated the memory effects of the recurrence intervals. We found that the conditional distributions of large and small recurrence intervals differ from each other and the conditional mean of the recurrence intervals scales as a power law of the preceding interval $\bar\tau(\tau_0)/\bar\tau \sim (\tau_0/\bar\tau)^\beta$ with $\beta>0$, indicating that the recurrence intervals have short-term correlations. Detrended fluctuation analysis and detrending moving average analysis further uncover that the recurrence intervals possess long-term correlations. We confirmed that the ``clustering'' of the volatility recurrence intervals is caused by the long-term correlations well known to be present in the volatility.

Our findings shed new lights on the behavior of large volatility and have potential applications in risk estimation of energy futures. Theoretically, these findings have important implications in the pricing and risk management of futures' derivatives such as options.

\section*{Acknowledgements}

This work was partly supported by National Natural Science Foundation of China (Grant No. 11075054), Shanghai Rising Star (Follow-up) Program (Grant No. 11QH1400800), Shanghai ``Chen Guang'' project (Grant No. 2010CG32), and the Fundamental Research Funds for the Central Universities.


\bibliographystyle{elsarticle-harv}
\bibliography{E:/Papers/Auxiliary/Bibliography_FullJournal}

\begin{thebibliography}{81}
\expandafter\ifx\csname natexlab\endcsname\relax\def\natexlab#1{#1}\fi
\expandafter\ifx\csname url\endcsname\relax
  \def\url#1{\texttt{#1}}\fi
\expandafter\ifx\csname urlprefix\endcsname\relax\def\urlprefix{URL }\fi

\bibitem[{Alessio et~al.(2002)Alessio, Carbone, Castelli, and
  Frappietro}]{Alessio-Carbone-Castelli-Frappietro-2002-EPJB}
Alessio, E., Carbone, A., Castelli, G., Frappietro, V., 2002. {Second-order
  moving average and scaling of stochastic time series}. European Physical
  Journal B 27, 197--200.

\bibitem[{Altmann and Kantz(2005)}]{Altmann-Kantz-2005-PRE}
Altmann, E.~G., Kantz, H., 2005. {Recurrence time analysis, long-term
  correlations, and extreme events}. Physical Review E 71, 056106.

\bibitem[{Arianos and Carbone(2007)}]{Arianos-Carbone-2007-PA}
Arianos, S., Carbone, A., 2007. {Detrending moving average algorithm: A
  closed-form approximation of the scaling law}. Physica A 382, 9--15.

\bibitem[{Bashan et~al.(2008)Bashan, Bartsch, Kantelhardt, and
  Havlin}]{Bashan-Bartsch-Kantelhardt-Havlin-2008-PA}
Bashan, A., Bartsch, R., Kantelhardt, J.~W., Havlin, S., 2008. {Comparison of
  detrending methods for fluctuation analysis}. Physica A 387, 5080--5090.

\bibitem[{Bogachev and Bunde(2008)}]{Bogachev-Bunde-2008-PRE}
Bogachev, M.~I., Bunde, A., 2008. {Memory effects in the statistics of
  interoccurrence times between large returns in financial record}. Physical
  Review E 78, 036114.

\bibitem[{Bogachev and Bunde(2009{\natexlab{a}})}]{Bogachev-Bunde-2009-PRE}
Bogachev, M.~I., Bunde, A., 2009{\natexlab{a}}. {Improved risk estimation in
  multifractal records: Application to the value at risk in finance}. Physical
  Review E 80, 026131.

\bibitem[{Bogachev and Bunde(2009{\natexlab{b}})}]{Bogachev-Bunde-2009-EPL}
Bogachev, M.~I., Bunde, A., 2009{\natexlab{b}}. {On the occurrence and
  predictability of overloads in telecommunication networks}. EPL (Europhysics
  Letters) 86, 66002.

\bibitem[{Bogachev and Bunde(2012)}]{Bogachev-Bunde-2012-EPL}
Bogachev, M.~I., Bunde, A., 2012. {Universality in the precipitation and river
  runoff}. EPL (Europhysics Letters) 97, 48011.

\bibitem[{Bogachev et~al.(2007)Bogachev, Eichner, and
  Bunde}]{Bogachev-Eichner-Bunde-2007-PRL}
Bogachev, M.~I., Eichner, J.~F., Bunde, A., 2007. {Effect of nonlinear
  correlations on the statistics of return intervals in multifractal data
  sets}. Physical Review Letters 99, 240601.

\bibitem[{Bogachev et~al.(2008)Bogachev, Eichner, and
  Bunde}]{Bogachev-Eichner-Bunde-2008-EPJST}
Bogachev, M.~I., Eichner, J.~F., Bunde, A., 2008. {The effects of
  multifractality on the statistics of return intervals}. European Physical
  Journal - Special Topics 161, 181--193.

\bibitem[{Bogachev et~al.(2009)Bogachev, Kireenkov, Nifontov, and
  Bunde}]{Bogachev-Kireenkov-Nifontov-Bunde-2009-NJP}
Bogachev, M.~I., Kireenkov, I.~S., Nifontov, E.~M., Bunde, A., 2009.
  {Statistics of return intervals between long heartbeat intervals and their
  usability for online prediction of disorders}. New Journal of Physics 11,
  063036.

\bibitem[{Bunde et~al.(2003)Bunde, Eichner, Havlin, and
  Kantelhardt}]{Bunde-Eichner-Havlin-Kantelhardt-2003-PA}
Bunde, A., Eichner, J.~F., Havlin, S., Kantelhardt, J.~W., 2003. {The effect of
  long-term correlations on the return periods of rare events}. Physica A 330,
  1--7.

\bibitem[{Bunde et~al.(2004)Bunde, Eichner, Havlin, and
  Kantelhardt}]{Bunde-Eichner-Havlin-Kantelhardt-2004-PA}
Bunde, A., Eichner, J.~F., Havlin, S., Kantelhardt, J.~W., 2004. {Return
  intervals of rare events in records with long-term persistence}. Physica A
  342, 308--314.

\bibitem[{Bunde et~al.(2005)Bunde, Eichner, Kantelhardt, and
  Havlin}]{Bunde-Eichner-Kantelhardt-Havlin-2005-PRL}
Bunde, A., Eichner, J.~F., Kantelhardt, J.~W., Havlin, S., 2005. {Long-term
  memory: A natural mechanism for the clustering of extreme events and
  anomalous residual times in climate records}. Physical Review Letters 94,
  048701.

\bibitem[{Cai et~al.(2009)Cai, Fu, Zhou, Gu, and
  Zhou}]{Cai-Fu-Zhou-Gu-Zhou-2009-EPL}
Cai, S.-M., Fu, Z.-Q., Zhou, T., Gu, J., Zhou, P.-L., 2009. {Scaling and memory
  in recurrence intervals of Internet traffic}. EPL (Europhysics Letters) 87,
  68001.

\bibitem[{Carbone(2009)}]{Carbone-2009-IEEE}
Carbone, A., 2009. {Detrending moving average algorithm: A brief review}.
  Science and Technology for Humanity (TIC-STH) IEEE, 691--696.

\bibitem[{Carbone and Castelli(2003)}]{Carbone-Castelli-2003-SPIE}
Carbone, A., Castelli, G., 2003. {Scaling properties of long-range correlated
  noisy signals: Appplication to financial markets}. Proceedings of the SPIE
  5114, 406--414.

\bibitem[{Carbone et~al.(2004{\natexlab{a}})Carbone, Castelli, and
  Stanley}]{Carbone-Castelli-Stanley-2004-PRE}
Carbone, A., Castelli, G., Stanley, H.~E., 2004{\natexlab{a}}. {Analysis of
  clusters formed by the moving average of a long-range correlated time
  series}. Physical Review E 69, 026105.

\bibitem[{Carbone et~al.(2004{\natexlab{b}})Carbone, Castelli, and
  Stanley}]{Carbone-Castelli-Stanley-2004-PA}
Carbone, A., Castelli, G., Stanley, H.~E., 2004{\natexlab{b}}. {Time-dependent
  Hurst exponent in financial time series}. Physica A 344, 267--271.

\bibitem[{Carbone and Stanley(2004)}]{Carbone-Stanley-2004-PA}
Carbone, A., Stanley, H.~E., 2004. {Directed self-organized critical patterns
  emerging from fractional Brownian paths}. Physica A 340, 544--551.

\bibitem[{Chen et~al.(2005)Chen, Hu, Carpena, Bernaola-Galvan, Stanley, and
  Ivanov}]{Chen-Hu-Carpena-BernaolaGalvan-Stanley-Ivanov-2005-PRE}
Chen, Z., Hu, K., Carpena, P., Bernaola-Galvan, P., Stanley, H.~E., Ivanov,
  P.~C., 2005. {Effect of nonlinear filters on detrended fluctuation analysis}.
  Physical Review E 71, 011104.

\bibitem[{Chen et~al.(2002)Chen, Ivanov, Hu, and
  Stanley}]{Chen-Ivanov-Hu-Stanley-2002-PRE}
Chen, Z., Ivanov, P.~C., Hu, K., Stanley, H.~E., 2002. {Effect of
  nonstationarities on detrended fluctuation analysis}. Physical Review E 65,
  041107.

\bibitem[{Clauset et~al.(2009)Clauset, Shalizi, and
  Newman}]{Clauset-Shalizi-Newman-2009-SIAMR}
Clauset, A., Shalizi, C.~R., Newman, M. E.~J., 2009. {Power-law distributions
  in empirical data}. SIAM Review 51, 661--703.

\bibitem[{Cunado et~al.(2010)Cunado, Gil-Alana, and Perez~de
  Gracia}]{Cunado-GilAlana-PerezDeGracia-2010-JFutM}
Cunado, J., Gil-Alana, L.~A., Perez~de Gracia, F., 2010. {Persistence in some
  energy futures markets}. Journal of Futures Markets 30, 490--507.

\bibitem[{Darling(1957)}]{Darling-1957-AMS}
Darling, D.~A., 1957. {The Kolmogorov-Smirnov, Cram{\'e}r-von Mises tests}.
  Annals of Mathematical Statistics 28, 823--838.

\bibitem[{Elder and Serletis(2008)}]{Elder-Serletis-2008-RFE}
Elder, J., Serletis, A., 2008. {Long memory in energy futures prices}. Review
  of Financial Economics 17, 146--155.

\bibitem[{Greco et~al.(2008)Greco, Sorriso-Valvo, Carbone, and
  Cidone}]{Greco-SorrisoValvo-Carbone-Cidone-2008-PA}
Greco, A., Sorriso-Valvo, L., Carbone, V., Cidone, S., 2008. {Waiting time
  distributions of the volatility in the Italian MIB30 index: Clustering or
  Poisson functions?} Physica A 387, 4272--4284.

\bibitem[{He and Chen(2011)}]{He-Chen-2011b-PA}
He, L.-Y., Chen, S.-P., 2011. {A new approach to quantify power-law
  cross-correlation and its application to crude oil markets}. Physica A 390,
  3806--3814.

\bibitem[{Hu et~al.(2001)Hu, Ivanov, Chen, Carpena, and
  Stanley}]{Hu-Ivanov-Chen-Carpena-Stanley-2001-PRE}
Hu, K., Ivanov, P.~C., Chen, Z., Carpena, P., Stanley, H.~E., 2001. {Effect of
  trends on detrended fluctuation analysis}. Physical Review E 64, 011114.

\bibitem[{Jeon et~al.(2010)Jeon, Moon, Oh, Yang, and
  Jung}]{Jeon-Moon-Oh-Yang-Jung-2010-JKPS}
Jeon, W., Moon, H.-T., Oh, G., Yang, J.-S., Jung, W.-S., 2010. {Return
  intervals analysis of the Korean stock market}. Journal of the Korean
  Physical Society 56, 922--925.

\bibitem[{Jiang et~al.(2013{\natexlab{a}})Jiang, Xie, Li, Podobnik, Zhou, and
  Stanley}]{Jiang-Xie-Li-Podobnik-Zhou-Stanley-2013-PNAS}
Jiang, Z.-Q., Xie, W.-J., Li, M.-X., Podobnik, B., Zhou, W.-X., Stanley, H.~E.,
  2013{\natexlab{a}}. {Calling patterns in human communication dynamics}.
  Proceedings of the National Academy of Sciences of the USA 110, submitted.

\bibitem[{Jiang et~al.(2013{\natexlab{b}})Jiang, Xie, and
  Zhou}]{Jiang-Xie-Zhou-2013-EE}
Jiang, Z.-Q., Xie, W.-J., Zhou, W.-X., 2013{\natexlab{b}}. {Testing the
  weak-form efficiency of the WTI crude oil futures market}. Energy Economics
  35, submitted.

\bibitem[{Jones and Kaul(1996)}]{Jones-Kaul-1996-JF}
Jones, C.~M., Kaul, G., 1996. {Oil and the stock markets}. Journal of Finance
  51, 463--491.

\bibitem[{Jung et~al.(2008)Jung, Wang, Havlin, Kaizoji, Moon, and
  Stanley}]{Jung-Wang-Havlin-Kaizoji-Moon-Stanley-2008-EPJB}
Jung, W.-S., Wang, F.-Z., Havlin, S., Kaizoji, T., Moon, H.~T., Stanley, H.~E.,
  2008. {Volatility return intervals analysis of the Japanese market}. European
  Physical Journal B 62, 113--119.

\bibitem[{Kaizoji and Kaizoji(2004)}]{Kaizoji-Kaizoji-2004a-PA}
Kaizoji, T., Kaizoji, M., 2004. {Power law for the calm-time interval of price
  changes}. Physica A 336, 563--570.

\bibitem[{Kotz and Nadarajah(2000)}]{Kotz-Nadarajah-2000}
Kotz, S., Nadarajah, S., 2000. {Extreme Value Distributions: Theory and
  Applications}. Imperial College Press, London.

\bibitem[{Laherr{\`e}re and Sornette(1998)}]{Laherrere-Sornette-1998-EPJB}
Laherr{\`e}re, J., Sornette, D., 1998. {Stretched exponential distributions in
  nature and economy: ``Fat tails'' with characteristic scales}. European
  Physical Journal B 2, 525--539.

\bibitem[{Lee et~al.(2006)Lee, Lee, and Rikvold}]{Lee-Lee-Rikvold-2006-JKPS}
Lee, J.~W., Lee, K.~E., Rikvold, P.~A., 2006. {Waiting-time distribution for
  Korean stock-market index KOSPI}. Journal of the Korean Physical Society 48,
  S123--S126.

\bibitem[{Li et~al.(2011)Li, Wang, Havlin, and
  Stanley}]{Li-Wang-Havlin-Stanley-2011-PRE}
Li, W., Wang, F.-Z., Havlin, S., Stanley, H.~E., 2011. {Financial factor
  influence on scaling and memory of trading volume in stock market}. Physical
  Review E 84, 046112.

\bibitem[{Liu et~al.(2009)Liu, Jiang, Ren, and
  Zhou}]{Liu-Jiang-Ren-Zhou-2009-PRE}
Liu, C., Jiang, Z.-Q., Ren, F., Zhou, W.-X., 2009. {Scaling and memory in the
  return intervals of energy dissipation rate in three-dimensional fully
  developed turbulence}. Physical Review E 80, 046304.

\bibitem[{Livina et~al.(2005)Livina, Havlin, and
  Bunde}]{Livina-Havlin-Bunde-2005-PRL}
Livina, V.~N., Havlin, S., Bunde, A., 2005. {Memory in the occurrence of
  earthquakes}. Physical Review Letters 95, 208501.

\bibitem[{Ludescher et~al.(2011)Ludescher, Tsallis, and
  Bunde}]{Ludescher-Tsallis-Bunde-2011-EPL}
Ludescher, J., Tsallis, C., Bunde, A., 2011. {Universal behaviour of
  interoccurrence times between losses in financial markets: An analytical
  description}. EPL (Europhysics Letters) 95, 68002.

\bibitem[{Ma et~al.(2010)Ma, Bartsch, Bernaola-Galv{\'a}n, Yoneyama, and
  Ivanov}]{Ma-Bartsch-BernaolaGalvan-Yoneyama-Ivanov-2010-PRE}
Ma, Q. D.~Y., Bartsch, R.~P., Bernaola-Galv{\'a}n, P., Yoneyama, M., Ivanov,
  P.~C., 2010. {Effect of extreme data loss on long-range correlated and
  anticorrelated signals quantified by detrended fluctuation analysis}.
  Physical Review E 81, 031101.

\bibitem[{Matsushita et~al.(2007)Matsushita, Gleria, Figueiredo, and
  Silva}]{Matsushita-Gleria-Figueiredo-Silva-2007-PLA}
Matsushita, R., Gleria, I., Figueiredo, A., Silva, S.~D., 2007. {Are pound and
  euro the same currency?} Physics Letters A 368, 173--180.

\bibitem[{Meng et~al.(2012)Meng, Ren, Gu, Xiong, Zhang, Zhou, and
  Zhang}]{Meng-Ren-Gu-Xiong-Zhang-Zhou-Zhang-2012-EPL}
Meng, H., Ren, F., Gu, G.-F., Xiong, X., Zhang, Y.-J., Zhou, W.-X., Zhang, W.,
  2012. {Effects of long memory in the order submission process on the
  properties of recurrence intervals of large price fluctuations}. EPL
  (Europhysics Letters) 98, 38003.

\bibitem[{Olla(2007)}]{Olla-2007-PRE}
Olla, P., 2007. {Return times for stochastic processes with power-law scaling}.
  Physical Review E 76, 011122.

\bibitem[{Pearson and Stephens(1962)}]{Pearson-Stephens-1962-Bm}
Pearson, E.~S., Stephens, M.~A., 1962. {The goodness-of-fit tests on $W_{N}^2$
  and $U_{N}^2$}. Biometrika 49, 397--402.

\bibitem[{Peng et~al.(1994)Peng, Buldyrev, Havlin, Simons, Stanley, and
  Goldberger}]{Peng-Buldyrev-Havlin-Simons-Stanley-Goldberger-1994-PRE}
Peng, C.-K., Buldyrev, S.~V., Havlin, S., Simons, M., Stanley, H.~E.,
  Goldberger, A.~L., 1994. {Mosaic organization of DNA nucleotides}. Physical
  Review E 49, 1685--1689.

\bibitem[{Podobnik et~al.(2009)Podobnik, Horvatic, Petersen, and
  Stanley}]{Podobnik-Horvatic-Petersen-Stanley-2009-PNAS}
Podobnik, B., Horvatic, D., Petersen, A.~M., Stanley, H.~E., 2009.
  {Cross-correlations between volume change and price change}. Proceedings of
  the National Academy of Sciences of the USA 106, 22079--22084.

\bibitem[{Qiu et~al.(2008)Qiu, Guo, and Chen}]{Qiu-Guo-Chen-2008-PA}
Qiu, T., Guo, L., Chen, G., 2008. {Scaling and memory effect in volatility
  return interval of the Chinese stock market}. Physica A 387, 6812--6818.

\bibitem[{Ren et~al.(2009{\natexlab{a}})Ren, Gu, and
  Zhou}]{Ren-Gu-Zhou-2009-PA}
Ren, F., Gu, G.-F., Zhou, W.-X., 2009{\natexlab{a}}. {Scaling and memory in the
  return intervals of realized volatility}. Physica A 388, 4787--4796.

\bibitem[{Ren et~al.(2009{\natexlab{b}})Ren, Guo, and
  Zhou}]{Ren-Guo-Zhou-2009-PA}
Ren, F., Guo, L., Zhou, W.-X., 2009{\natexlab{b}}. {Statistical properties of
  volatility return intervals of Chinese stocks}. Physica A 388, 881--890.

\bibitem[{Ren and Zhou(2008)}]{Ren-Zhou-2008-EPL}
Ren, F., Zhou, W.-X., 2008. {Multiscaling behavior in the volatility return
  intervals of Chinese indices}. EPL (Europhysics Letters) 84, 68001.

\bibitem[{Ren and Zhou(2010{\natexlab{a}})}]{Ren-Zhou-2010-NJP}
Ren, F., Zhou, W.-X., 2010{\natexlab{a}}. {Recurrence interval analysis of
  high-frequency financial returns and its application to risk estimation}. New
  Journal of Physics 12, 075030.

\bibitem[{Ren and Zhou(2010{\natexlab{b}})}]{Ren-Zhou-2010-PRE}
Ren, F., Zhou, W.-X., 2010{\natexlab{b}}. {Recurrence interval analysis of
  trading volumes}. Physical Review E 81, 066107.

\bibitem[{Sadorsky(1999)}]{Sadorsky-1999-EE}
Sadorsky, P., 1999. {Oil price shocks and stock market activity}. Energy
  Economics 21, 449--469.

\bibitem[{Saichev and Sornette(2006)}]{Saichev-Sornette-2006-PRL}
Saichev, A., Sornette, D., 2006. {``Universal'' distribution of interearthquake
  times explained}. Physical Review Letters 97, 078501.

\bibitem[{Santhanam and Kantz(2008)}]{Santhanam-Kantz-2008-PRE}
Santhanam, M.~S., Kantz, H., 2008. {Return interval distribution of extreme
  events and long-term memory}. Physical Review E 78, 051113.

\bibitem[{Serletis and Rosenberg(2007)}]{Serletis-Rosenberg-2007-PA}
Serletis, A., Rosenberg, A.~A., 2007. {The Hurst exponent in energy futures
  prices}. Physica A 380, 325--332.

\bibitem[{Serletis and Rosenberg(2009)}]{Serletis-Rosenberg-2009-CSF}
Serletis, A., Rosenberg, A.~A., 2009. {Mean reversion in the US stock market}.
  Chaos, Solitons \& Fractals 40, 2007--2015.

\bibitem[{Serletis(2008)}]{Serletis-2008-CSF}
Serletis, D., 2008. {Effect of noise on fractal structure}. Chaos, Solitons \&
  Fractals 38, 921--924.

\bibitem[{Shang et~al.(2009)Shang, Lin, and Liu}]{Shang-Lin-Liu-2009-PA}
Shang, P.-J., Lin, A.-J., Liu, L., 2009. {Chaotic SVD method for minimizing the
  effect of exponential trends in detrended fluctuation analysis}. Physica A
  388, 720--726.

\bibitem[{Shao et~al.(2012)Shao, Gu, Jiang, Zhou, and
  Sornette}]{Shao-Gu-Jiang-Zhou-Sornette-2012-SR}
Shao, Y.-H., Gu, G.-F., Jiang, Z.-Q., Zhou, W.-X., Sornette, D., 2012.
  {Comparing the performance of FA, DFA and DMA using different synthetic
  long-range correlated time series}. Scientific Reports 2, 835.

\bibitem[{Smirnov(1948)}]{Smirnov-1948-AMS}
Smirnov, N.~V., 1948. {Table for estimating the goodness of fit of empirical
  distributions}. Annals of Mathematical Statistics 19, 279--281.

\bibitem[{Stephens(1964)}]{Stephens-1964-Bm}
Stephens, M.~A., 1964. {The distribution of the goodness-of-fit statistic,
  $U_N^2$. II}. Biometrika 51, 393--397.

\bibitem[{Stephens(1970)}]{Stephens-1970-JRSSB}
Stephens, M.~A., 1970. {Use of the Kolmogorov-Smirnov, Cram{\'e}r-Von Mises and
  related statistics without extensive tables}. Journal of the Royal
  Statistical Society B 32~(1), 115--122.

\bibitem[{Stephens(1974)}]{Stephens-1974-JASA}
Stephens, M.~A., 1974. {EDF statistics for goodness of fit and some
  comparisons}. Journal of the American Statistical Association 69, 730--737.

\bibitem[{Tabak and Cajueiro(2007)}]{Tabak-Cajueiro-2007-EE}
Tabak, B.~M., Cajueiro, D.~O., 2007. {Are the crude oil markets becoming weakly
  efficient over time? A test for time-varying long-range dependence in prices
  and volatility}. Energy Economics 29, 28--36.

\bibitem[{Vandewalle and Ausloos(1998)}]{Vandewalle-Ausloos-1998-PRE}
Vandewalle, N., Ausloos, M., 1998. {Crossing of two mobile averages: A method
  for measuring the roughness exponent}. Physical Review E 58, 6832--6834.

\bibitem[{Varotsos et~al.(2005)Varotsos, Sarlis, Tanaka, and
  Skordas}]{Varotsos-Sarlis-Tanaka-Skordas-2005-PRE}
Varotsos, P.~A., Sarlis, N.~V., Tanaka, H.~K., Skordas, E.~S., 2005. {Some
  properties of the entropy in the natural time}. Physical Review E 71, 032102.

\bibitem[{Wang and Wang(2012)}]{Wang-Wang-2012-CIE}
Wang, F., Wang, J., 2012. {Statistical analysis and forecasting of return
  interval for SSE and model by lattice percolation system and neural network}.
  Computers \& Industrial Engineering 62, 198--205.

\bibitem[{Wang et~al.(2007)Wang, Weber, Yamasaki, Havlin, and
  Stanley}]{Wang-Weber-Yamasaki-Havlin-Stanley-2007-EPJB}
Wang, F., Weber, P., Yamasaki, K., Havlin, S., Stanley, H.~E., 2007.
  {Statistical regularities in the return intervals of volatility}. European
  Physical Journal B 55, 123--133.

\bibitem[{Wang et~al.(2006)Wang, Yamasaki, Havlin, and
  Stanley}]{Wang-Yamasaki-Havlin-Stanley-2006-PRE}
Wang, F.-Z., Yamasaki, K., Havlin, S., Stanley, H.~E., 2006. {Scaling and
  memory of intraday volatility return intervals in stock markets}. Physical
  Review E 73, 026117.

\bibitem[{Wang et~al.(2008)Wang, Yamasaki, Havlin, and
  Stanley}]{Wang-Yamasaki-Havlin-Stanley-2008-PRE}
Wang, F.-Z., Yamasaki, K., Havlin, S., Stanley, H.~E., 2008. {Indication of
  multiscaling in the volatility return intervals of stock markets}. Physical
  Review E 77, 016109.

\bibitem[{Wang et~al.(2009)Wang, Yamasaki, Havlin, and
  Stanley}]{Wang-Yamasaki-Havlin-Stanley-2009-PRE}
Wang, F.-Z., Yamasaki, K., Havlin, S., Stanley, H.~E., 2009. {Multifactor
  analysis of multiscaling in volatility return intervals}. Physical Review E
  79, 016103.

\bibitem[{Xu et~al.(2005)Xu, Ivanov, Hu, Chen, Carbone, and
  Stanley}]{Xu-Ivanov-Hu-Chen-Carbone-Stanley-2005-PRE}
Xu, L.~M., Ivanov, P.~C., Hu, K., Chen, Z., Carbone, A., Stanley, H.~E., 2005.
  {Quantifying signals with power-law correlations: A comparative study of
  detrended fluctuation analysis and detrended moving average techniques}.
  Physical Review E 71, 051101.

\bibitem[{Xu et~al.(2009)Xu, Shang, and Kamae}]{Xu-Shang-Kamae-2009-CSF}
Xu, N., Shang, P.-J., Kamae, S., 2009. {Minimizing the effect of exponential
  trends in detrended fluctuation analysis}. Chaos, Solitons \& Fractals 41,
  311--316.

\bibitem[{Yamasaki et~al.(2005)Yamasaki, Muchnik, Havlin, Bunde, and
  Stanley}]{Yamasaki-Muchnik-Havlin-Bunde-Stanley-2005-PNAS}
Yamasaki, K., Muchnik, L., Havlin, S., Bunde, A., Stanley, H.~E., 2005.
  {Scaling and memory in volatility return intervals in financial markets}.
  Proceedings of the National Academy of Sciences of the USA 102, 9424--9428.

\bibitem[{Yamasaki et~al.(2006)Yamasaki, Muchnik, Havlin, Bunde, and
  Stanley}]{Yamasaki-Muchnik-Havlin-Bunde-Stanley-2006-inPFE}
Yamasaki, K., Muchnik, L., Havlin, S., Bunde, A., Stanley, H.~E., 2006.
  {Scaling and memory in return loss intervals: Application to risk
  estimation}. In: Takayasu, H. (Ed.), {Practical Fruits of Econophysics}.
  Springer-Verlag, Berlin, pp. 43--51.

\bibitem[{Young(1977)}]{Young-1977-JHcCc}
Young, I.~T., 1977. {Proof without prejudice: Use of the Kolmogorov-Smirnov
  test for the analysis of histograms from flow systems and other sources}.
  Journal of Histochemistry \& Cytochemistry 25, 935--941.

\bibitem[{Zhang et~al.(2010)Zhang, Wang, and Shao}]{Zhang-Wang-Shao-2010-ACS}
Zhang, J.-H., Wang, J., Shao, J.-G., 2010. {Finite-range contact process on the
  market return intervals distributions}. Advances in Complex Systems 13,
  643--657.

\end{thebibliography}







\end{document}